\pdfoutput=1

\documentclass[twocolumn]{aastex62}
\received{November 13, 2018}
\accepted{January 15, 2019}

\usepackage{amsmath}
\usepackage{amssymb}
\usepackage{graphicx,graphics}
\usepackage{rotating}
\usepackage{natbib}
\usepackage{color}
\usepackage{comment}
\usepackage{apjfonts}

\usepackage{listings}
\usepackage{hyperref}
\hypersetup
{
    pdfauthor={M.~Kokubo et al.},
    pdftitle={Kokubo et al. 2019},
}

\shorttitle{A Long-Duration Luminous Type IIn Supernova KISS15s}
\shortauthors{M.~Kokubo et al.}

\begin{document}

\title{A Long-Duration Luminous Type IIn Supernova KISS15s: Strong Recombination Lines from the Inhomogeneous Ejecta-CSM Interaction Region and Hot Dust Emission from Newly Formed Dust \footnote{This work is based in part on observations obtained with the Apache Point Observatory 3.5-meter telescope (PI: M.~Kokubo), which is owned and operated by the Astrophysical Research Consortium.}}

\correspondingauthor{Mitsuru Kokubo}
\email{mkokubo@astr.tohoku.ac.jp}

\author{Mitsuru Kokubo}
\altaffiliation{JSPS Fellow}
\affil{Astronomical Institute, Tohoku University, 6-3 Aramaki-Aza-Aoba, Aoba-ku, Sendai, Miyagi 980-8578, Japan}

\author{Kazuma Mitsuda}
\affil{Astronomical Institute, Tohoku University, 6-3 Aramaki-Aza-Aoba, Aoba-ku, Sendai, Miyagi 980-8578, Japan}

\author{Tomoki Morokuma}
\affil{Institute of Astronomy, the University of Tokyo, 2-21-1 Osawa, Mitaka, Tokyo 181-0015, Japan}
\affil{Kavli Institute for the Physics and Mathematics of the Universe (WPI), Institutes for Advanced Study, University of Tokyo, Kashiwa, Chiba 277-8583, Japan}

\author{Nozomu Tominaga}
\affil{Department of Physics, Faculty of Science and Engineering,
Konan University, Kobe, Hyogo 658-8501, Japan}
\affil{Kavli Institute for the Physics and Mathematics of the Universe (WPI), Institutes for Advanced Study, University of Tokyo, Kashiwa, Chiba 277-8583, Japan}

\author{Masaomi Tanaka}
\affil{Astronomical Institute, Tohoku University, 6-3 Aramaki-Aza-Aoba, Aoba-ku, Sendai, Miyagi 980-8578, Japan}

\author{Takashi~J. Moriya}
\affil{Division of Theoretical Astronomy, National Astronomical Observatory of Japan, National Institutes of Natural Sciences, 2-21-1 Osawa, Mitaka, Tokyo 181-8588, Japan}

\author{Peter Yoachim}
\affil{Department of Astronomy, University of Washington, Box 351580, Seattle, WA 98195-1580, USA}

\author{\v{Z}eljko Ivezi\'{c}}
\affil{Department of Astronomy, University of Washington, Box 351580, Seattle, WA 98195-1580, USA}

\author{Shigeyuki Sako}
\affil{Institute of Astronomy, the University of Tokyo, 2-21-1 Osawa, Mitaka, Tokyo 181-0015, Japan}

\author{Mamoru Doi}
\affil{Institute of Astronomy, the University of Tokyo, 2-21-1 Osawa, Mitaka, Tokyo 181-0015, Japan}
\affil{Research Center for the Early Universe, Graduate School of Science, The University of Tokyo, Hongo, 7-3-1, Bunkyo-ku, Tokyo, 113-0033, Japan}



\begin{abstract}

We report the discovery of an SN~1988Z-like type IIn supernova KISS15s found in a low-mass star-forming galaxy at redshift $z=0.038$ during the course of the Kiso Supernova Survey (KISS).
KISS15s shows long-duration optical continuum and emission line light curves, indicating that KISS15s is powered by a continuous interaction between the expanding ejecta and dense circumstellar medium (CSM).
The H$\alpha$ emission line profile can be decomposed into four Gaussians of narrow, intermediate, blue-shifted intermediate, and broad velocity width components, with a full width at half maximum of $\lesssim 100$, $\sim 2,000$, and $\sim14,000$ km~s${}^{-1}$ for the narrow, intermediate, and broad components, respectively.
The presence of the blue-shifted intermediate component, of which the line-of-sight velocity relative to the systemic velocity is about $-5,000$~km~s${}^{-1}$, suggests that the ejecta-CSM interaction region has an inhomogeneous morphology and anisotropic expansion velocity.
We found that KISS15s shows increasing infrared continuum emission, which can be interpreted as hot dust thermal emission of $T \sim 1,200$~K from newly formed dust in a cool, dense shell in the ejecta-CSM interaction region.
The progenitor mass-loss rate, inferred from bolometric luminosity, is $\dot{M} \sim 0.4~M_{\odot}~\text{yr}^{-1}~(v_{w}/40~\text{km}~\text{s}^{-1})$, where $v_{w}$ is the progenitor's stellar wind velocity.
This implies that the progenitor of KISS15s was a red supergiant star or a luminous blue variable that had experienced a large mass-loss in the centuries before the explosion.

\end{abstract}




\section{Introduction}

Type IIn supernovae (SNe IIn) are a rare class of core collapse SNe (CC SNe) \citep[$\sim$ 7\% of all CC SNe;][]{li11,smi11b,shi17} that show strong interactions between SN~ejecta and dense circumstellar medium (CSM) produced by progenitor mass-loss episodes prior to the SN~explosion \citep{che81,sch90}.
SNe IIn are generally characterized by strong, narrow hydrogen and helium emission lines on top of broad- and intermediate-width emission lines, accompanied by slowly fading highly luminous blue continua \citep[e.g.,][]{sch90,ric02}.

Emission line profiles in SNe IIn provide rich information on the SN ejecta and the CSM.
The strong narrow lines associated with SNe IIn ($v\sim 10-100$ km~s${}^{-1}$) are thought to originate from the unshocked CSM outside of the forward shock, which is photoionized by radiation from a cool dense shell (CDS) between forward and reverse shocks \citep[][and references therein]{che94,smi08,smi09,tad13,jae15,chu18}.
The intermediate emission lines ($v\sim$1,000 km~s${}^{-1}$) may originate from the CDS \citep[e.g.,][]{smi08,smi17b} or radiative shock regions of dense clumpy wind \citep{chu94,chu18}, although it can also be produced by the broadening of intrinsically narrow lines by multiple thermal electron scattering in the opaque CSM \citep{chu01,des09}.
The broad emission lines ($v\gtrsim$10,000 km~s${}^{-1}$) may be attributed to shocked and/or unshocked photoionized ejecta \citep{chu94,smi08,smi17b}; however, multiple electron scattering events in the opaque ejecta/CSM can contribute to the production of very broad emission lines \citep{chu01,des09}.
The absence of a P Cygni absorption feature in many of the SNe~IIn indicates that the CSM formed around the progenitors of SNe~IIn is optically thick enough to obscure the underlying SN photosphere expansion \citep[e.g.,][]{cha12}.
It has been suggested that the optical continuum emission of long-duration SNe IIn can be produced from the CDS region; however, there may be a second photosphere in the SN ejecta that can be seen if the CDS develops clumps and/or becomes optically thin \citep{smi08}.

Another interesting observational property of SNe IIn is late-time mid-infrared (MIR) excess emission at $\sim 3-4$~$\mu$m observed in several SNe IIn \citep[][]{fox11,fox13,sza18}.
The MIR emission can readily be identified as hot dust thermal emission at dust temperatures of $T_{d} \sim 1,000$~K, which can be produced from either IR echoes from radiatively heated preexisting dust grains in the unshocked CSM or newly formed dust in the CDS \citep{fox13,gal14,sar18,chu18}.
Therefore, the dust IR emission of SNe IIn can provide further observational constraints on the CSM properties, once the responsible emission mechanism is identified.
To disentangle which of the emission mechanisms is more important for the IR excess at late-time spectral energy distributions of SNe IIn, it is crucial to follow the temporal evolution of the IR light curves from early- to late-time \citep{sza18}.

Because the photometric and spectral properties of SNe IIn inevitably reflect the CSM formed by stellar mass loss episodes of the progenitor stars several hundreds of years before the SN explosions, close examination of SNe IIn enables observational probing of the final stages in the evolution of massive stars \citep[e.g.,][and references therein]{smi17}.
In this paper, we report the discovery of an SN~1988Z-like type IIn supernova KISS15s found in a low-mass star-forming galaxy at redshift $z = 0.038$ in the course of the Kiso Supernova Survey (KISS). 
We present 800~days of broad-band optical light curves and optical spectra of KISS15s.
KISS15s has been continuously detected in archival Near-Earth Object Wide-field Infrared Survey Explorer (NEOWISE) W1- and W2-band MIR data since $\sim 7$~days after the first optical detection.
Thus, NEOWISE data enable one to follow the dust IR emission light curves of KISS15s from the very early epochs.

The remainder of this paper is organized as follows.
Section~\ref{sec:obs} describes the photometric and spectroscopic observations for KISS15s.
Section~\ref{sec:result} presents detailed analyses of the broadband light curves and emission line profiles and comparisons of the observed properties with known SNe IIn in the literature.
In Section~\ref{sec:discussion}, we discuss implications for the progenitor and CSM properties of KISS15s derived from analyses of the light curves and spectra.
Conclusions are summarized in Section~\ref{sec:conclusions}.

\section{Observations and data reduction}
\label{sec:obs}

We describe the optical-IR broad-band photometric observations for KISS15s in Section~\ref{sec:broadband_obs}, which includes optical data from Kiso/Kiso Wide Field Camera (KWFC) Section~\ref{kiss15s_phot}), SkyMapper (Section~\ref{sec:skymapper}), Panoramic Survey Telescope and Rapid Response System (Pan-STARRS) (Section~\ref{sec:panstarrs}), Mayall/Kitt Peak Ohio State Multi-Object Spectrograph (KOSMOS) (Section~\ref{obs:mayall}), and Blanco/DECam (Section~\ref{obs:decam}), and IR data from WISE/NEOWISE (Section~\ref{obs:wise}) and Nayuta/NIC (Section~\ref{sec:nayuta_nic_obs}).
In Section~\ref{obs:sepc}, we describe optical spectroscopic data from Nayuta/Line Imager and Split Spectrograph (LISS) (Section~\ref{sec:nayuta_spec}) and ARC3.5-m/Dual Imaging Spectrograph (DIS) (Section~\ref{data:dis}).
Finally, we provide an estimate of the dust extinction inside of the host galaxy in Section~\ref{note_on_host_extinction}.

Throughout the paper we assume a Wilkinson Microwave Anisotropy Probe (WMAP) three-year Lambda cold dark matter ($\Lambda$CDM) cosmology with $\Omega_{\Lambda}$ = 0.73, $\Omega_{M}$ = 0.27, and $H_0$ = 73~km~s${}^{-1}~\text{Mpc}^{-1}$ \citep{spe07}.
Barycentric corrections for time and radial velocity were applied using {\tt barycorr} \citep{eas10,wri14,kan18}.
The modified Julian date (MJD) time stamp is expressed as the barycentric MJD (using the barycentric dynamical time scale).

\subsection{Broad-band photometry}
\label{sec:broadband_obs}

\subsubsection{Discovery and follow-up photometry with Kiso/KWFC}
\label{kiss15s_phot}

\begin{figure*}[tbp]
\center{
\includegraphics[clip, width=6.8in]{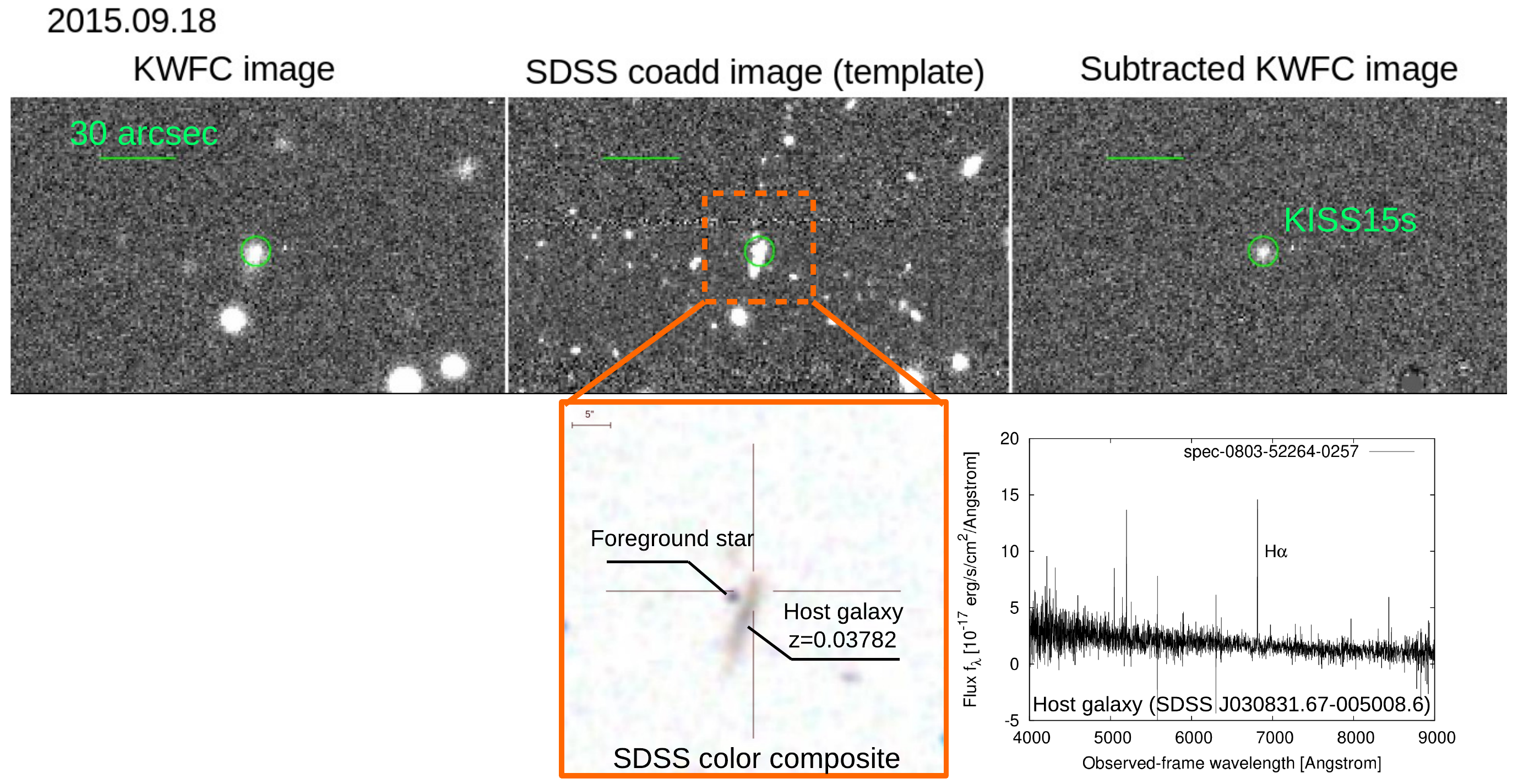}
}
 \caption{Top figures show a KWFC $g$-band image obtained 2015 September 18 (left), the remapped reference SDSS Stripe~82 $g$-band coadd image (middle), and the subtracted KWFC image (right) of KISS15s.
 North is up and east is left.
 The position of KISS15s is indicated by circles.
 Bottom figures show the SDSS color composite image (middle) and the SDSS spectrum of the host galaxy of KISS15s ($z=0.03782$), taken directly from SDSS DR12 SkyServer.
 }
 \label{fig:image1}
\end{figure*}

\begin{deluxetable}{lcccc}
\tablecolumns{5}
\tablewidth{0pc}
\tablecaption{Kiso/KWFC photometry of KISS15s. \label{obslog_kiso}}
\tablehead{
\colhead{MJD} & \colhead{Date} & \colhead{mag.} & \colhead{error in mag.} & \colhead{band}   }
\startdata
57276.636 & 2015-09-11 & 19.65 & 0.08 & $g$ \\
57283.784 & 2015-09-18 & 19.61 & 0.06 & $g$ \\
57308.679 & 2015-10-13 & 19.69 & 0.06 & $g$ \\
\nodata \\
\enddata
\tablecomments{Galactic extinction is uncorrected. The complete listing of
this table is available in the online edition.}
\end{deluxetable}

KISS15s was discovered on 2015 September 18.78 UT (MJD=57283.78) as a $g\sim$19.6 mag transient (Figure~\ref{fig:image1}) during the KISS \citep[][]{morokuma14}, using the 1.05-m Kiso Schmidt Telescope located at Kiso Observatory in Japan; the telescope is equipped with a $2.2{}^{\circ}\times 2.2{}^{\circ}$ WFC \citep[the KWFC;][]{sak12}.
KISS had performed $g$-band observations for selected sky regions within the Sloan Digital Sky Survey (SDSS) Legacy Survey region \citep{yor00}, and searched for optical transients through the point spread function (PSF)-matched image subtraction technique, using SDSS $g$-band mosaic images as the reference frames \citep[see][for details of the KISS pipeline]{morokuma14}.
Re-examination of the previous KWFC images revealed KISS15s detection in the image obtained on 2015 September 11.63 UT (MJD=57276.63), although the PSF profile of KISS15s was partially affected by bad pixels in this first detection image.
After the discovery of KISS15s, we carried out follow-up imaging observations with the Kiso/KWFC for $g$-, $r$-, and $i$-bands.

We noticed that KISS15s is identical to an optical transient, PS15bva, independently discovered by the Pan-STARRS Survey for Transients \citep[][]{kai10} \footnote{\href{https://star.pst.qub.ac.uk/ps1threepi/psdb/}{https://star.pst.qub.ac.uk/ps1threepi/psdb/}}.
The discovery date of PS15bva by Pan-STARRS is 2015 August 30 (MJD=57264.56)\footnote{The record is available at the Open Supernova Catalog \citep{gui17}:\href{https://sne.space/sne/PS15bva/}{https://sne.space/sne/PS15bva/}}, 12~days earlier than the first detection by KISS. 
No spectroscopic follow-up observation for KISS15s = PS15bva has been reported.
This work presents the first spectroscopic classification of this object.

KISS15s is located at 
$\alpha=03{}^{h}08{}^{m}31{}^{s}.640$, $\delta=-00{}^{\circ}50'05.55''$ (J2000)\footnote{KISS15s is recorded in the Gaia public data release 2 \citep{gaia18} as source\_id = 3265523970849514880 at a sky coordinate $\alpha = 47.13178$~deg, $\delta = -0.83495$~deg; however, the Gaia sky position determination of KISS15s may be affected by the underlying host galaxy light.}, $0.53''$ west and $3.05''$ north of the center of the star-forming galaxy
 \href{http://skyserver.sdss.org/dr12/en/tools/explore/Summary.aspx?id=1237663783144259952}{SDSS~J030831.67-005008.6}, at a redshift of $z=$0.03782 measured by the SDSS.
Although the SDSS database assigns a redshift of $z=0.03794$ for a point-source object on the northeast side of SDSS~J030831.67-005008.6 (Figure~\ref{fig:image1}), this object is probably a foreground Galactic star with an SDSS spectrum contaminated by emission lines from SDSS~J030831.67-005008.6.
Because KISS15s is located on the edge of the extended structure of SDSS~J030831.67-005008.6, we identify SDSS~J030831.67-005008.6 as the host galaxy of KISS15s.
Follow-up spectroscopy for KISS15s (described in Section~\ref{obs:sepc}) confirms that the redshift of KISS15s is consistent with the redshift of this galaxy.
A luminosity distance ($d_{L}$) and distance modulus (DM) of the host galaxy corrected for the Virgo infall + Great Attractor + Shapley supercluster local flow is $d_{L} = 156$~Mpc and $\text{DM} = m-M = 35.97$~mag, respectively\footnote{$d_{L}$ and $m-M$ were taken from the NASA/IPAC Extragalactic Database (NED)}.
The angular-spatial size scale is 0.720 kpc/$''$; thus, the galactocentric offset of KISS15s is $\sim$ 2.23 kpc.
The Galactic extinction toward KISS15s was obtained from the NASA/IPAC Extragalactic Database (NED) as $E(B-V)=0.053$ mag \citep[$R_{V} = 3.1$;][]{fit99,sch11}; $A_{u}, A_{g}, A_{r}, A_{i}, A_{z}, A_{J}, A_{H}$, and $A_{K} = 0.255, 0.199, 0.137, 0.102$, $0.076$, $0.043$, $0.027$, and $0.018$~mag, respectively.
The Galactic extinction coefficients of WISE W1- and W2-bands are $A_{\text{W1}}=0.011$ and $A_{\text{W2}}=0.007$~mag, according to the \cite{fit99} extinction curve.

The KISS data reduction pipeline only provides rough estimates of the $g$-band magnitude of the transients.
To extract the $g$-, $r$-, and $i$-band photometric measurements of KISS15s, we reanalyzed the KISS imaging data, starting from the raw FITS files.
Each of the KISS Kiso/KWFC images was obtained with an exposure time of 180~s without dithering.
After applying an overscan, bias, and pixel-flat and illumination-flat corrections \citep[see][for details of the KWFC data reduction]{kok16}, the images were processed by {\tt SExtractor} \citep{ber96} to produce aperture photometry catalogs of field stars around KISS15s.
For each image, the aperture diameter size was set to twice the seeing full width at half maximum (FWHM) size to maximize the signal-to-noise (S/N) ratio of the photometric measurements and
minimize the effects of systematic centering errors \citep[e.g.,][]{mig99}.
Then the photometric zero-point of each image was determined by taking a 3$\sigma$-clipping weighted average of the differences between the instrumental magnitudes in the SExractor catalog and the SDSS magnitudes of field stars retrieved from the SDSS Stripe82 database \citep[see Appendix of][]{ani14}.

Then template matching and image subtraction were applied to the KWFC images, using $WCSRemap$ and High-order Transform of PSF and Template Subtraction ($HOTPANTS$) version 5.1.10 software, written by A.~Becker\footnote{\href{https://github.com/acbecker/hotpants}{https://github.com/acbecker/hotpants}} \citep{bec15}.
SDSS Stripe~82 deep coadd images, constructed by \cite{ani14} from all imaging data obtained on or before 2005 December 1 \citep[][]{yor00,fri08}, were used as pre-SN template (=reference) images.
The SDSS Stripe~82 coadd images are much deeper and have higher resolution than KWFC images \citep[a 50~\% completeness limit for point sources of $23.6, 24.6, 24.2, 23.7$, and $22.3$~mag for the $u$, $g$, $r$, $i$, and $z$ bands, respectively, with a median seeing size of $\sim 1''.1$;][]{ani14}.
$WCSRemap$ was used to produce astrometrically remapped reference images that aligned exactly with individual KWFC images.
Then $HOTPANTS$ was used to align the PSF and flux scale of the reference image with those of the KWFC image and to subtract the reference image from the KWFC image.
SExtractor aperture photometry for KISS15s was applied to the subtracted KWFC images.
It should be noted that because the host galaxy light is almost completely removed from the subtracted KWFC images, the photometric measurements of KISS15s derived here, as described, are free from contamination of the underlying host galaxy flux.

\begin{figure*}[tbp]
\center{
\includegraphics[clip, width=5.3in]{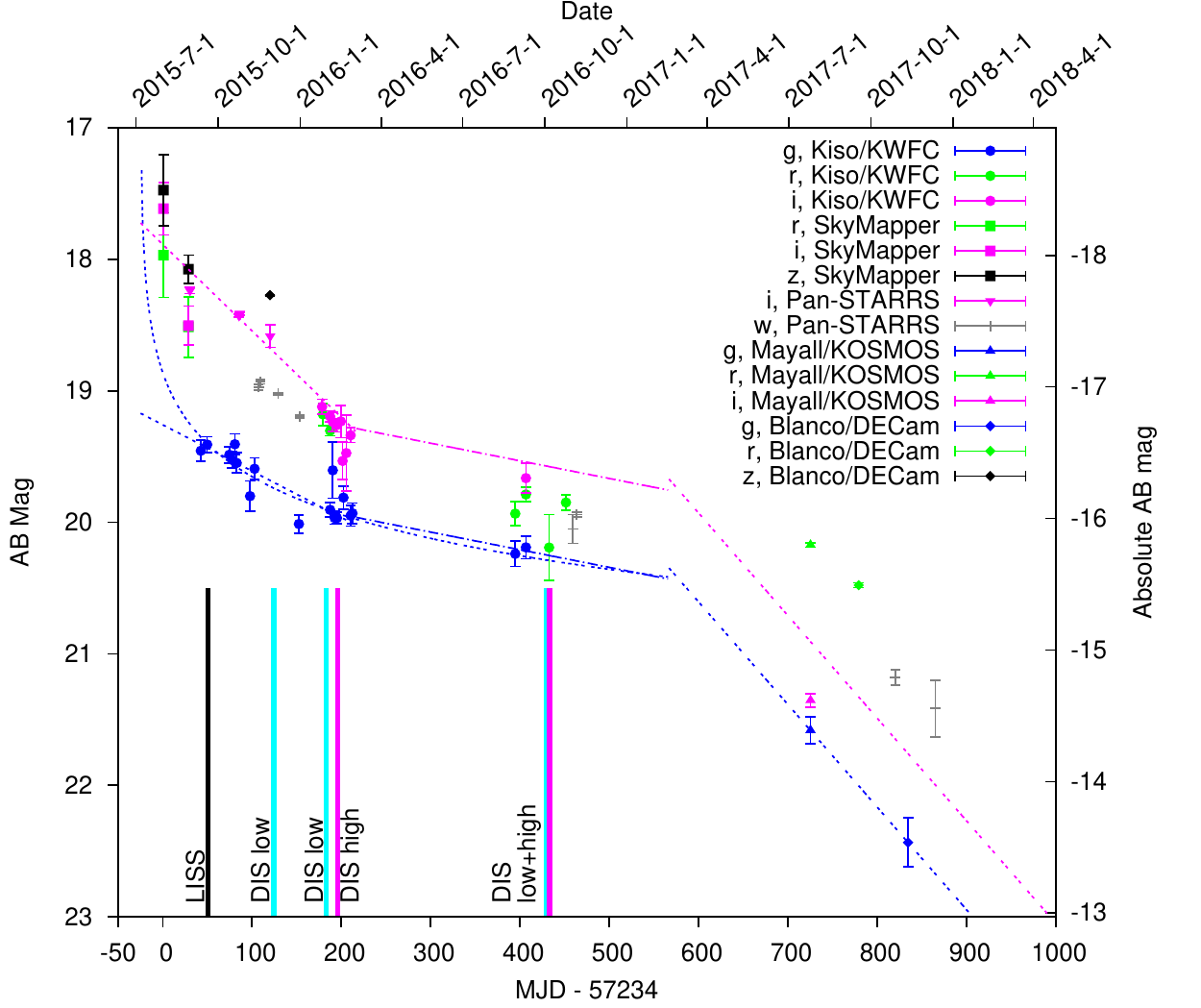}
\includegraphics[clip, width=5.3in]{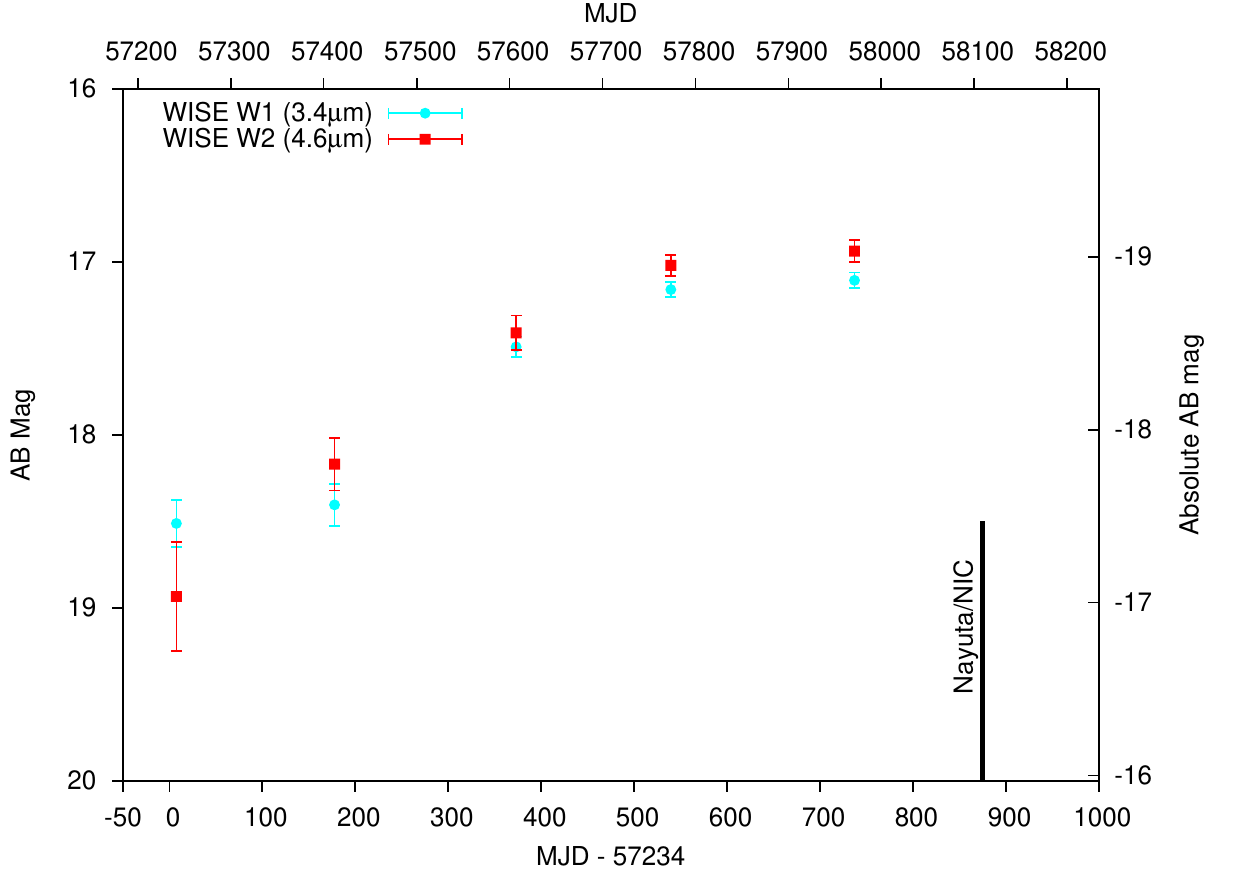}
}
 \caption{Top: Image subtraction photometry light curves of KISS15s, with corrections for Galactic extinction in $g$-, $r$-, $i$-, and $z$-band data.
 The origin of the time axis is taken to be the date of the first detection by SkyMapper, 2015 July 31 (MJD=57234; Section~\ref{sec:skymapper}).
 The time unit is in the observer's frame.
 A power-law light curve model for the early epoch $g$-band light curve and broken-line $g$- and $i$-band light curve models are also shown (see Section~\ref{sec:powerlaw_fitting} for details).
 The $r$-band excess in late epochs is probably due to flux contamination of the strong H$\alpha$ emission line.
 The epochs of optical spectroscopic observations with Nayuta/LISS and ARC3.5-m/DIS (low- and high-resolution grisms) are denoted by vertical bars.
 Bottom: W1- and W2-band photometry for the difference NEOWISE coadded images after 2015 August 5. Galactic extinction is corrected.
 The epoch of the Nayuta/NIC $J$-, $H$-, $K_{s}$-band observation (MJD=58108.9) is denoted by a vertical bar.
 }
 \label{fig:lightcurve_5}
\end{figure*}

Finally, photometric measurements obtained with the same filter each day were combined by taking a weighted mean.
Table~\ref{obslog_kiso} presents the $g$-, $r$-, and $i$-band measurements for KISS15s from the Kiso/KWFC image subtraction photometry; Figure~\ref{fig:lightcurve_5} shows the light curves corrected for Galactic extinction.

\subsubsection{Early photometry obtained with the SkyMapper Survey}
\label{sec:skymapper}

\begin{deluxetable}{lcccc}
\tablecolumns{5}
\tablewidth{0pc}
\tablecaption{SkyMapper photometry of KISS15s ($>3\sigma$ detection). \label{obslog_skymapper}}
\tablehead{
\colhead{MJD} & \colhead{Date} & \colhead{mag.} & \colhead{error in mag.} & \colhead{band}   }
\startdata
57234.7870 & 2015-07-31 &  18.11 & 0.32 & $r$\\
57234.7873 & 2015-07-31 &  17.72 & 0.20 & $i$\\
57234.7877 & 2015-07-31 &  17.55 & 0.27 & $z$\\\hline
57262.7524 & 2015-08-28 &  18.65 & 0.23 & $r$\\
57262.7527 & 2015-08-28 &  18.61 & 0.15 & $i$\\
57262.7531 & 2015-08-28 &  18.15 & 0.10 & $z$\\
\enddata
\tablecomments{Galactic extinction is uncorrected.}
\end{deluxetable}

\begin{figure}[tbp]
\center{
\includegraphics[clip, width=1.3in]{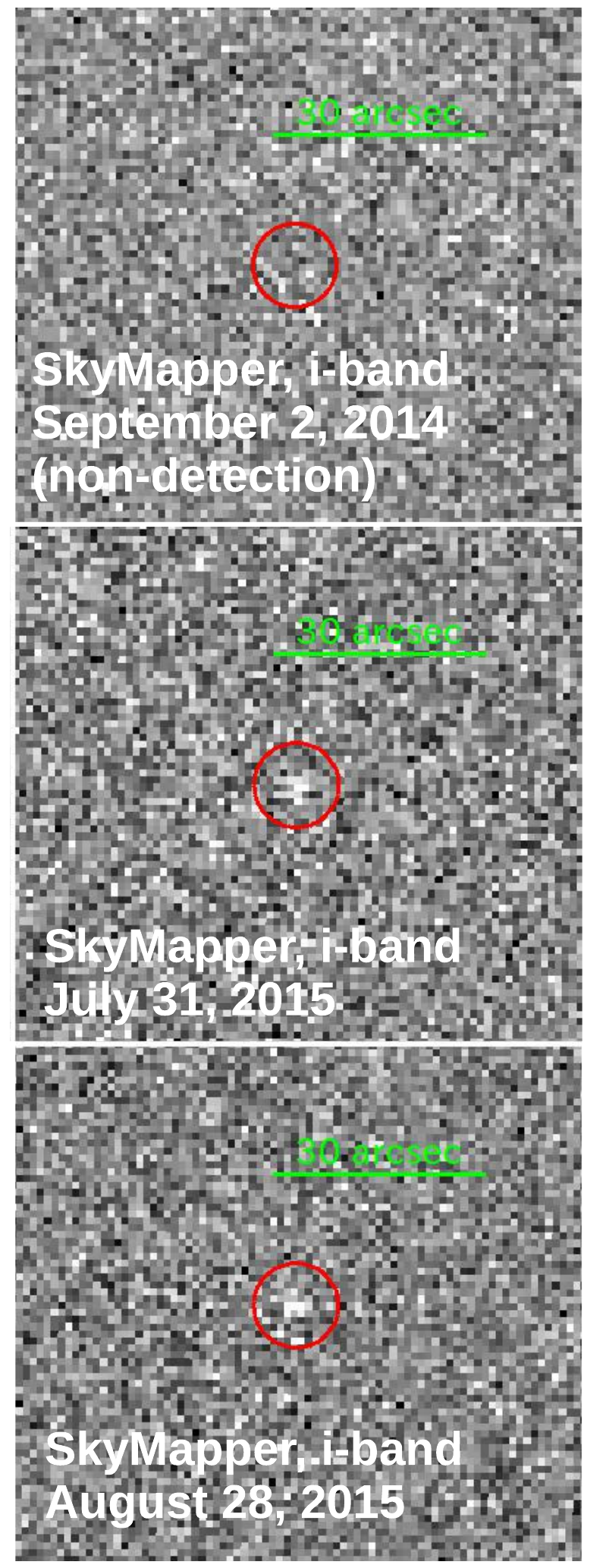}
}
 \caption{SkyMapper $i$-band reference-subtracted images at the position of KISS15s.
 The position of KISS15s determined by the KWFC observations is marked by a circle. The images are $2 \times 2$-binned for clarity.
 }
 \label{fig:skymapper_iband}
\end{figure}

SkyMapper is a fully automated 1.3 m optical telescope at Siding Spring Observatory in Australia \citep{sca17}.
First Data Release (DR1.1) of the SkyMapper Southern Survey \citep{wol18}\footnote{DR1 includes only images from the Shallow Survey; \href{http://skymapper.anu.edu.au/data-release/dr1/}{http://skymapper.anu.edu.au/data-release/dr1/}, \doi{10.4225/41/593620ad5b574}} contains $u$-, $g$-, $r$-, $i$-, and $z$-band imaging data at the position of KISS15s obtained on MJD = 56899.78 (2014 August 30), 56902.77 (2014 September 2), 57234.78 (2015 July 31), and 57262.75 (2015 August 28).

First, we searched for catalogued photometry data at the position of KISS15s in the SkyMapper photometry table ({\tt dr1.master}), and found that the SkyMapper pipeline detected KISS15s in $r$-, $i$-, and $z$-band images on 2015 August 28 (MJD = 57262.75), named SMSS~030831.61-005005.8.
This means that the SkyMapper data provide optical photometric measurements at earlier epochs than the Kiso/KWFC first detection.
The PSF magnitudes reported in the photometry table are $r = 18.225 \pm 0.041$~mag, $i=18.462 \pm 0.110$~mag, and $z=18.125 \pm 0.053$~mag.
It should be noted that the SkyMapper images were obtained with short exposure times to achieve 10$\sigma$ magnitude limits of $\sim 18.0$~mag for all wavelength bands \citep{wol18}; thus, the photometric errors are larger than those for Kiso/KWFC photometry, although KISS15s is brighter at the epoch of the SkyMapper observation.

To further check whether KISS15s was detected in much earlier SkyMapper images and to carry out difference image photometry, we downloaded $600'' \times 600''$ SkyMapper cut-out images centered on KISS15s.
$HOTPANTS$ image subtraction was applied to the SkyMapper images, with SDSS Stripe~82 coadd images as a reference.
Then, using the Python module {\tt Photutils} \citep{pho17}, we applied forced aperture photometry for the difference images with the circular aperture centered on the sky coordinate of KISS15s.
Magnitude zero-points were calculated by comparing measured instrumental magnitudes of field stars around KISS15s with the SDSS magnitudes. 
Finally, photometric measurements of each filter obtained on the same day were binned by taking weighted averages.

Table~\ref{obslog_skymapper} reports the $>3~\sigma$ detection SkyMapper photometry data points for KISS15s.
We found that KISS15s was barely detected in $r$-, $i$-, and $z$-band images on 2015 July 31 (Figure~\ref{fig:skymapper_iband}).
Only a single exposure per each filter was obtained at this epoch, in which the exposure times were 40, 5, 5, 10, and 20~s for $u$-, $g$-, $r$-, $i$-, and $z$-bands, respectively.
KISS15s was not detected in $u$- and $g$-band images obtained on the same day, giving magnitude 5$\sigma$ lower limits of $u>18.09$~mag and $g>17.88$~mag.
The SkyMapper's first detection date is earlier than the first detection by KISS survey data by 42~days; thus, SkyMapper provides the earliest optical photometry data of KISS15s.

On SkyMapper's second detection date, 2015 August 28, KISS15s was detected in $r$-, $i$-, and $z$-bands, but not in $u$- and $g$-bands, giving magnitude 5$\sigma$ lower limits of $u>18.69$~mag and $g>18.98$~mag.
At this epoch, two exposures were obtained per each filter; the total exposure times were 80, 10, 10, 20, and 40~s for $u$-, $g$-, $r$-, $i$-, and $z$-bands, respectively.
The last non-detection by SkyMapper observations was 2014 September 2, roughly 1~year before the first detection date of 2015 July 31.

SkyMapper $r$-, $i$-, and $z$-band photometry data of KISS15s are shown in Figure~\ref{fig:lightcurve_5}, along with Kiso/KWFC data.

\subsubsection{Pan-STARRS $i$-band photometry}
\label{sec:panstarrs}

\begin{deluxetable}{lcccc}
\tablecolumns{5}
\tablewidth{0pc}
\tablecaption{Pan-STARRS photometry of KISS15s. \label{obslog_panstarrs}}
\tablehead{
\colhead{MJD} & \colhead{Date} & \colhead{mag.} & \colhead{error in mag.} & \colhead{band}   }
\startdata
57264.5795 & 2015-08-30 & 18.34 & 0.02 & $i$\\
57319.4943 & 2015-10-24 & 18.53 & 0.01 & $i$\\
57320.4304 & 2015-10-25 & 18.51 & 0.01 & $i$\\
57354.4335 & 2015-11-28 & 18.69 & 0.08 & $i$\\\hline
57341.3750 & 2015-11-15 & 18.97 & 0.02 & $w$\\
57343.4161 & 2015-11-17 & 18.93 & 0.02 & $w$\\
57363.3764 & 2015-12-07 & 19.02 & 0.01 & $w$\\
57387.3332 & 2015-12-31 & 19.20 & 0.01 & $w$\\
57693.4538 & 2016-11-01 & 20.05 & 0.11 & $w$\\
57697.4347 & 2016-11-05 & 19.94 & 0.02 & $w$\\
58054.5285 & 2017-10-28 & 21.18 & 0.06 & $w$\\
58099.3644 & 2017-12-12 & 21.42 & 0.22 & $w$\\
\enddata
\tablecomments{Galactic extinction is uncorrected.}
\end{deluxetable}

The Pan-STARRS 3$\pi$ Survey detected KISS15s as of 2015 August 30 (referred to as PS15bva in the Pan-STARRS Survey for Transients database).
Give that single-exposure Pan-STARRS images are not publicly available, we retrieved Pan-STARRS photometry data of KISS15s directly from the Open Supernova Catalog \citep{gui17}.
Transient photometry was performed on difference images, using reference images created by stacking frames in each band between 2010 and 2012 \footnote{\href{https://star.pst.qub.ac.uk/ps1threepi/psdb/}{https://star.pst.qub.ac.uk/ps1threepi/psdb/}}.
KISS15s was imaged in $i$- and $w$-bands, in which the $i$-band is similar to that of the SDSS system and the $w$-band is approximately a $g$+$r$+$i$-band \citep{ton12}.
Table~\ref{obslog_panstarrs} lists the Pan-STARRS $i$- and $w$-band photometry data of KISS15s, where multiple photometry data obtained within a day are binned by taking the weighted average.
The $i$- and $w$-band light curves are shown in Figure~\ref{fig:lightcurve_5}; in the figure, the $i$-band magnitude is corrected for the Galactic extinction of $A_{i}=0.102$~mag, however the $w$-band magnitude is not corrected, because the extinction estimate for the $w$-band is uncertain.

\subsubsection{Mayall 4-m/KOSMOS optical photometry}
\label{obs:mayall}

\begin{deluxetable}{lcccc}
\tablecolumns{5}
\tablewidth{0pc}
\tablecaption{Mayall/KOSMOS photometry of KISS15s. \label{obslog_mayall}}
\tablehead{
\colhead{MJD} & \colhead{Date} & \colhead{AB mag.} & \colhead{error in mag.} & \colhead{band}   }
\startdata
57959.454 & 2017-07-25 & 21.78 & 0.10 & $g$ \\
57959.456 & 2017-07-25 & 20.31 & 0.02 & $r$ \\
57959.459 & 2017-07-25 & 21.46 & 0.05 & $i$ \\
\enddata
\tablecomments{Galactic extinction is uncorrected.}
\end{deluxetable}

We obtained single epoch $g$-, $r$-, and $i$-band photometry for KISS15s with the Mayall 4-m Telescope in KOSMOS imaging mode \citep{martini14} on 2017 July 25 UT (MJD=57959.5).
KOSMOS was used with a 2k4k E2V charge-coupled device (CCD) with a pixel scale of $0''.292$~pixel${}^{-1}$.
The exposure time for each band was 180~s.
The seeing condition on the observing night was $\sim 5$~pixel = $1''.5$.

Mayall/KOSMOS images were analyzed in the same way as the Kiso/KWFC data, using $HOTPANTS$ and $g$-, $r$-, and $i$-band references created from SDSS Stripe82 coadd data.
Aperture photometry for the reference-subtracted images was performed using {\tt Photutils}.
The Mayall/KOSMOS image subtraction photometry of KISS15s is summarized in Table~\ref{obslog_mayall}.
The Mayall/KOSMOS $g$- and $i$-band data are well below the power-law extrapolation of the Kiso/KWFC and SkyMapper photometry data (Figure~\ref{fig:lightcurve_5}).
The significant drop in optical continuum emission indicates that the heating sources of the continuum emission region, which may be the innermost part of KISS15s, lose most of their kinetic energy (see Section~\ref{sec:powerlaw_fitting} for details).

Mayall/KOSMOS $r$-band photometry shows an excess compared to $g$- and $i$-band data, and has a consistent brightness with the last Kiso/KWFC $r$-band observation.
The apparent $r$-band excess relative to the $g$- and $i$-bands is probably due to flux contamination from a persistently strong H$\alpha$ emission line (Section~\ref{obs:sepc}).
It should be noted that the Mayall/KOSMOS $r$-band filter transmission is relatively enhanced at wavelengths around $\lambda_\text{obs} \sim 6800$~\AA\ compared to that of the original SDSS $r$-band filter (adopted by the Kiso/KWFC).
The enhanced H$\alpha$ emission line equivalent width at later epochs (Section~\ref{obs:sepc}) and the enhanced filter response can explain the $r$-band excess.
We note that a similar $r$-band excess relative to the other bands due to the contamination of the strong H$\alpha$ emission line was also observed in luminous type IIn SN~2010jl \citep{fra14}.

\subsubsection{Blanco 4-m/DECam optical photometry}
\label{obs:decam}

\begin{deluxetable}{lcccc}
\tablecolumns{5}
\tablewidth{0pc}
\tablecaption{Blanco/DECam DECaLS photometry of KISS15s. \label{obslog_decam}}
\tablehead{
\colhead{MJD} & \colhead{Date} & \colhead{AB mag.} & \colhead{error in mag.} & \colhead{band}   }
\startdata
57354.148 & 2015-11-28 & 18.35 & 0.01 & $z$ \\
58013.384 & 2017-09-17 & 20.62 & 0.02 & $r$ \\
58068.701 & 2017-11-11 & 22.63 & 0.19 & $g$ \\
\enddata
\tablecomments{Galactic extinction is uncorrected.}
\end{deluxetable}

KISS15s was serendipitously detected in $g$-, $r$-, and $z$-band images obtained through the DECam Legacy Survey \citep[DECaLS DR7;][]{dey18} \footnote{Specifically, the DECam Legacy Survey of the SDSS Equatorial Sky (PI: D. Schlegel and A. Dey); \href{http://legacysurvey.org/}{http://legacysurvey.org/}}.
DECaLS uses a 3~deg${}^{2}$ FoV prime focus optical imager DECam, which is mounted on the Blanco 4-m telescope at Cerro Tololo Inter-American Observatory (CTIO) and includes 62 2k4k CCDs at $0''.263$~pixel${}^{-1}$ resolution.
The $g$-, $r$-, and $z$-band images of KISS15s were obtained on 2017 November 11-12, 2017 September 17, and 2015 November 28, respectively.
Two images were obtained per each filter at detector positions of S28 and N29 \citep[see e.g.,][]{sha15}; the exposure times of $g$-, $r$-, $z$-band images were 80 and 166~s, 65 and 58~s, and 85 and 88~s, respectively.
The seeing FWHMs were $\sim 3.3-4.6$~pixels corresponding $0''.9-1''.2$.

We downloaded the images and associated inverse-variance maps from the DECam Legacy Survey web page.
The DECaLS images were analyzed in the same way as Kiso/KWFC data, using $HOTPANTS$ and $g$-, $r$-, and $z$-band references created from SDSS Stripe82 data.
Aperture photometry for the reference-subtracted images was performed using {\tt Photutils}.
The DECaLS image subtraction photometry of KISS15s is summarized in Table~\ref{obslog_decam}, where a weighted average of the two photometric measurements for each filter is reported.
As shown in Figure~\ref{fig:lightcurve_5}, the DECaLS $g$-band measurement confirms a faster rate of decline at the late epochs ($\gtrsim$ 600~days since discovery) compared to the slower temporal evolution at the early epochs.

As with Mayall/KOSMOS $r$-band photometry (Section~\ref {obs:mayall}), DECaLS $r$-band photometry also shows an excess, compared to $g$- and $i$-band Mayall/KOSMOS data in late epochs.
The DECam $r$-band filter transmission is relatively enhanced at wavelengths around $\lambda_\text{obs} \sim 6800$~\AA\ compared to the original SDSS $r$-band filter.
Thus, the apparent $r$-band excess of the DECaLS measurement is probably due to flux contamination from the strong H$\alpha$ emission line.

\subsubsection{NEOWISE W1- and W2-band infrared photometry}
\label{obs:wise}

\begin{figure*}[tbp]
\center{
\includegraphics[clip, width=6.8in]{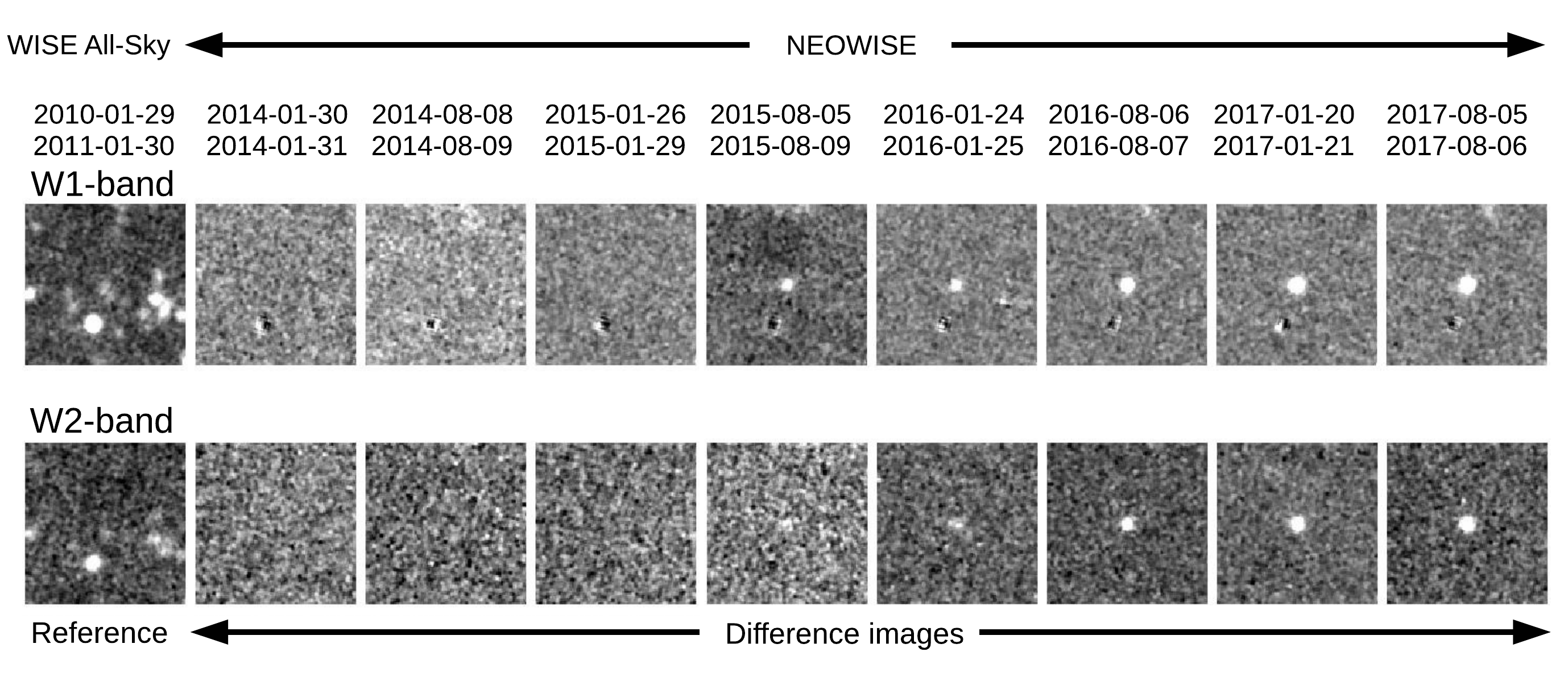}
}
 \caption{WISE/NEOWISE reference (leftmost column) and difference (the other columns) coadded images at the position of KISS15s. The image size is $0.03^{\circ} \times 0.03^{\circ}$. Upper panels show W1-band images, and the lower panels show W2-band images. KISS15s is detected in the images after 2015 August 5, both in the W1- and W2-bands.
 }
 \label{fig:image_wise}
\end{figure*}

\begin{deluxetable}{lcccc}
\tablecolumns{5}
\tablewidth{0pc}
\tablecaption{NEOWISE photometry of KISS15s. \label{obslog_neowise}}
\tablehead{
\colhead{mid. MJD} & \colhead{mid. date} & \colhead{AB mag.} & \colhead{error in mag.} & \colhead{band}   }
\startdata
57241.5 & 2015-08-07 & 18.52 & 0.14 & W1 \\
57411.8 & 2016-01-24 & 18.41 & 0.12 & W1 \\
57607.1 & 2016-08-07 & 17.50 & 0.06 & W1 \\
57773.6 & 2017-01-20 & 17.17 & 0.04 & W1 \\
57971.2 & 2017-08-06 & 17.12 & 0.05 & W1 \\\hline
57241.5 & 2015-08-07 & 18.94 & 0.32 & W2 \\
57411.8 & 2016-01-24 & 18.18 & 0.15 & W2 \\
57607.1 & 2016-08-07 & 17.42 & 0.10 & W2 \\
57773.6 & 2017-01-20 & 17.03 & 0.06 & W2 \\
57971.2 & 2017-08-06 & 16.94 & 0.06 & W2 \\\hline
\enddata
\tablecomments{Galactic extinction is uncorrected.}
\end{deluxetable}

The NEOWISE reactivation mission has been scanning the entire sky, using a space-based infrared telescope in the MIR wavelength range, specifically W1 (3.4 $\mu$m) and W2 (4.6 $\mu$m) bands, since late 2013 as an extension of the original mission, the WISE All-Sky Survey \citep{wri10,mai14}.
The NEOWISE survey visits each sky region roughly twice per year (each visit consists of 9-16 exposures) and thus is useful for examining the long-term MIR light curve evolution of astronomical transients.
In this work, we used the NEOWISE 2018 Data Release, which includes all W1- and W2-band images and associated photometry catalogs acquired before the end of 2017.

First, we searched for a point source at the position of KISS15s in the NEOWISE-R Single Exposure (L1b) Source Table available at the NASA/IPAC Infrared Science Archive, and found that KISS15s was detected in a NEOWISE W1-band single exposure image obtained on 2015 August 5 (MJD=57239.47; frame id = 62453b177).
KISS15s has continuously been detected since the first detection both in the W1- and W2-bands.
The first detection of KISS15s by NEOWISE was 1 month prior to the first detection by Kiso/KWFC (Section~\ref{kiss15s_phot}), and $\sim 7$~days after the detection by SkyMapper (Section~\ref{sec:skymapper}).

Then we examined whether KISS15s appeared in WISE/NEOWISE images by creating a coadded image for each visit for each band, using an online version of the WISE/NEOWISE Coadder ICORE \citep{mas09,mas13} \footnote{\href{http://irsa.ipac.caltech.edu/applications/ICORE/}{http://irsa.ipac.caltech.edu/applications/ICORE/}}.
Images of $0.03{}^{\circ} \times 0.03{}^{\circ}$ (at $1''.0$~pixel${}^{-1}$) coadd intensity and standard deviation at the position of KISS15s were created for each band for each NEOWISE visit ($\sim$ 10 exposures per visit).
Images with a {\tt qual\_frame}>=5, a moon separation angle > 20$^{\circ}$, and {\tt ssa\_sep} (distance from South Atlantic Anomaly edge) > 0$^{\circ}$ were used as input.
A simple area weighting was used for its interpolation, converting the native input pixel scale of $2''.75$~pixel${}^{-1}$ to the output pixel scale of $1''.0$~pixel${}^{-1}$ \citep[Section~7 of][]{mas13}.
Aside from the NEOWISE data, W1- and W2-band images obtained during the epochs of the WISE all-sky survey (in 2010-2011; $\sim$ 40 exposures) were coadded into W1- and W2-band referenced images, respectively; then the reference images were subtracted from the single visit coadd images to create difference images.
The residual background in the difference image was estimated by taking a median of the pixel fluxes, and the estimated residual background was subtracted from the difference image.

Figure~\ref{fig:image_wise} shows the WISE/NEOWISE reference and difference images at the position centered on KISS15s.
No signal was detected in the difference images before the NEOWISE visit on August 5, 2015.
Since then, KISS15s has appeared continuously as a point source.
We can see that the W2-band fluxes were only barely detected in August 2015 (at the $\sim$ 3.43$\sigma$ level).
Aperture photometry was applied to the difference images using {\tt Photutils}, in which a fixed standard circular aperture with $8''.25$ radius centered on the sky coordinate of KISS15s was used.
The same aperture photometry was also applied to the associated variance images to estimate the photometric errors of the aperture fluxes.
To account for the spatial correlation of pixel noise, the photometric errors of the aperture fluxes derived from the photometry of the variance map were inflated by a factor of 2.75 \citep[= the ratio of the input to output pixel scale; Section~13 of][]{mas13}.
To determine the magnitude zero points, first we downloaded $0.3{}^{\circ} \times 0.3{}^{\circ}$ ALLWISE W1- and W2-band images around KISS15s and measured the circular aperture instrumental magnitudes of field stars.
The mean difference between the instrumental and Vega magnitudes of field stars catalogued in the ALLWISE point source catalog was calculated as a magnitude zero point for each of the W1- and W2-bands.
The magnitude offsets between Vega and AB magnitude systems were taken from \cite{jar11} as $m_{W1, \text{AB}}=m_{W1, \text{Vega}} + 2.699$ mag and $m_{W2, \text{AB}}=m_{W2, \text{Vega}} + 3.339$ mag.

The bottom panel of Figure~\ref{fig:lightcurve_5} shows the W1- and W2-band difference image photometry light curves of KISS15s after 2015 August 5 (Table~\ref{obslog_neowise}).
Interestingly, the W1- and W2-band MIR fluxes increased over time from 2015 to 2017, as opposed to the optical light curves that have steadily decreased since discovery (Figure~\ref{fig:lightcurve_5}).
The late time excess emission at IR wavelengths observed in SNe~IIn are commonly interpreted as new dust formation in the CDS region, or as dust IR echoes from the CSM heated by X-ray and ultraviolet (UV) radiation from the ejecta-CSM interaction \citep[e.g.,][]{ger02,fox10,fox11,str12,fra14,sar18,sza18}.
Actually the W1-W2 band color of KISS15s during the late phase (W1-W2 $\simeq$ 0~AB~mag) indicates that this IR emission component has a high color temperature of $T_{\text{dust}} \sim 1,200$~K, which is close to the sublimation temperature of graphite/silicate dust grains \citep[e.g.,][]{fra14}.
More detailed analyses of the IR light curves are presented in Section~\ref{sec:ir_lightcurves}.

\subsubsection{Nayuta/NIC $J$, $H$, $K_{s}$-band near-infrared photometry}
\label{sec:nayuta_nic_obs}

\begin{figure}[tbp]
\center{
\includegraphics[clip, width=3.2in]{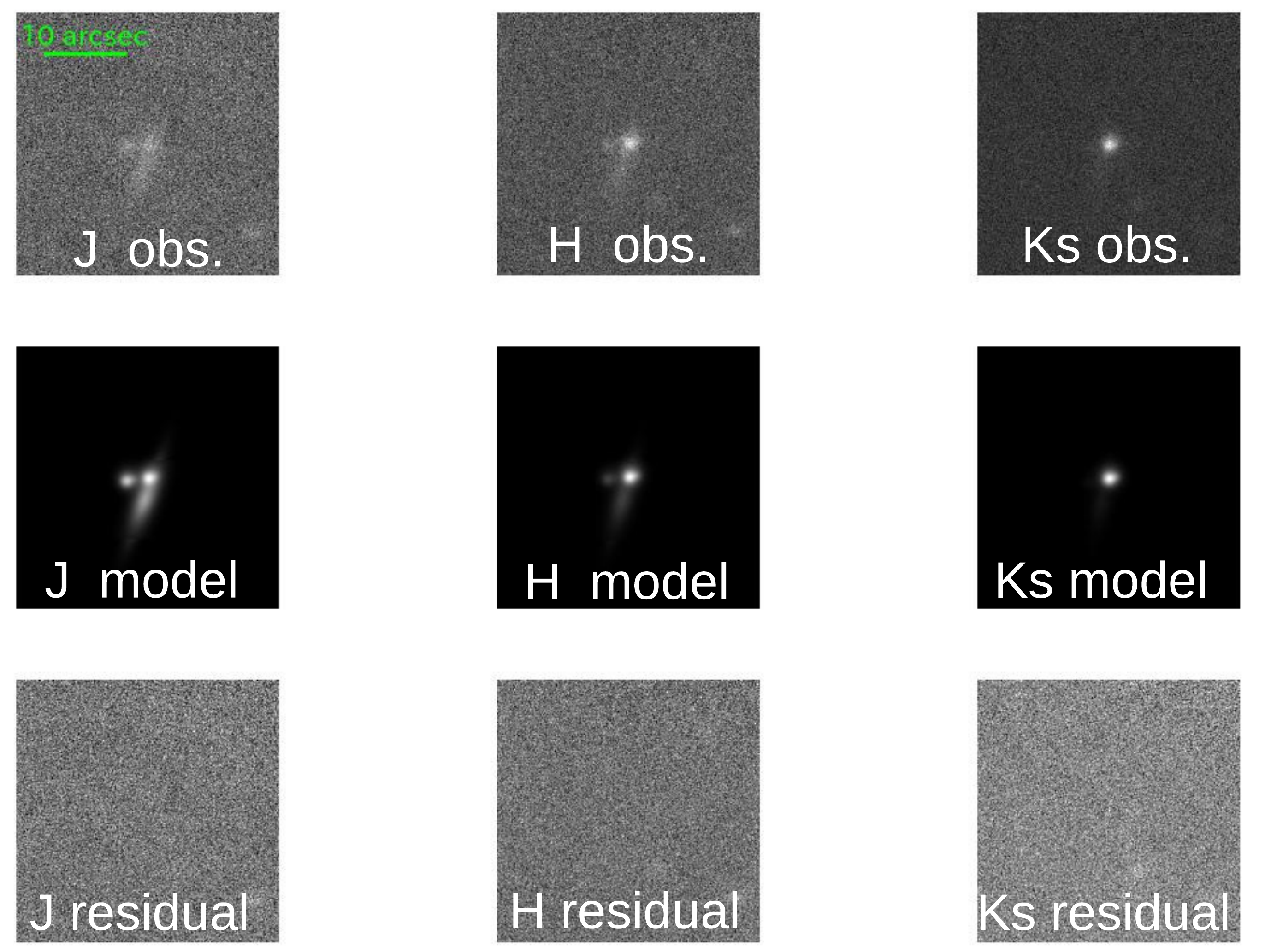}
}
 \caption{Top: Nayuta/NIC $J$-, $H$-, $K_{s}$-band images centered on KISS15s obtained on 2017 December 21-22. Middle: The best-fit GALFIT model images. The model is composed of two point sources (KISS15s and the foreground Galactic star) and an exponential disk (the host galaxy). Bottom: the fitting residual images.
 }
 \label{fig:image_nic}
\end{figure}

\begin{deluxetable}{lcccc}
\tablecolumns{5}
\tablewidth{0pc}
\tablecaption{Nayuta/NIC GALFIT modeling photometry of KISS15s. \label{obslog_nic}}
\tablehead{
\colhead{mid. MJD} & \colhead{Date} & \colhead{AB mag.} & \colhead{error in mag.} & \colhead{band}   }
\startdata
58108.9  & 2017-12-21/22 &  20.33 & 0.08 & $J$\\
58108.9  & 2017-12-21/22 &  19.27 & 0.02 & $H$\\
58108.9  & 2017-12-21/22 &  18.04 & 0.02 & $K_s$\\
\enddata
\tablecomments{Galactic extinction is uncorrected.}
\end{deluxetable}

We performed $J$-, $H$-, and $K_{s}$-band (1.24, 1.66, and 2.16~$\mu$m, respectively) imaging observations for KISS15s on 2017 December 21-22 UT (MJD = 58108.9 at mid exposure) with the Nayuta telescope, using the Nishi-Harima infrared camera (NIC).
NIC is composed of three 1024$\times$1024 pixel HAWAII arrays; each of the three arms has a $2'.7 \times 2'.7$ field-of-view, with a pixel scale of $0''.16$~pixel${}^{-1}$, and is capable of obtaining $J$, $H$, and $K_{s}$-band images simultaneously.
KISS15s and five relatively brighter field stars were imaged simultaneously on each image for differential photometry.
Most of the images were obtained with a 10-point circular dithering pattern (radius: $10''$), using an exposure time of 150~s per frame and adopting Fowler sampling with eight readouts at the beginning and end of the exposure (readout noise $\sim$10~e${}^{-1}$).
The total on-source exposure time was 215 min.
All images obtained during the run were shifted and added to generate a final image using {\tt SWARP} \citep{ber02}.

We performed two-dimensional fitting of the KISS15s and the host galaxy to derive the NIR magnitudes of KISS15s.
We first modeled the PSFs of the coadded $J$-, $H$-, and $K_{s}$-band images by fitting three Gaussians to a simultaneously imaged nearby field star,  2MASS~J03083221-0050320.
Then, each of the NIR images of KISS15s + foreground star + host galaxy were fitted with a model composed of two PSFs for KISS15s, the foreground star, and one inclined exponential disk for the host galaxy.
The magnitude zero points of the NIR images were determined by comparing the 2MASS point source magnitudes \citep{str06} and PSF-fitting instrumental magnitude measurements for five nearby field stars simultaneously imaged with KISS15s.
The Vega magnitudes were converted into AB magnitudes using offsets between AB and Vega magnitude systems of 0.91, 1.39, and 1.85~mag for $J$-, $H$-, and $K_{s}$-bands, respectively \citep{bla07}.
The PSF fitting was performed using IDL routine {\tt MPFIT2DFUN}\footnote{\href{https://www.physics.wisc.edu/~craigm/idl/down/mpfit2dfun.pro}{https://www.physics.wisc.edu/~craigm/idl/down/mpfit2dfun.pro}.},  developed by Craig B. Markwardt; multi-component model fitting was performed using {\tt GALFIT} \citep[Version 3.0.5;][]{pen02}.

Figure~\ref{fig:image_nic} shows the best-fitting GALFIT model of $J$-, $H$-, and $K_{s}$-band images.
The PSF magnitudes of KISS15s are summarized in Table~\ref{obslog_nic}.
KISS15s is brighter in the longer wavelength bands, suggesting that the NIR bands are dominated by the same host dust component revealed by NEOWISE W1- and W2-band photometry.

\subsection{Optical Spectroscopy}
\label{obs:sepc}

\begin{deluxetable*}{lllccc}
\tablecolumns{5}
\tablewidth{0pc}
\tablecaption{Log of Nayuta/LISS and ARC3.5-m/DIS long-slit spectroscopy of KISS15s. \label{obslog_dis}}
\tablehead{
\colhead{MJD} & \colhead{Date} & \colhead{Inst.-Grating} & \colhead{Slit} & \colhead{CCD binning} & \colhead{$\sigma_\text{inst}$}\\
\colhead{} & \colhead{} & \colhead{} & \colhead{} & \colhead{(spatial $\times$ dispersion)} & \colhead{(km~s${}^{-1}$)}
}
\startdata
57284.7 & 2015-09-19 & LISS-very low  & $2''.0$ & $2 \times 2$ & 1490 \\\hline
57358.2 & 2015-12-02 & DIS-B400/R300  & $2''.0$ & $1 \times 1$ & 141.6 \\
57417.1 & 2016-01-30 & DIS-B400/R300  & $1''.5$ & $2 \times 1$ & 111.6 \\
57665.4 & 2016-10-04 & DIS-B400/R300  & $2''.0$ & $1 \times 1$ & 551.9 (b), 148.6 (n) \\\hline
57430.1 & 2016-02-12 & DIS-B1200/R1200& $0''.9$ & $1 \times 1$ & 20.8 \\
57665.4 & 2016-10-04 & DIS-B1200/R1200& $0''.9$ & $1 \times 1$ & 150.4 (b), 18.4 (n) \\
\enddata
\tablecomments{The slit direction is set to $PA = 158.5^{\circ}$. The velocity dispersions of the instrumental line profiles $\sigma_\text{inst}$ are evaluated by fitting Gaussian functions to the arc lamp spectra at $\lambda_\text{obs} = 6900 - 7100$~\AA. The instrumental line profiles on 2016 October 4 have two components: a narrow core (``n'') and a broad wing (``b''), which may be due to defocusing of the spectrograph. For the spectra obtained on 2016 October 4, we assume $\sigma_\text{inst} = \sigma_\text{inst}(\text{b})$ (Section~\ref{append:inst_profile}).}
\end{deluxetable*}

\subsubsection{Nayuta/LISS very low-resolution spectroscopy}
\label{sec:nayuta_spec}

Following the Kiso/KWFC discovery of KISS15s on 2015 September 18.78 UT, we carried out optical spectroscopy measurements for rapid classification of the transient on 2015 September 19.7 UT (MJD = 57284.7) with the 2.0 m Nayuta telescope, using an optical spectrograph, LISS \citep{has14}.
LISS employs a back-illuminated fully depleted-type Hamamatsu 2k1k CCD.
We used a grism with a linear dispersion of 12.4~\AA~pixel${}^{-1}$ and a $2''.0$ width long-slit to obtain a very low-resolution optical spectrum for KISS15s.
The observed-frame wavelength range was 4,000-10,000~\AA, and the total on-source exposure time was 65 min.
The spatial scale of LISS is $0''.244$~pixel${}^{-1}$, and the instrumental broadening is estimated to be $\sigma_{\lambda} \sim 30$~\AA.
The slit direction was set to a position angle (PA) of 158.5${}^{\circ}$ to align the slit spatial direction to the semi-major axis of the host galaxy\footnote{PA = 0 deg corresponds to the North-South, PA increasing from North towards East.}.
The spectra were extracted using an aperture size of $2''.44$.

The spectrophotometric standard star G191-B2B was observed for the flux calibration.
Because the atmospheric extinction coefficients (mag/airmass values as a function of wavelength) at NHAO are not well constrained, we did not apply the airmass correction for the KISS15s spectrum.
The airmass values during the observations were 1.320-1.248 for KISS15s and 1.530-1.518 for G191-B2B; thus, the bluest part of the KISS15s spectrum may be overestimated by $\sim 0.1$~mag relative to the reddest part.
Data reduction, including overscan, bias, flat correction, cosmic ray rejections with the use of $L.A.Cosmic$ \citep[{\tt lacos\_spec};][]{dok01}, and wavelength and flux calibrations, were carried out using IRAF.
The LISS spectrum was corrected for telluric absorption features in the wavelength ranges of $6,860-6,890$~\AA\ and $7,600-7,630$~\AA\ (O${}_2$ bands) by dividing the KISS15s spectrum by the normalized combined spectrum of the standard star. 
The spectrum was corrected for Galactic extinction using the \cite{fit99} extinction curve.

Finally, the absolute flux of the LISS spectrum was scaled to match the Galactic extinction-corrected $i$-band magnitude of KISS15s obtained from broadband photometry.
We chose the $i$-band as the reference wavelength band, because the $i$-band is less affected by the H$\alpha$ emission line and blue continuum emission from the star-forming host galaxy than the $g$- and $r$-bands.
To perform the spectrophotometry, we first interpolated the $i$-band magnitude of KISS15s at the epoch of Nayuya/LISS spectroscopy from the $i$-band broken-line light curve model (Figure~\ref{fig:lightcurve_5}; see Section~\ref{sec:powerlaw_fitting} for details).
Then the filter-convolved $i$-band magnitude of the raw LISS spectrum was calculated using {\tt Speclite}\footnote{
\href{https://github.com/dkirkby/speclite}{https://github.com/dkirkby/speclite}. Spectral response curves of SDSS \citep{doi10} and WISE \citep{wri10} filters are included in the {\tt Speclite} module.}.
The spectrophotometric scaling factor was calculated from the difference between the broadband and spectroscopic $i$-band magnitudes.

\begin{figure*}[tbp]
\center{
\includegraphics[clip, width=6.0in]{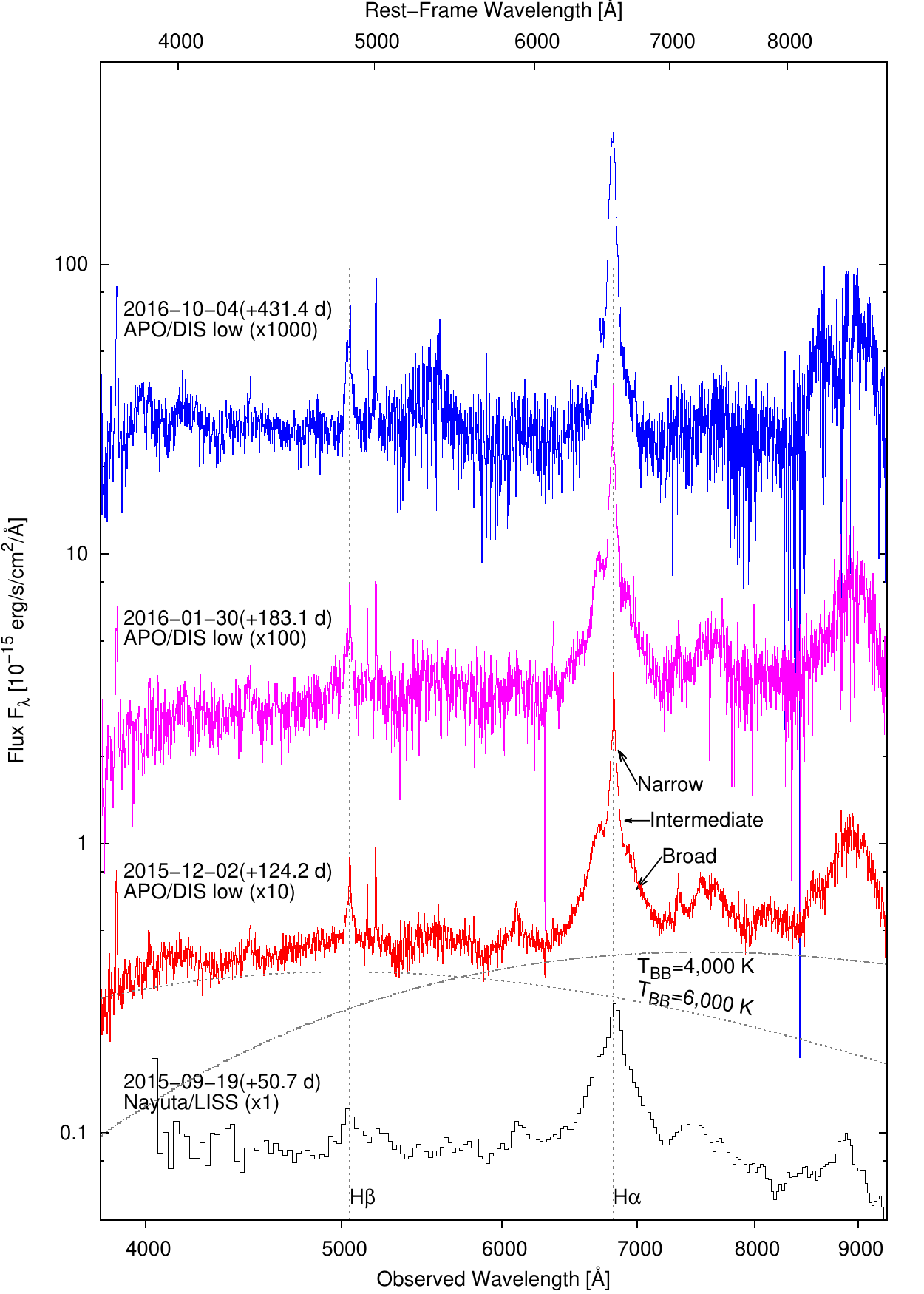}
}
 \caption{Nayuta/LISS very low-resolution spectrum and ARC3.5-m/DIS spectra of KISS15s. 
 The ARC3.5-m/DIS spectra obtained on 2015 December 2, 2016 January 30, and 2016 October 4 are scaled by factors of 10, 100, and 1,000 in flux, respectively.
 The observation dates since the first detection date (MJD = 57234) are indicated at the left-band side.
 The Galactic extinction is corrected.
 The wavelengths of H$\beta$ and H$\alpha$ emission lines are indicated by dotted vertical lines, assuming a redshift of $z=0.03782$.
 Arbitrarily scaled black-body spectra with a temperature of $T_{\text{BB}}=4,000$ and $6,000$ K are also shown.
 The Galactic extinction-corrected spectra of KISS15s shown in this figure are available as the Data behind the Figure.
 }
 \label{fig:spec_low}
\end{figure*}

Figure~\ref{fig:spec_low} presents the Nayuta/LISS spectrum of KISS15s.
Based on the redshift information ($z=0.03782$) of the host galaxy SDSS~J030831.67-005008.6 taken from the SDSS database, a strong, resolved broad emission line at the observed wavelength of $\lambda_{\text{obs}} \sim 6800$\AA\ can readily be identified as the H$\alpha$ emission line related to KISS15s.
The H$\beta$ emission line, \ion{He}{1} emission line ($\lambda$5876), and \ion{Ca}{2} IR triplet ($\lambda \lambda$8498, 8542, 8662) are also barely detectable in the Nayuta/LISS spectrum with a consistent redshift.
Thus, the Nayuta/LISS spectrum confirms that KISS15s actually belongs to the galaxy SDSS~J030831.67-005008.6.
The broad H$\alpha$ line does not show any P Cygni features, as is the case for other SNe IIn.

\begin{figure*}[tbp]
\center{
\includegraphics[clip, width=3.2in]{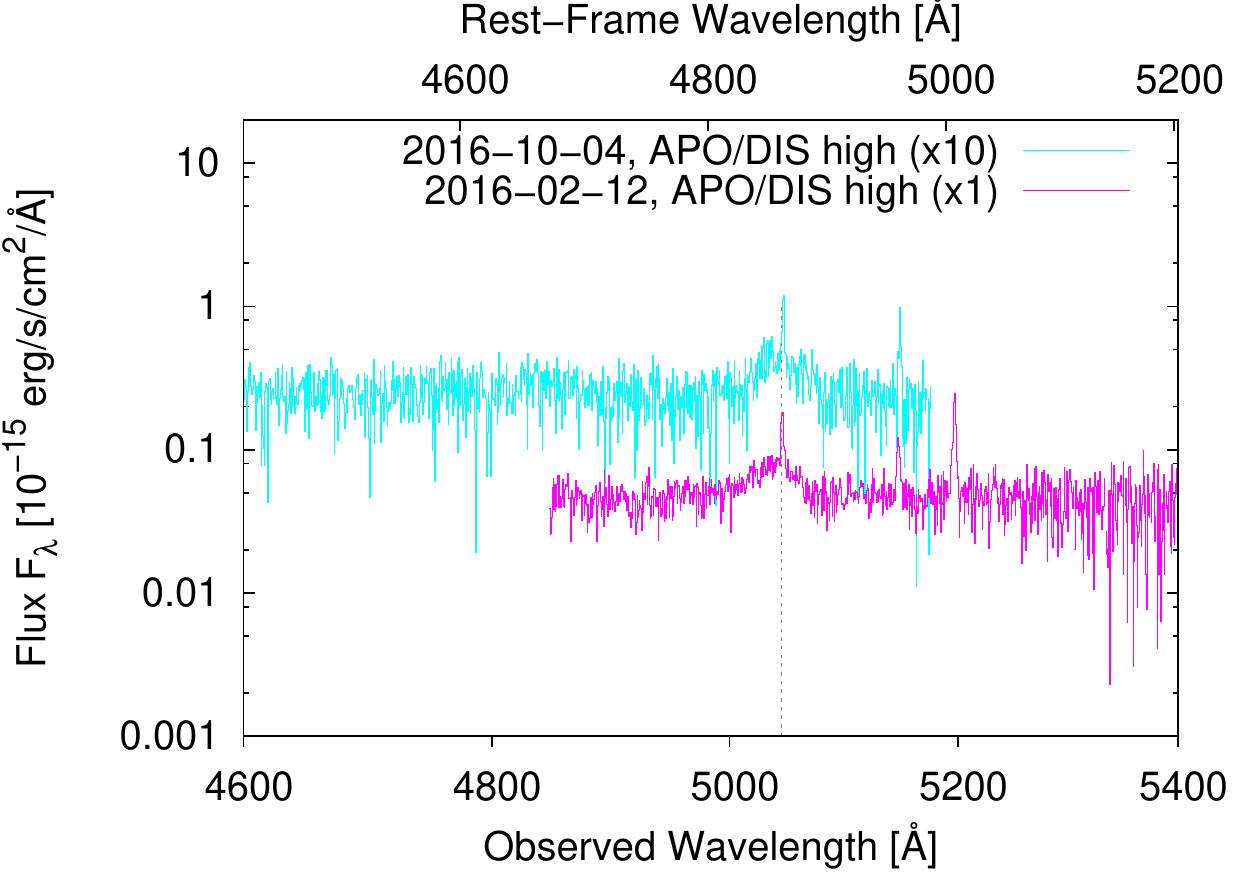}
\includegraphics[clip, width=3.2in]{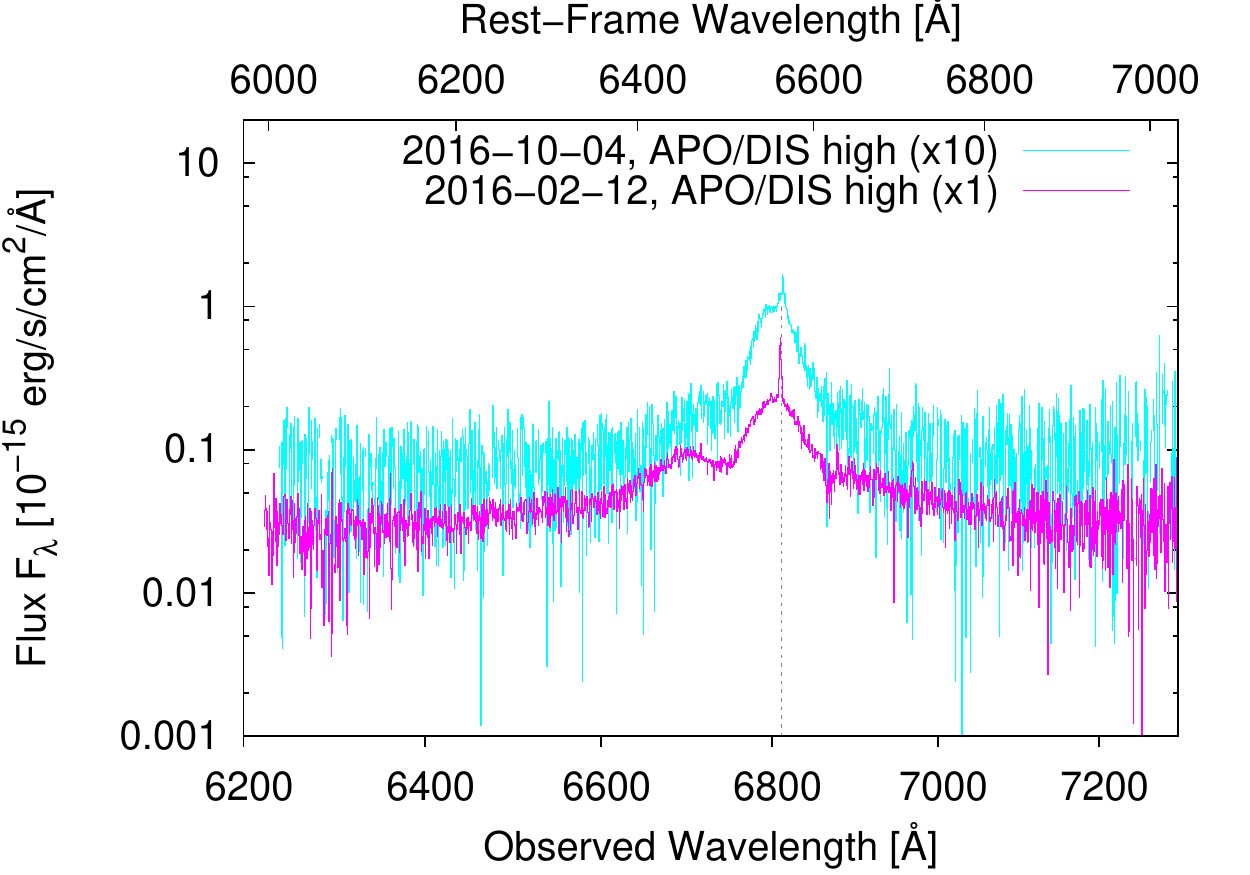}
}
 \caption{
 ARC3.5-m/DIS high-resolution spectra of KISS15s in the wavelength regions of H$\beta$ and H$\alpha$.
 The Galactic extinction is corrected.
 The wavelengths of the H$\beta$ and H$\alpha$ emission lines are denoted by dotted vertical lines, assuming a source redshift of $z=0.03782$.
 The spectra obtained on 2016 October 4 are scaled by a factor of 10 in flux for clarity.
 Note that these high-resolution spectra are not spectro-photometrically calibrated.
 The Galactic extinction-corrected spectra of KISS15s shown in this figure are available as the Data behind the Figure.
 }
 \label{fig:spec_high}
\end{figure*}

\subsubsection{ARC3.5-m/DIS low- and high-resolution spectroscopy}
\label{data:dis}

At later epochs, we obtained three low-resolution spectra and two high-resolution spectra for KISS15s using the DIS mounted on the ARC~3.5-meter telescope at the Apache Point Observatory (APO).
DIS is a dual channel imaging-spectrograph, composed of a Marconi CCD42-20-0-310 on the blue side and an E2V CCD42-20-1-D21 on the red side\footnote{\href{https://www.apo.nmsu.edu/arc35m/Instruments/DIS}{https://www.apo.nmsu.edu/arc35m/Instruments/DIS}.}.
Table~\ref{obslog_dis} lists the log of the ARC3.5-m/DIS observations of KISS15s.

The first epoch ARC3.5-m/DIS low-resolution optical spectrum of KISS15s was obtained on 2015 December 2 UT (MJD=57358.2).
A standard low-resolution grating setup of B400/R300 for blue/red channels, 1$\times$1 on-chip binning, and a $2''.0$ long-slit were used.
The spatial scales were $0''.40$~pixel${}^{-1}$ and $0''.42$~pixel${}^{-1}$ for the blue and red channels, respectively.
The total exposure time was 45 min (three exposures of 15 min each), and the slit direction was set to PA=158.5${}^{\circ}$.
A spectrophotometric standard star GD50 was also observed on the same day for the flux calibration.

The second epoch ARC3.5-m/DIS low-resolution spectrum was obtained on 2016 January 30 UT (MJD = 57417.1), using the same instrumental setup as the first observation, except for the use of a $1''.5$ width long-slit and 2$\times$1  binning.
The observation may have been affected to some extent by cirrus clouds.

The third epoch ARC3.5-m/DIS spectrum was obtained on 2016 February 12 UT (MJD=57430.1) using a high spectral resolution mode: $0''.9$ width long-slit, 1$\times$1 binning, a no-order sorting filter, and B1200/R1200 gratings.
The B1200 and R1200 grisms were tilted in their housings so that the target wavelengths (5040\AA\ and 6800\AA, respectively) were located towards the center of the detector.
As summarized in Table~\ref{obslog_dis}, the instrumental broadening of the high-resolution mode was evaluated to be $\sigma_\text{inst} \sim 20$~km~s${}^{-1}$ in the H$\alpha$ emission wavelength region of KISS15s.
The total exposure time was about 74 min, and the slit direction was set to PA=158.5${}^{\circ}$.
The spectrophotometric standard star GD50 was observed on the same day for flux calibration.

The fourth epoch ARC3.5-m/DIS spectrum was obtained on 2016 October 4 (MJD = 57665.4), using both low- and high-spectral resolution modes and the same instrumental configurations as the previous observations.
The total exposure times were 60 and 66.7 min for the low- and high-resolution spectroscopy mode, respectively; however, cloudy weather conditions prevented us from achieving the planned S/N objective.

Data reduction of ARC3.5-m/DIS spectra was carried out using IRAF.
Simple aperture extraction was performed, in which the extraction aperture was set to $\sim 2''.4$ to minimize flux contamination from the host galaxy.
The relative airmass correction was applied using the APO atmospheric extinction coefficients compiled by J. Davenport \footnote{\href{http://astronomy.nmsu.edu:8000/apo-wiki/wiki/DIS}{http://astronomy.nmsu.edu:8000/apo-wiki/wiki/DIS}}.
All spectra were corrected for Galactic extinction using the \cite{fit99} extinction curve, under the assumption of $R_V=3.1$.
The sensitivity functions derived from the data of the spectrophotometric standard stars obtained on the first (low-resolution) and third (high-resolution) DIS observation epochs were used for the flux calibration.
Note that due to the differences in the grism tilt angles between observations runs, the fourth epoch blue-channel DIS spectrum could not be accurately calibrated.
For each DIS spectrum, we corrected for telluric absorption features in the wavelength ranges of $6,860-6,890$ and $7,600-7,630$~\AA\ (O${}_2$ bands) by dividing the KISS15s spectrum by a continuum-normalized spectrophotometric standard star spectrum.

Finally, the absolute flux scales of DIS spectra were calibrated by spectrophotometry.
Because the $i$-band is less affected by blue continuum emission from the star-forming host galaxy (Section~\ref{sec:hostproperties}) and the strong emission lines compared to $g$ and $r$-bands, we used the broken-line $i$-band light curve model (Figure~\ref{fig:lightcurve_5}; see Section~\ref{sec:powerlaw_fitting}) as the reference magnitude for spectrophotometry measurements.
High-resolution mode spectra did not cover a sufficient wavelength ranges; thus, we applied spectrophotometry only to low-resolution mode spectra. 
Specifically, the observed spectra were convolved with the SDSS $i$-band filter transmission curve using {\tt Speclite}; the spectrophotometric calibration factor for each spectrum was derived from the ratio of the $i$-band power-law fitted flux and the filter-convolved flux.

Figure~\ref{fig:spec_low} shows low-resolution ARC3.5-m/DIS spectra.
At early epochs, the DIS spectra showed consistent spectral features with the lower-resolution Nayuta/LISS spectrum. 
The broad H$\alpha$, H$\beta$, and \ion{Ca}{2} IR triplet emission lines are clearly visible.
In addition, an intermediate velocity width component can be identified on top of the broad H$\alpha$ emission line in both DIS and Nayuta/LISS spectra.
The broad component ($v_{\text{FWHM}} \sim 14,000$~km~s${}^{-1}$) had weakened considerably by 2016 October 4; however, the intermediate component ($v_{\text{FWHM}} \sim 2,000$~km~s${}^{-1}$) retained its high luminosity during the observations.

Several narrow emission lines, e.g., [\ion{O}{2}]$\lambda$3727, H$\beta$, [\ion{O}{3}]$\lambda\lambda$4959,5007, and H$\alpha$, are also clearly visible in the DIS spectra.
As discussed in Section~\ref{balmerfit}, these narrow components are probably due to the emission from extended \ion{H}{2} regions in the host galaxy and thus are not directly related to the SN explosion of KISS15s.
As for the narrow component, the different slit widths and variable seeing conditions during our observations were believed to be the main causes of the variation in the narrow line flux among DIS spectra.
As discussed in detail in Section~\ref{append:inst_profile}, the DIS red-arm spectrum obtained on 2016 October 4 has complex instrumental broadening profiles; this may have an effect on the apparent weakness of the narrow emission line in Figure~\ref{fig:spec_low}.

In Figure~\ref{fig:spec_low}, black-body spectra with temperatures of $T_{\text{BB}}=4,000$ and $6,000$ K are also shown for comparison.
It is clear that the observed optical spectra of KISS15s cannot be explained by a single temperature black-body spectrum.
This may imply that the optical light from KISS15s is affected by the host galaxy's dust extinction, as discussed in Section~\ref{note_on_host_extinction}.

Figure~\ref{fig:spec_high} presents high-resolution ARC3.5-m/DIS spectra.
Although we obtained the high-resolution spectra for the purpose of detecting a narrow P Cygni feature due to the CSM moving at the wind velocity \citep[expected to be $v_{w}\sim 10-100$~km~s${}^{-1}$;][]{mil10}, we found no P Cygni feature in the H$\alpha$ emission line profile of KISS15s.
Therefore, it was impossible to determine the wind velocity $v_w$ of the progenitor of KISS15s from these spectroscopic observations.
As mentioned above, the DIS red-arm spectrum obtained on 2016 October 4 suffers from complex instrumental broadening (Section~\ref{append:inst_profile}).
Thus, we interpreted the apparent line profile variation in the narrow emission line component between high-resolution spectra on February 12 and October 4, 2016, as artificial.

The multi-component H$\alpha$ emission line profile of KISS15s revealed by the ARC3.5-m/DIS confirms that this object is an SN IIn with strong ejecta-CSM interaction.
Particularly, the long-duration optical continuum light curves, IR emission excess, and complex spectral properties of KISS15s all suggest that KISS15s belongs to the ``1988Z-like'' subclass of SNe IIn \citep[e.g.,][]{str12,tad13,smi17}.

\subsection{Note on the dust extinction in the host galaxy}
\label{note_on_host_extinction}

\begin{figure*}[tbp]
\center{
\includegraphics[clip, width=6.4in]{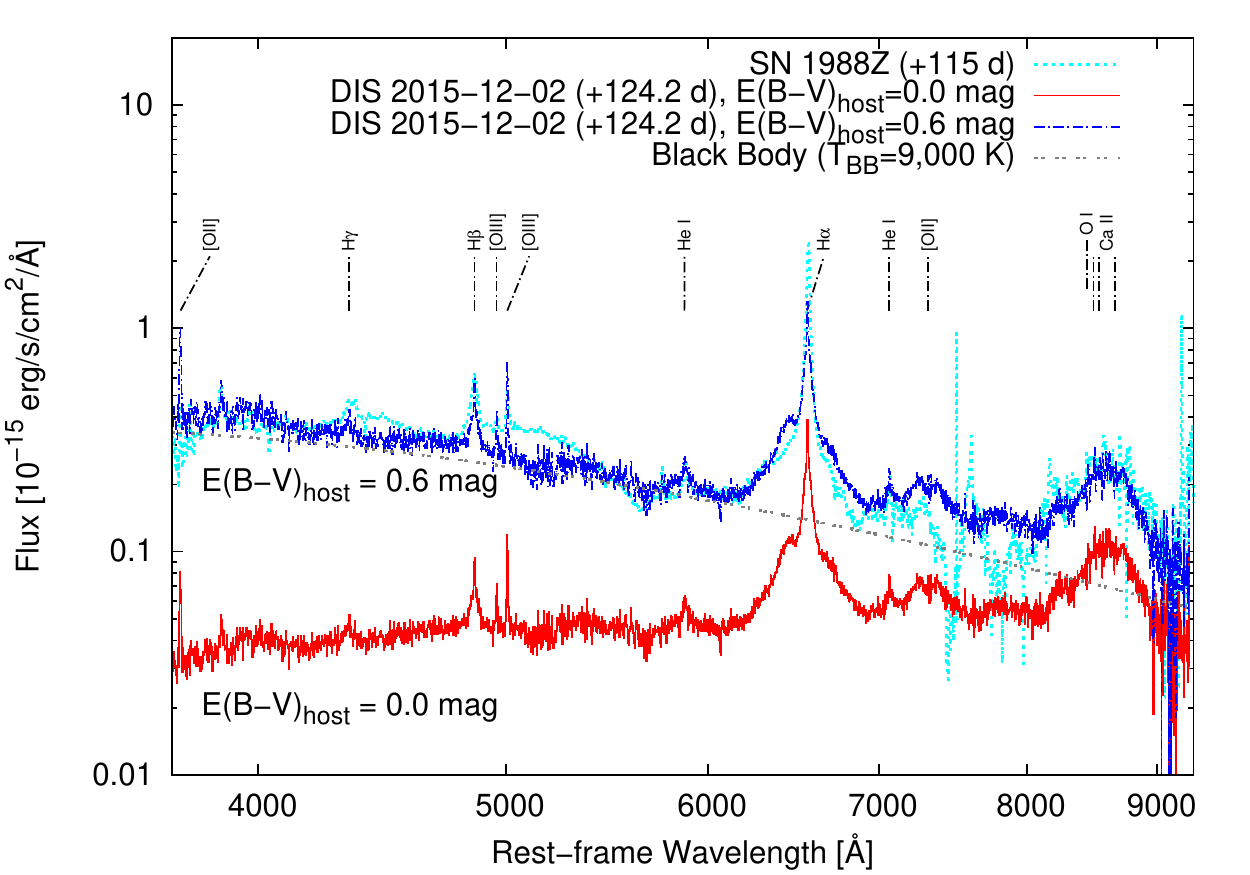}
}
 \caption{
 DIS low-resolution spectrum of KISS15s obtained on 2015 December 2 (+124.2~days since the first detection date), compared to a spectrum of SNIIn 1988Z at +115~days \citep[$z=0.022$;][]{tur93}.
 Also shown is the dereddened KISS15s spectrum assuming SMC-like extinction of $E(B-V)_{\text{host}}=0.6$ mag in the host galaxy. 
 A black-body spectrum with a temperature of $T_{BB}=9,000$ K is also shown for comparison.
 Note that the narrow emission lines seen in the KISS15s spectrum are probably from \ion{H}{2} regions in the host galaxy and are not directly related to KISS15s emission.
 }
 \label{fig:spec_comparison}
\end{figure*}

In previous studies, the strengths of the equivalent widths (EWs) of interstellar \ion{Na}{1} D $\lambda\lambda 5889, 5895$ absorption lines have often been used as the host galaxy extinction estimator for SNe through the relation of $E(B-V)_{\text{host}}=0.16 \times$ EW(\ion{Na}{1} D) \citep[][]{tur03,tad13}.
In the DIS spectrum of KISS15s, no clear absorption feature from \ion{Na}{1} D is detected.
However, it should be noted that the EW measurements of \ion{Na}{1} D absorption lines are claimed to provide poor constraints on reddening \citep{poz11}.
Moreover, circumstellar dusts related to the progenitor activity of KISS15s and newly formed dust in the ejecta-CSM interaction region may not be traced by gas absorption lines.
Therefore, the possibility of the presence of substantial dust extinction for the optical emission of KISS15s cannot readily be ruled out.

One estimate of the amount of dust extinction can be obtained by matching the KISS15s spectrum to those of other SNe IIn of a similar kind.
As pointed out in Section~\ref{data:dis}, the long duration optical light curves and the broad emission line profile of KISS15s are very similar to SN~1988Z (and its analogues), a typical SN IIn with strong ejecta-CSM interactions \citep[e.g.,][]{tur93,are99,str12,smi17}.
Figure~\ref{fig:spec_comparison} compares the DIS spectrum of KISS15s obtained on 2015 December 2 (+124~days since SkyMapper's first detection) to a +115~day spectrum of SN~1988Z downloaded from 
WISEREP \citep[][]{yar12}\footnote{The Weizmann Interactive Supernova data REPository --- WISeREP; \href{https://wiserep.weizmann.ac.il/spectra/list}{https://wiserep.weizmann.ac.il/spectra/list}.}.
In the same figure, we also plotted a DIS spectrum of KISS15s corrected for putative host galaxy dust extinction, on the assumption of Small Magellanic Cloud (SMC)-like extinction \citep[$R_{V}=2.93$;][]{pei92} with 
\begin{equation}
E(B-V)_{\text{host}}=0.6~\text{mag} 
\end{equation}
in the host galaxy of KISS15s.
We find that the overall spectral shape of KISS15s well matches that of SN~1988Z when the SMC-like extinction correction is applied.
Figure~\ref{fig:spec_comparison} shows that the spectral shape of the dereddened continuum of KISS15s can reasonably be explained by a single black-body of $T_{BB} \sim 9,000$~K.

In this work, we assumed that KISS15s is actually affected by the host galaxy extinction of $E(B-V)_{\text{host}}=0.6~\text{mag}$.
The extinction is either due to interstellar dust in the host galaxy or circumstellar dust around KISS15s.
$E(B-V)_{\text{host}}=0.6~\text{mag}$ corresponds to the observed-frame extinction of $A_{u, \text{host}}$, $A_{g, \text{host}}$, $A_{r, \text{host}}$, $A_{i, \text{host}}$, $A_{z, \text{host}}$, $A_{J, \text{host}}$, $A_{H, \text{host}}$, $A_{K, \text{host}}$, $A_{\text{W1}, \text{host}}$, and $A_{\text{W2}, \text{host}}$ = $2.972$, $2.183$, $1.530$, $1.148$, $0.870$,  $0.477$, $0.286$, $0.173$, $0.079$, and $0.052$~mag, respectively, on the assumption of the $R_{V}=2.93$ SMC-like extinction curve of \cite{pei92}.
This implies that the intrinsic emission of KISS15s is several magnitudes brighter than the observed emission, and that KISS15s is intrinsically a luminous SN ($M_g = m_{g} - \text{DM}- A_{g} - A_{g, \text{host}} \simeq -18.8$~mag at MJD = 57234; see Figure~\ref{fig:lightcurve_5}), which is $\sim 1$~mag brighter than SN~1988Z \citep[$M_{B}\sim-18$ mag;][]{tur93} at around the discovery epoch.

\section{Analyses}
\label{sec:result}

\subsection{Broad-band light curves}
\label{phot_lightcurve}

In this subsection we examine the details of the temporal evolution of the optical and IR luminosity of KISS15s, and compare the properties with those of other 1988Z-like SNe IIn.

\subsubsection{Optical light curves}
\label{sec:powerlaw_fitting}

\begin{figure}[tbp]
\center{
\includegraphics[clip, width=3.3in]{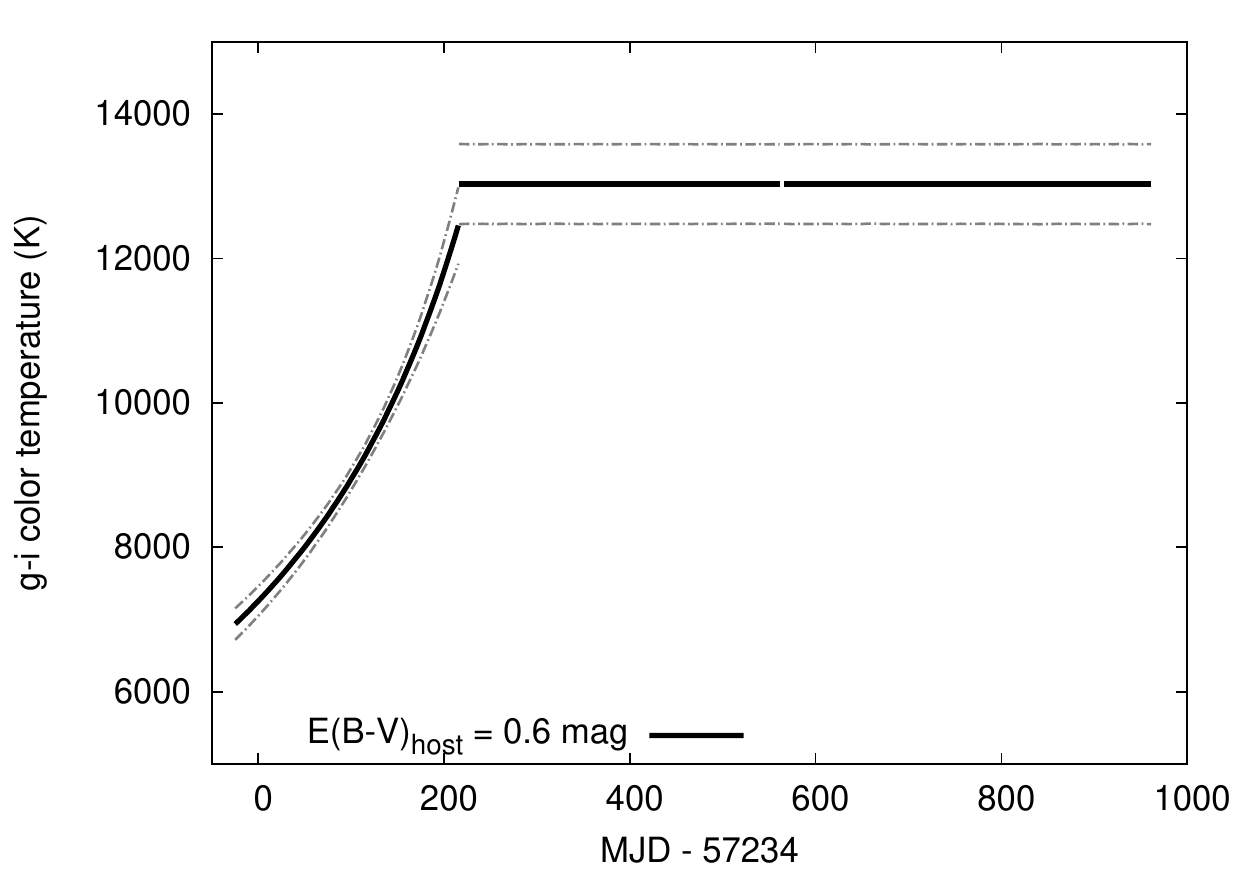}
}
 \caption{$g-i$ color temperature with 1$\sigma$ uncertainty calculated from $g$- and $i$-band broken-line light curve models (Section~\ref{sec:powerlaw_fitting}). The $g-i$ color is corrected for the host galaxy dust extinction of $E(B-V)_{\text{host}}=0.6$~mag.
 }
 \label{fig:lc_opt_bb_tmp}
\end{figure}

\begin{figure*}[tbp]
\center{
\includegraphics[clip, width=6.3in]{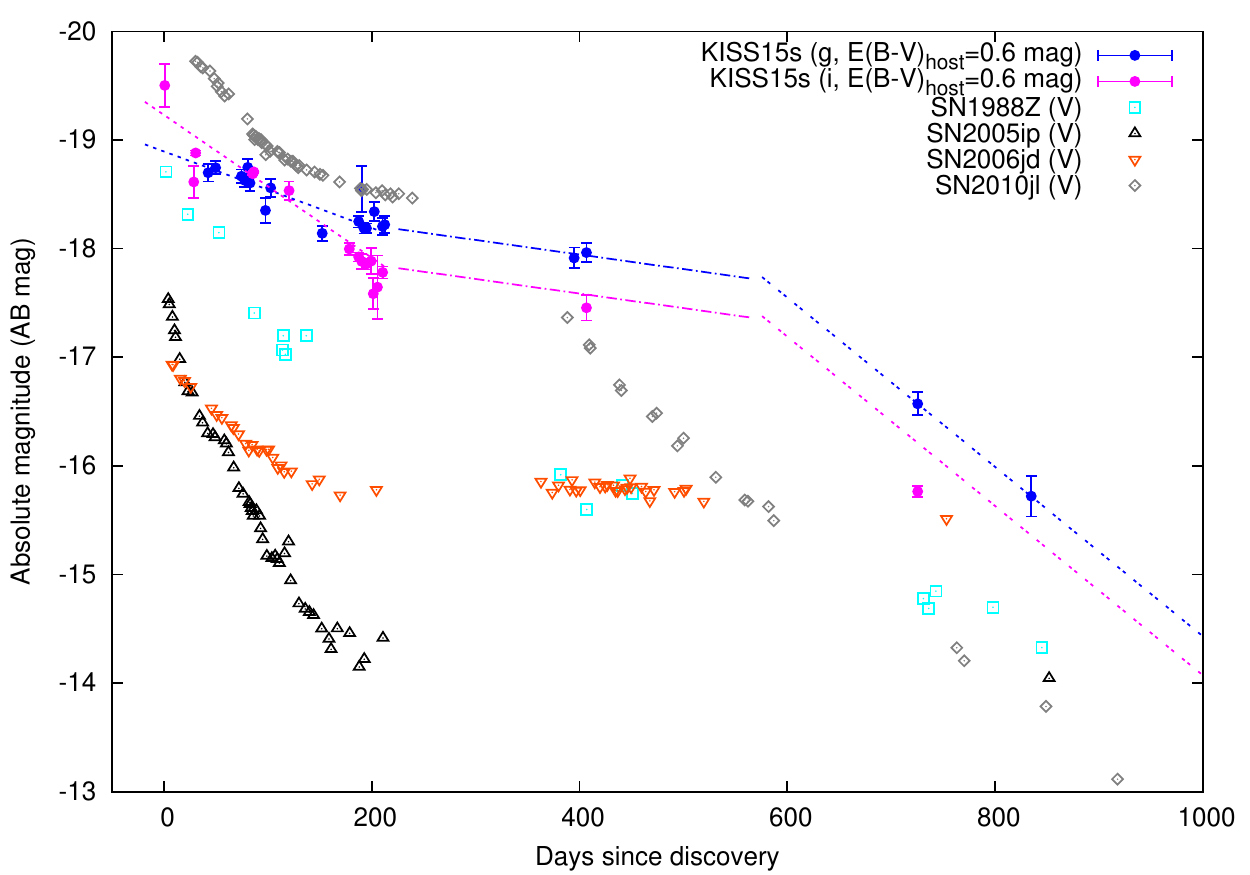}
}
 \caption{Comparisons of the optical absolute magnitudes between KISS15s and other 1988Z-like SNe IIn. The magnitudes of KISS15s are corrected for the host galaxy extinction of $E(B-V)_{\text{host}} = 0.6$~mag. The $V$-band light curves of SNe IIn, SN~1988Z \citep{tur93}, SN~2005ip, SN2006jd \citep{str12}, and SN2010jl \citep{fra14} are taken from the Open Supernova Catalog \citep{gui17}.
 Galactic extinction is corrected, and the host galaxy extinction is uncorrected.
 The horizontal axis is the observed-frame days since discovery.
 $V$-band Vega magnitudes are converted into AB magnitudes assuming an AB-Vega magnitude offset of 0.02~mag \citep{bla07}.
 For KISS15s, the broken-line light curve models are also shown (Section~\ref{sec:powerlaw_fitting}).
 }
 \label{fig:lc_comparisons}
\end{figure*}

The $g$- and $i$-band measurements at early epochs (MJD < 57800) revealed that the optical light curves of KISS15s showed a slow decline for $\sim 500-600$~days since the first detection by SkyMapper at MJD = 57234 (Figure~\ref{fig:lightcurve_5}).
After that, KISS15s experienced a rapid decline in luminosity, as discussed in Section~\ref{obs:mayall}.
Although the $r$- and $z$-bands are contaminated by strong H$\alpha$ and \ion{Ca}{2} IR triplet emission, respectively, the $g$- and $i$-band fluxes are dominated by optical continuum emission.

If the $g$-band luminosity light curve at MJD < 57800 is assumed to be a power-law:
\begin{equation}
L_{g}\propto t^{\alpha},
\label{g_light curve}
\end{equation}
the observed magnitude can be expressed as
\begin{equation}
g = c_g - 2.5 \alpha \log({\rm MJD} - {\rm MJD}_0),
\label{g_mag_fit}
\end{equation}
where $\alpha$ is the power-law index in Equation~\ref{g_light curve}, $c_{g}$ is the $g$-band magnitude at ${\rm MJD} = {\rm MJD}_0 + 1$, and ${\rm MJD}_0$ is a model parameter corresponding to the SN explosion date.
By fitting Equation~\ref{g_mag_fit} to the Galactic extinction-corrected $g$-band light curve of KISS15s at MJD < 57800, we obtain $c_g = 17.29^{-0.76}_{+0.50}$~mag, $\alpha = -0.45^{-0.11}_{+0.08}$, and ${\rm MJD}_0 = 57209.2_{-36.1}^{+22.5}$.
${\rm MJD}_0 = 57209.2$ corresponds to 2015 July 6, which implies that the explosion date of KISS15s is probably $\sim 1$ month before the first detection by SkyMapper on MJD=57234.
The best-fit power-law model is shown in Figure~\ref{fig:lightcurve_5}.

Next, to evaluate the temporal evolution of the $g-i$ color, we constructed broken-line light curve models for the $g$- and $i$-bands as follows.
First, the light curve was divided into three parts ($\text{MJD} < 57450$, $57450 < \text{MJD} < 57800$, and $57800 < \text{MJD}$), and a linear regression line was fitted separately to each.
At $\text{MJD} < 57450$, linear regression lines for the $g$- and $i$-band magnitude light curves were calculated as the light curve model.
At $57450 < \text{MJD} < 57800$, the $g$-band light curve was modeled by a linear regression line of $g$-band data between MJD = 57400 and 57800.
In the same way, the $g$-band light curve at $57800 < \text{MJD}$ was modeled by a linear regression line of the two late-time $g$-band measurements from Mayall/KOSMOS and Blanco/DECam.
Given that $i$-band data are too sparse to directly estimate $g-i$ color at $\text{MJD} > 57450$, we simply assumed a constant $g-i$ color fixed to an average value evaluated from the data between MJD = 57420 and 57450.
The Galactic extinction-corrected $g-i$ color at $57420  < \text{MJD} < 57450$ was evaluated as $g-i = 0.67 \pm 0.03~\text{mag}$; we assumed that this color remained constant at $\text{MJD} > 57450$:
\begin{equation}
g-i = 0.67 \pm 0.03~\text{mag} \ \ \ \ \ (\text{MJD} > 57450)
\end{equation}

The best-fit linear light curve models are shown in Figure~\ref{fig:lightcurve_5}.
The $r$-band shows a clear excess relative to the $g$- and $i$-band light curves, particularly in late epochs ($> 600$~days since discovery), indicating a stronger flux contribution from the H$\alpha$ emission in the $r$-band at later epochs.
With the $g$- and $i$-band broken-line light curve models, the $g-i$ color corrected for the host galaxy extinction of $E(B-V)_{\text{host}} = 0.6$~mag and the corresponding $g-i$ color temperature were calculated; the $g-i$ black-body color as a function of black-body temperature was calculated using {\tt Speclite} and assuming a source redshift of $z=0.03782$.
Then the temperatures of the $g-i$ colors of KISS15s were searched.
Figure~\ref{fig:lc_opt_bb_tmp} shows the $g-i$ color temperature as a function of time, along with the uncertainty due to broken-line light curve modeling.
On the assumption of a SMC-like host galaxy extinction of $E(B-V)_{\text{host}} = 0.6$~mag (Section~\ref{note_on_host_extinction}), the $g-i$ color at $\text{MJD} > 57450$ becomes $g-i = -0.36 \pm 0.03$~mag, and the $g-i$ color temperature is
\begin{equation}
T_{\text{BB,opt}} = 13,000~\text{K} \pm 600~\text{K}\ \ \ \ \ \text{(MJD > 57450)},
\end{equation}
as shown in Figure~\ref{fig:lc_opt_bb_tmp}.

The slow time evolutions of the optical light curves suggest that the CSM interaction is the dominant source of the optical luminosity of KISS15s (rather than the radioactive decay of ${}^{56}$Ni and ${}^{56}$Co), similar to other SNe IIn.
The rapid decline in luminosity in late epochs ($>$ 600 days since discovery) implies that the volume of the continuum emission region begins to suddenly decrease at this epoch.
Figure~\ref{fig:lc_comparisons} compares the absolute magnitude optical light curves of KISS15s with other 1988Z-like SNe IIn\footnote{Discovery dates of SN~1988Z \citep{tur93}, SN~2005ip, SN~2006jd \citep{str12}, SN~2010jl \citep{sto11}, and KISS15s (this work) are MJD = 47507.5, 53679.16, 54020.54, 55478.64, and 57234, respectively.}.
In Figure~\ref{fig:lc_comparisons}, the magnitudes are corrected for the putative host galaxy extinction of $E(B-V)_{\text{host}} = 0.6$~mag.
The long-duration optical light curve of KISS15s is a common characteristic of the SN~1988Z-like subclass of SN IIn \citep{str12,smi17b,smi17}.
Figure~\ref{fig:lc_comparisons} shows that after host galaxy extinction correction, KISS15s is one of the most luminous and longest duration SN IIn among the well-observed samples of SNe IIn.
The 1988Z-like SNe IIn share remarkably similar early- and late-phase photometric and spectroscopic properties at the optical wavelengths; in fact, KISS15s shows emission line profiles similar to SN~1988Z, the prototype of the 1988Z-like SNe IIn (see Figure~\ref{fig:spec_comparison}).
The sudden drop in the continuum emission at the late epoch probed by $g$- and $i$-band photometry of KISS15s may correspond to a similar luminosity drop observed in SN~2010jl at $\sim 320$~days since discovery. 
In the case of SN~2010jl, the sudden luminosity drop has been interpreted such that the shocked shell escapes the dense (bipolar) CSM at the time of the transition \citep{fra14,mor14b}; the same scenario may be the case for the optical luminosity drop observed in KISS15s.

\subsubsection{IR light curves}
\label{sec:ir_lightcurves}

\begin{figure*}[tbp]
\center{
\includegraphics[clip, width=6.3in]{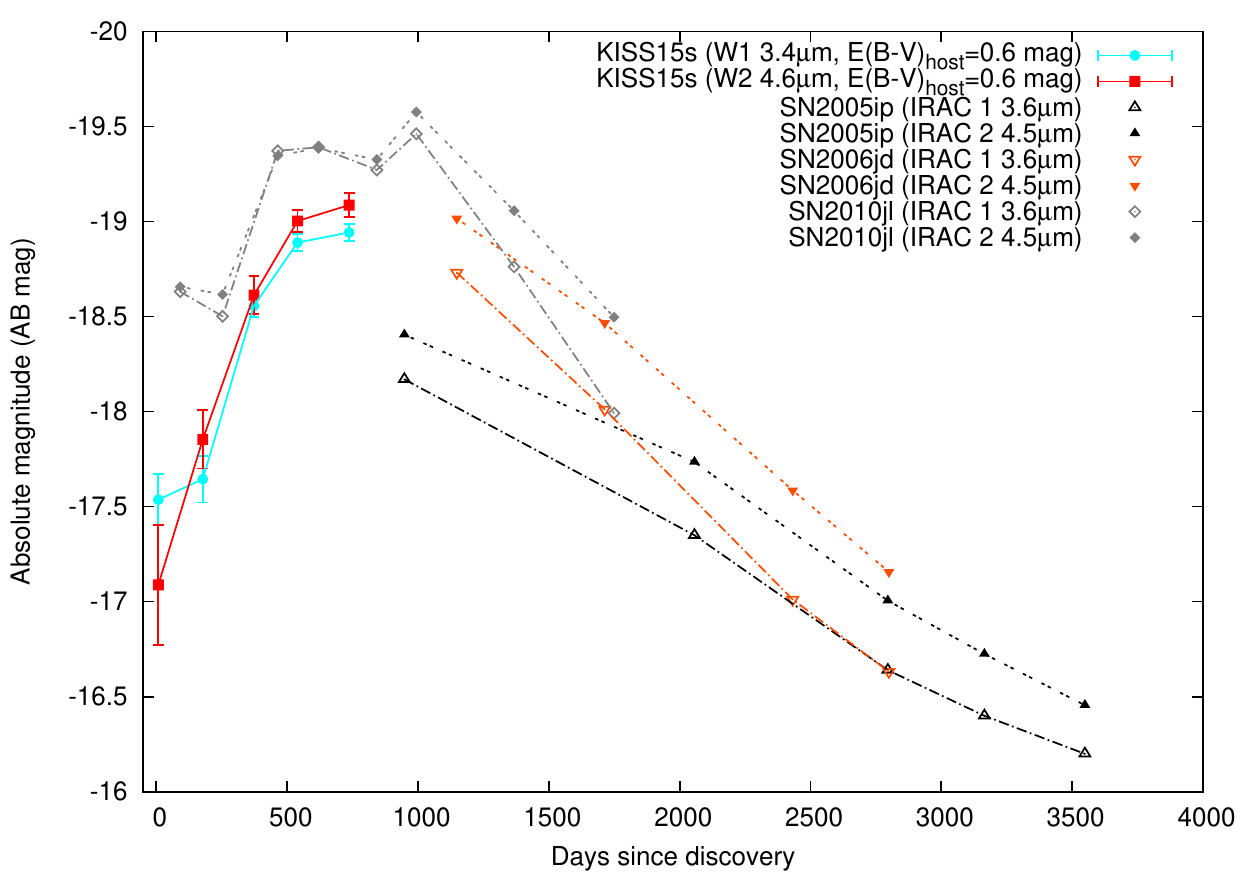}
}
 \caption{Comparisons of the IR absolute magnitudes between KISS15s and other 1988Z-like SNe IIn. The Spirzer/IRAC 3.6 and 4.5~$\mu$m band light curves of SNe IIn, SN~2005ip \citep{fox10,fox13,sza18}, SN2006jd \citep{fox11,fox13,sza18}, and SN2010jl \citep{and11,fox13,fra14,sza18}, taken from the Open Supernova Catalog \citep{gui17}.
 The host galaxy extinction of $E(B-V)_{\text{host}} = 0.6$~mag is corrected.
 The horizontal axis is the observed-frame days since discovery.
 The Vega magnitudes of the Spirzer/IRAC 3.6 and 4.5~$\mu$m bands are converted into AB magnitudes assuming  AB-Vega magnitude offsets of 2.78 and 3.26~mag, respectively \citep{irac15}.
 }
 \label{fig:lc_ir_comparisons}
\end{figure*}

\begin{figure}[tbp]
\center{
\includegraphics[clip, width=3.3in]{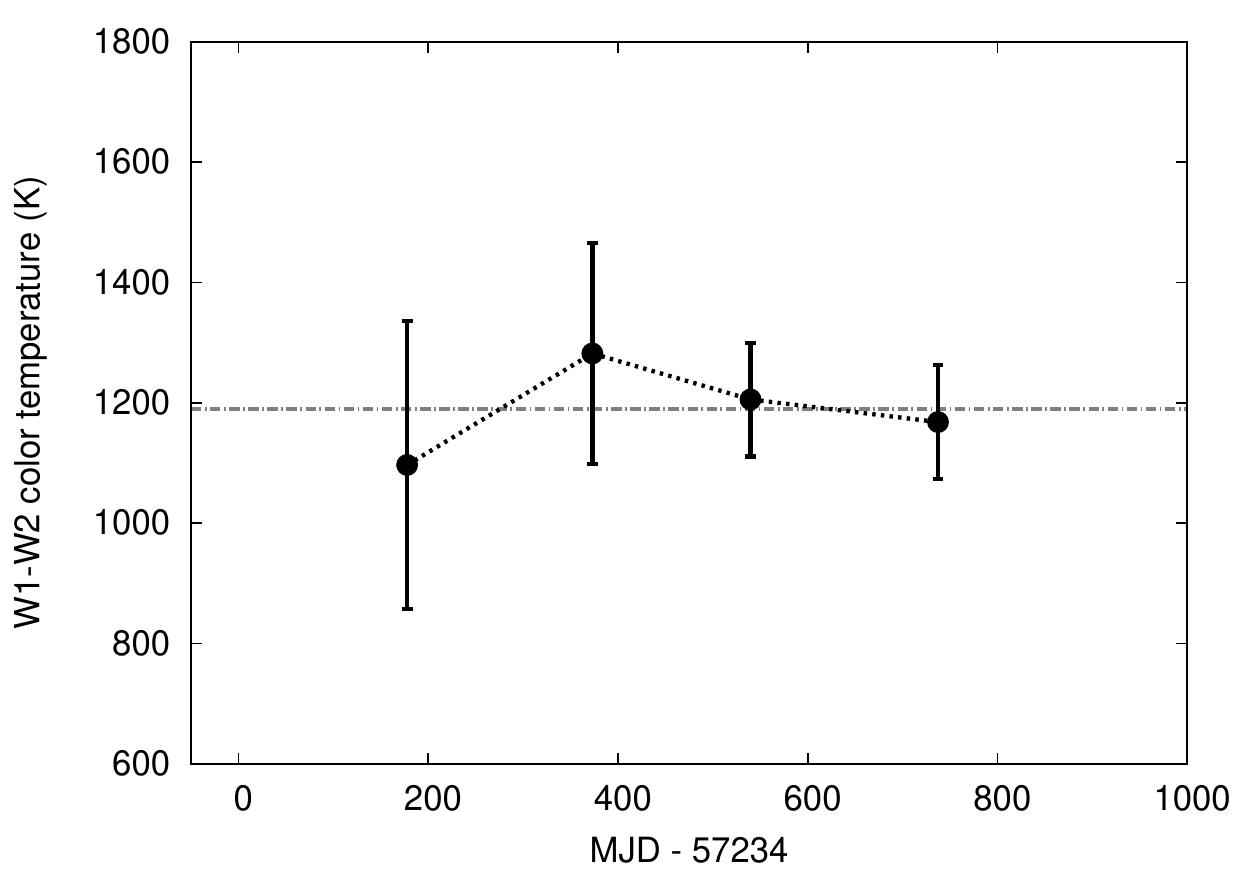}
}
 \caption{
 WISE W1 - W2 color temperatures as a function of observed epochs. The photometry data are corrected for the host galaxy extinction of $E(B-V)_{\text{host}}=0.6$~mag.
 The first epoch W1- and W2-band fluxes at 7.5~days (Figure~\ref{fig:wise_sed}) are dominated by the optical black-body component, and are excluded from the plot.
 The weighted average of the W1-W2 color temperature is 1,190~K, as denoted by a horizontal line.
 }
 \label{fig:wise_color_temperature}
\end{figure}

\begin{figure*}[tbp]
\center{\includegraphics[clip, width=6.3in]{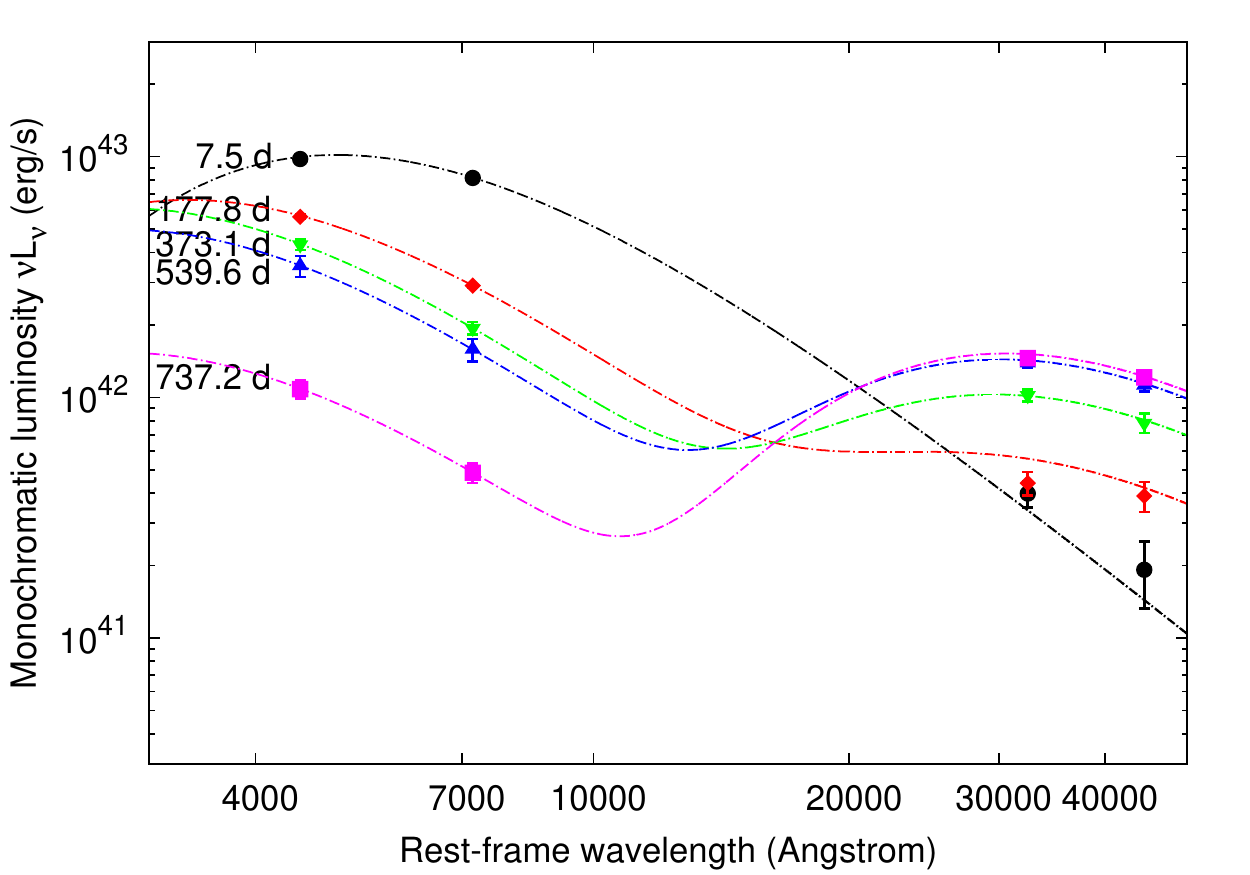}
}
 \caption{
 $g$-, $i$-, W1-, and W2-band spectral energy distribution (SED) of KISS15s at the epochs of the NEOWISE observations (indicated as the days since discovery, defined as MJD$-$57234).
 The optical points are interpolated by the broken-line light curve models shown in Figure~\ref{fig:lightcurve_5}.
 The photometry data are dereddened by assuming SMC-like dust extinction in the host galaxy of $E(B-V)_{\text{host}} = 0.6$~mag. 
Optical + IR black-body model spectra, composed of an optical black-body with temperature $T_{\text{BB,opt}}$ estimated from the $g-i$ color (Figure~\ref{fig:lc_opt_bb_tmp}) and IR black-body with temperature $T_{\text{BB, IR}} = 1,190$~K (Figure~\ref{fig:wise_color_temperature}) are also shown.
 }
 \label{fig:wise_sed}
\end{figure*}

\begin{figure}[tbp]
\center{
\includegraphics[clip, width=3.3in]{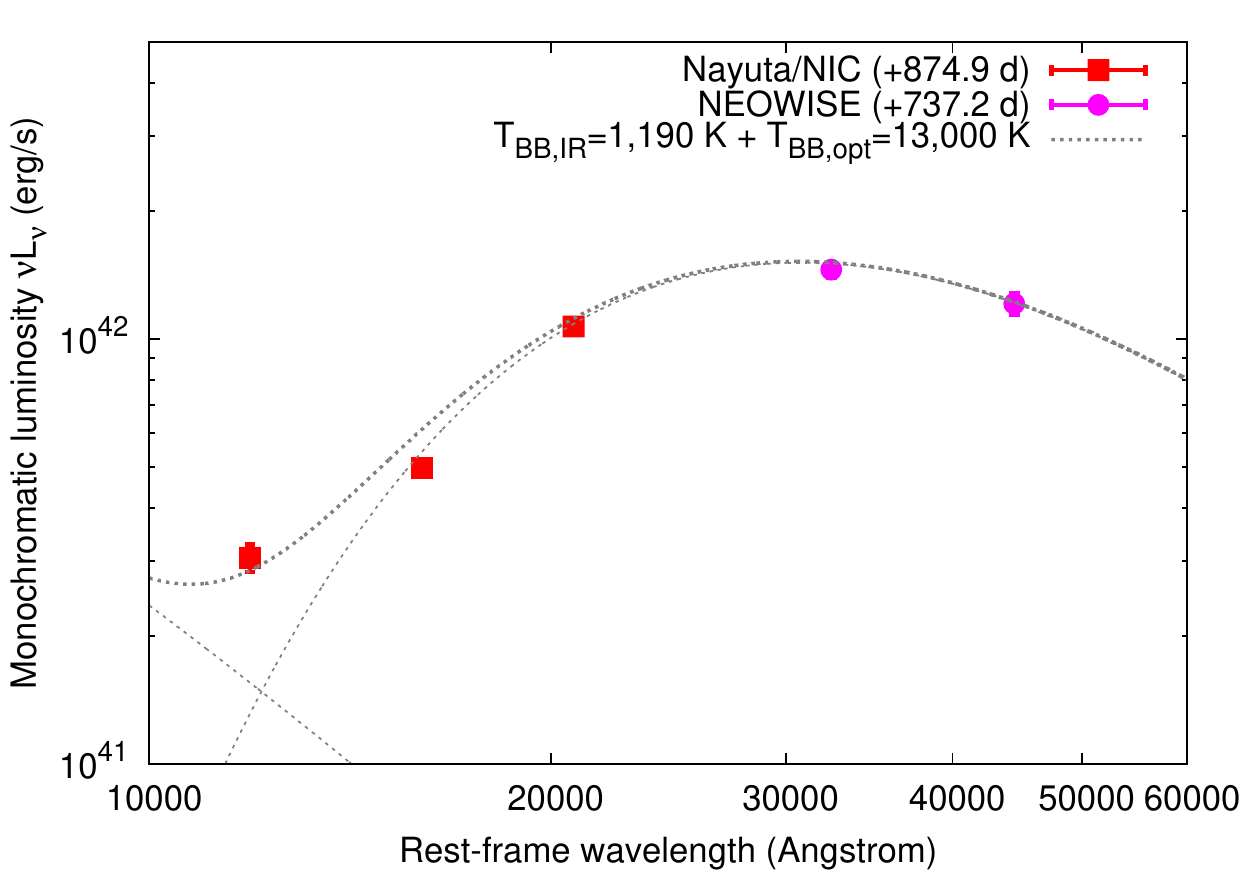}
}
 \caption{IR broad-band SED. $J$-, $H$-, and $K_{s}$-band data from Nayuta/NIC were obtained on 2017 December 21-22 ($\sim$ 874.9~days since discovery).
 W1- and W2-band data from NEOWISE obtained on 2017 August 5-6 ($\sim$ 737.2~days since discovery; same as those shown in Figure~\ref{fig:wise_sed}) are also shown for comparison. 
 The photometry data are corrected for the host galaxy extinction of $E(B-V)_{\text{host}}=0.6$~mag.
 The dotted lines are the same black-body model shown in Figure~\ref{fig:wise_sed} (737.2~days).
 }
 \label{fig:nayuta_nic}
\end{figure}

\begin{figure}[tbp]
\center{
\includegraphics[clip, width=3.3in]{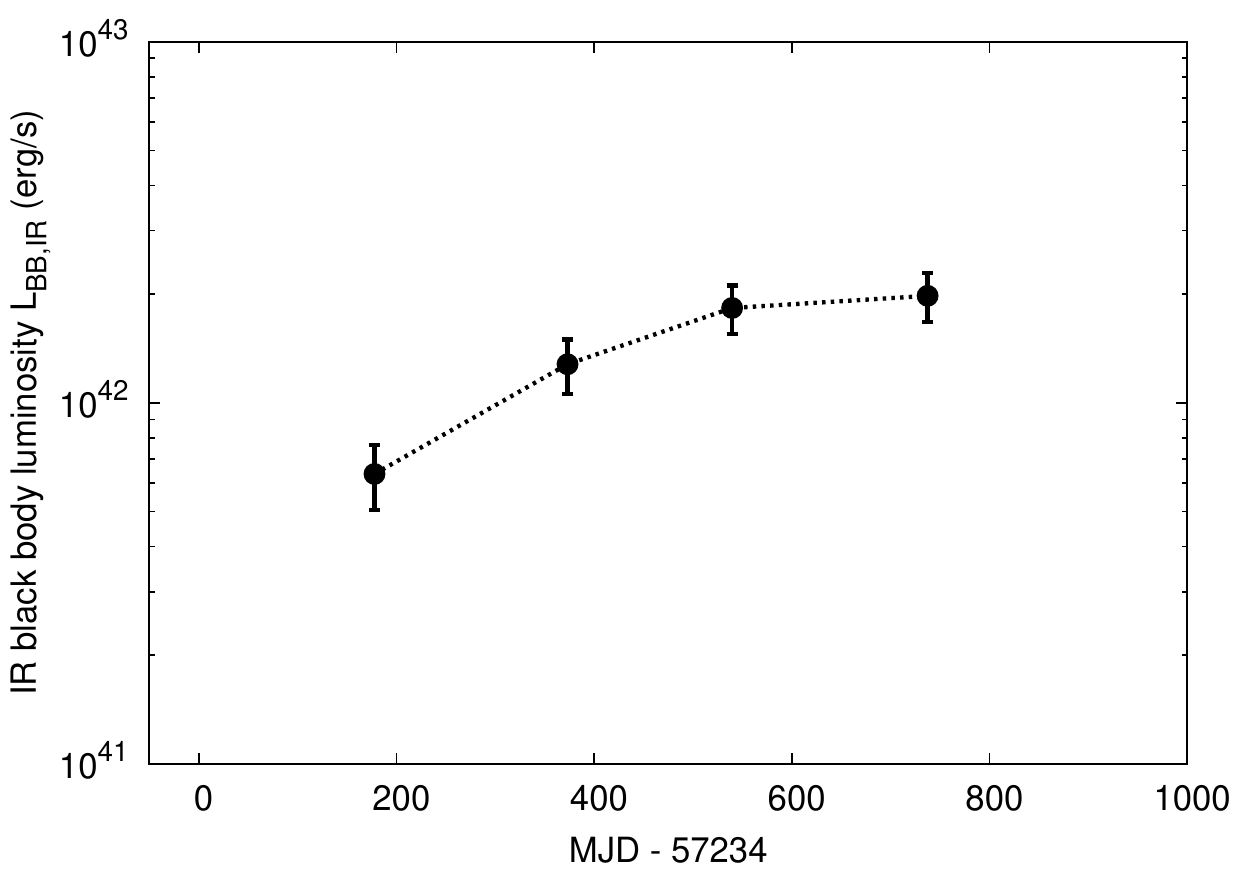}
\includegraphics[clip, width=3.3in]{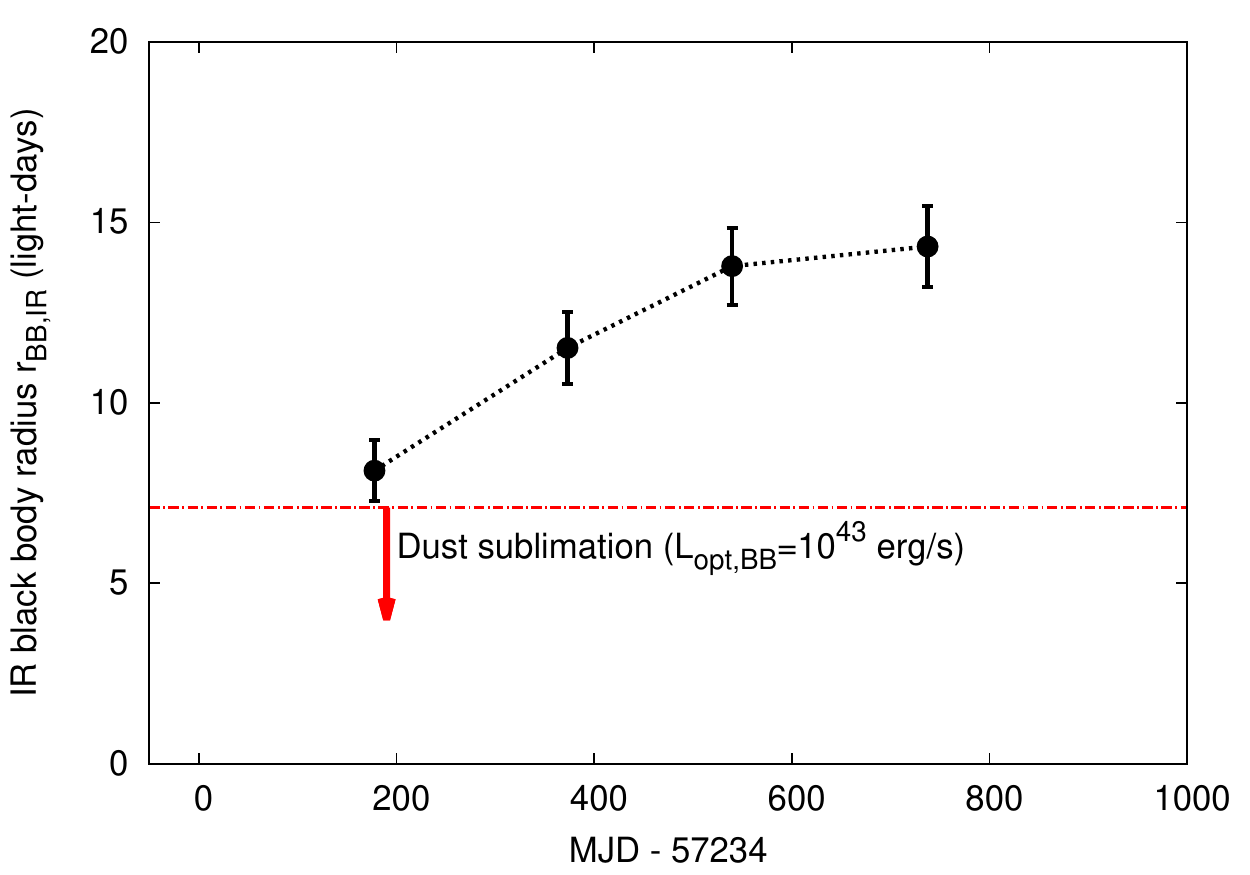}
\includegraphics[clip, width=3.3in]{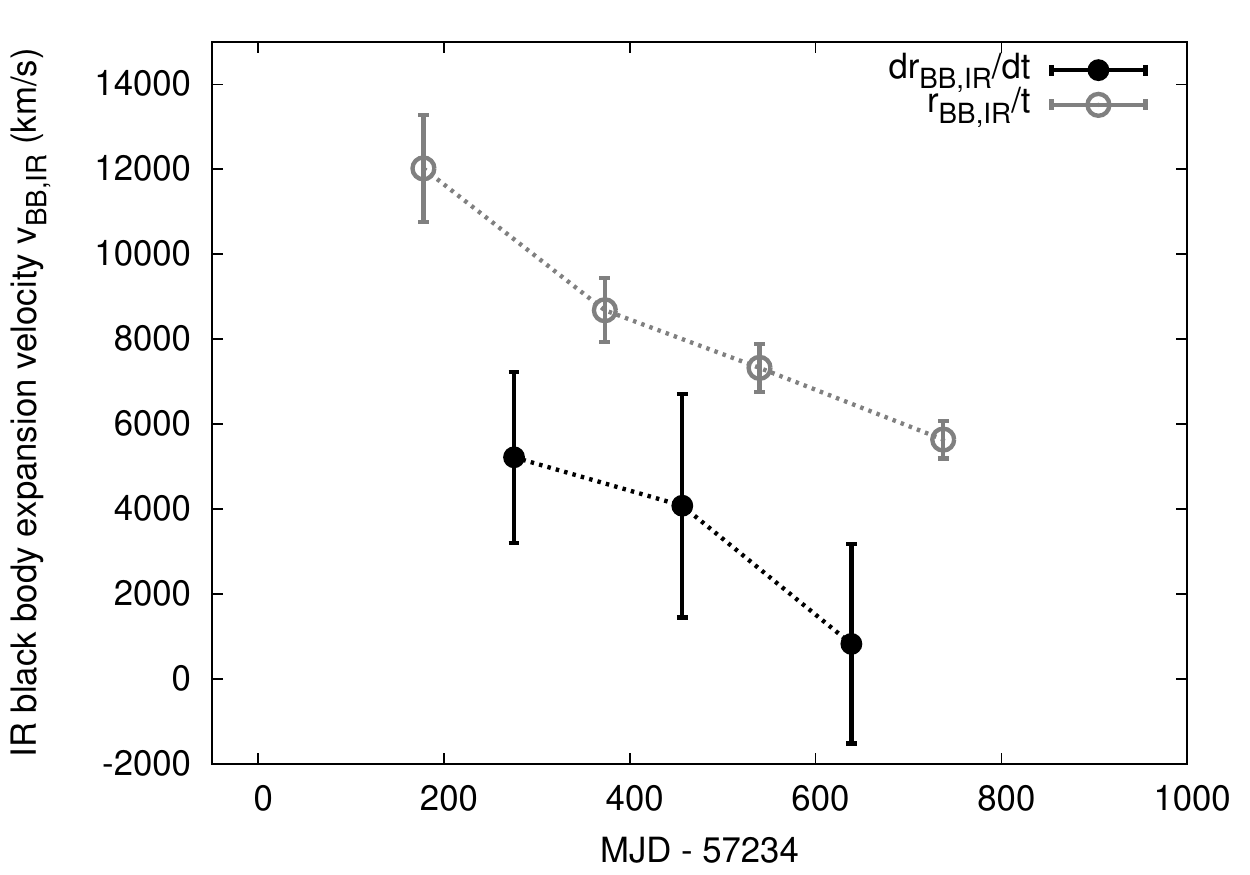}
}
 \caption{Top: the IR black-body luminosities on the assumption of $T_{\text{BB, IR}} = 1,190$~K $\pm 60$~K. The first epoch NEOWISE measurement (7.5~days since discovery) is dominated by the optical black-body component of $T_{\text{BB}} \sim 7,400$~K (Figure~\ref{fig:wise_sed}) and is excluded from the plot. 
 The photometry data are corrected for the host galaxy extinction of $E(B-V)_{\text{host}}=0.6$~mag.
 Middle: the black-body radius $r_{\text{BB, IR}}$ calculated using Equation~\ref{eqn:bb_radius}. 
 The horizontal line indicates a dust sublimation radius $r_{\text{sub}}$ in the case of $L_{\text{opt, BB}}=10^{43}~\text{erg}~\text{s}^{-1}$ (Equation~\ref{eqn:sub}). 
 Bottom: inferred expansion velocity of the black-body radius $v_{\text{dust}}$ (Equations~\ref{eqn:linear_velocity} and \ref{eqn:diff_velocity}).
 }
 \label{fig:wise_luminosity}
\end{figure}

The $3-4~\mu$m IR emission of SNe~IIn can readily be identified as hot dust thermal emission \citep{smi09,fox10,mae13,gal14,sza18}.
Graphite/silicate dust grains sublimate at temperatures of $T_{\text{sub}} = 1400-1800$~K \citep[e.g.,][]{guh89,bas18}, and the $3-4~\mu$m IR emission corresponds to dust thermal emission close to the dust sublimation temperature.
Figure~\ref{fig:lc_ir_comparisons} compares the W1- and W2-band light curves of KISS15s with several years of Spitzer/IRAC band 1 (3.6~$\mu$m) and band 2 (4.5~$\mu$m) light curves for other SN~1988Z-like SNe IIn (SN~2005ip, SN~2006jd, and SN~2010jl) taken from the literature.
Compared to the Spitzer/IRAC sample of SNe IIn, NEOWISE data of KISS15s enable the temporal evolution of the IR emission to be followed from the very early epochs ($\sim 7$~days after the first detection).
The peak IR luminosity of KISS15s is similar to those of other SN~1988Z-like SNe~IIn, suggesting that the IR emission mechanism in KISS15s is identical to them.

As for SN~2010jl, although the significant IR excess above the extrapolated UV-optical emission at <~1~year since discovery is most likely due to the IR echo of the initial flash of the SN by preexisting dust in the CSM \citep{sar18}, the late-time IR emission associated with dust-induced increasingly blueshifted line profiles are due to new dust formation in the CDS \citep{and11,mae13,gal14,sza18}.
As discussed below, the first W1- and W2-band data points of KISS15s (at 7.5~days since discovery) are consistent with the Rayleigh-Jeans tail of the optical black-body emission, as evidenced by the blue W1-W2 color.
We found that the W1- and W2-band dust thermal emission light curves of KISS15s increase smoothly, without any early-phase excess at <~1~year, as opposed to SN~2010jl.
The smooth IR light curves of KISS15s indicates that the early-phase IR echo is unimportant for KISS15s and that the IR emission is mainly from newly formed dust in the CDS, suggested as the origin of hot dust thermal emissions of other SNe IIn \citep[e.g.,][]{fox10,fox11,mae13,gal14,fra14}.

Figure~\ref{fig:wise_color_temperature} shows the temporal evolution of the $W1- W2$ color temperature of the IR emission component of KISS15s at 177.5, 373.1, 539.6, and 737.2~days since discovery, where the W1-W2 black-body colors of various temperatures were calculated using {\tt Speclite}, assuming a source redshift of $z=0.03782$.
Here, we also assumed that the flux contribution from the Rayleigh-Jeans tail of the optical black-body emission is not significant at these epochs, and the observed W1-W2 color was directly used to calculate the color temperature.
The W1-W2 color of KISS15s shows little time variation, and the color temperature remains almost constant during the observations at $\sim 1,200$~K.
The weighted average of the W1-W2 color temperature is 
\begin{equation}
T_{\text{BB, IR}} = 1,190~\text{K}~(\pm 60~\text{K}).
\end{equation}
It should be noted that dust grains formed inside of SN ejecta observed in type II-P SNe generally have dust temperatures of at most several hundreds of Kelvin \citep{sza18}; thus, the persistently high dust temperature of KISS15s can definitely be attributed to the newly formed dust grains in the CDS, as observed in other SNe IIn.
The estimated dust temperature is slightly lower than the sublimation temperature of silicate and graphite grains \citep[$T_\text{sub} = 1,400$ and $1,800$~K, respectively; e.g.,][]{bas18}, indicating that both of these dust grain species are likely contributors to the IR continuum emission of KISS15s.

In contrast to the increasing IR light curves, the optical light curves of KISS15s decrease over time during the observations.
Figure~\ref{fig:wise_sed} presents the $g$-, $r$-, W1-, and W2-band optical-to-IR broad-band spectral energy distribution (SED) of KISS15s at the epochs of NEOWISE W1- and W2-band observations, where the optical data points at $g$- and $i$-bands are interpolated by the broken-line light curve models shown in Figures~\ref{fig:lightcurve_5} and \ref{fig:lc_comparisons}.
The photometry data in Figure~\ref{fig:wise_sed} are corrected for the host galaxy extinction of $E(B-V)_{\text{host}} = 0.6$~mag.
In Figure~\ref{fig:wise_sed}, the best-fit optical + IR black-body model spectrum for the SED of each epoch is also shown, where the time-dependent optical black-body temperature is from Figure~\ref{fig:lc_opt_bb_tmp}; the IR black-body is assumed to have a constant temperature of $T_{\text{BB, IR}} = 1,190~\text{K}$ and is scaled to the W2-band magnitudes.
As for the continuum emission, the IR emission increases as the optical emission fades, and the $3-4~\mu$m IR luminosity is comparable to the optical luminosity at about 2~years after the discovery.
This observation is consistent with the suggestion that new dust formation in the SN environment is enhanced several hundreds of days after the SN explosion \citep[e.g.,][]{gal14,sar18}.
Figure~\ref{fig:wise_sed} shows that after the correction of $E(B-V)_{\text{host}} = 0.6$~mag, the earliest-phase IR luminosities (7.5 days since discovery) can fully be attributed to Rayleigh–Jeans tail emission in the optical black-body emission of $T_{\text{BB}} = 7,400$~K (see Figure~\ref{fig:lc_opt_bb_tmp}).
This provides additional evidence that KISS15s is heavily reddened by SN-related/unrelated dust in the host galaxy. 
In addition, the non-detection of the IR black-body component at the earliest phase (7.5~days since discovery) excludes the possibility of a large IR flux contribution from IR echoes of the SN shock breakout radiation by preexisting dust in close proximity to KISS15s at this epoch.

Figure~\ref{fig:nayuta_nic} presents the IR broad-band SED from Nayuta/NIC $J$-, $H$-, and $K_{s}$-band data on 2017 December 21-22 (874.9~days since discovery) and from NEOWISE W1- and W2-band data on 2017 August 5-6 (737.2~days since discovery).
Although these data were not strictly simultaneously obtained, we can conclude that the overall IR SED shape is consistent with the $T_{\text{BB,IR}}=1,190$~K black-body SED.
The perfect black-body IR SED shape of KISS15s suggests that the typical radius $a$ of the dust grains responsible for the IR continuum is much larger than the observed IR wavelengths ($2 \pi a \gtrsim \lambda_{\text{rest}}$; i.e., $a \gtrsim 0.3-0.6$~$\mu$m).
The $J$-band photometry shows a clear excess from the IR black-body SED, suggesting that the short IR wavelengths are affected by the flux contamination from unmodeled host galaxy components, the Rayleigh-Jeans tail of the optical black-body of KISS15s (Figure~\ref{fig:wise_sed}), and also from strong emission lines in the $J$-band (\ion{He}{1}, Pa$\gamma$, and Pa$\beta$).

The constant temperature and the increasing luminosity indicate that the surface area of the dust thermal emission region is expanding.
Figure~\ref{fig:wise_luminosity} shows the integrated IR black-body luminosity as a function of observation epoch, $L_{\text{BB, IR}}(t)$, calculated from the W2-band magnitudes of KISS15s by assuming a constant black-body temperature of $T_{\text{BB, IR}} = 1,190$~K.
If we assume that the dust distribution is approximated as a homogeneous thin shell, the black-body radius can be defined as \citep{fox10,fox11}
\begin{eqnarray}
r_{\text{BB, IR}}(t) &=& \left(\frac{L_{\text{BB, IR}}(t)}{4\pi\sigma_{\text{sb}}T_{\text{BB, IR}}^{4}}\right)^{1/2} \nonumber\\
&=& 14.4~\text{light-days} \times \left( \frac{L_{\text{BB, IR}}(t)}{2\times 10^{42}~\text{erg~s}^{-1}} \right)^{1/2} \left( \frac{T_{\text{BB, IR}}}{1,190~\text{K}} \right)^{-2},
\label{eqn:bb_radius}
\end{eqnarray}
where $\sigma_{\text{sb}}$ is Stefan-Boltzmann's constant, and $r_{\text{BB, IR}}(t)$ is increasing over time.
The numerical calculation results of $r_{\text{BB, IR}}(t)$ are shown in the middle panel of Figure~\ref{fig:wise_luminosity}.
Note that the actual dust radius can be a few times larger than $r_{\text{BB, IR}}$ if the dust distribution is inhomogeneous.
The bottom panel of Figure~\ref{fig:wise_luminosity} shows the expansion velocities of the IR black-body radius, $v_{\text{dust}}$.
Two definitions of the velocities are possible: one is 
\begin{equation}
v_{\text{dust}} = \frac{r_{\text{BB, IR}}(t)}{t} = \frac{ r_{\text{BB, IR}}(\text{MJD}) }{\text{MJD} - \text{MJD}_{0}}
\label{eqn:linear_velocity}
\end{equation}
and the other is
\begin{equation}
v_{\text{dust}} = \frac{dr_{\text{BB, IR}}(t)}{dt} \simeq \frac{ \Delta r_{\text{BB, IR}}(\text{MJD}) }{\Delta \text{MJD}},
\label{eqn:diff_velocity}
\end{equation}
where the former represents the hypothetical linear velocity required for the dust shell to reach $r_{\text{BB, IR}}(t)$ at a given $t$, and the latter is the dust shell velocity between two adjacent epochs.
We assume an explosion date of $\text{MJD}_{0} = 57209.2$ (Section~\ref{sec:powerlaw_fitting}) with Equation~\ref{eqn:linear_velocity}.

It should be noted that the derived $r_{\text{BB, IR}}(t)$ is more than an order of magnitude smaller than the light-crossing distance since the SN explosion of KISS15s ($\text{MJD}_{0} \sim 57209.2$; Section~\ref{sec:powerlaw_fitting}).
This suggests that the IR echo model is not the main source of the late-phase NIR excess emission.
The estimated dust shell velocity $v_{\text{dust}}$ is consistent with the ejecta and/or shock front velocities inferred from the velocity width of broad/intermediate H$\alpha$ emission line components (see Sections~\ref{sec:spec_lightcurve} and \ref{sec:massloss}), suggesting that the dust grains are located in the proximity of the ejecta-CSM interaction region or CDS.
These results, the late-time IR excess, the hot dust temperature, and the inferred expansion velocity, are consistent with the scenario that the hot dust thermal emission of KISS15s is due to newly formed dust grains in the CDS.

To examine whether the dust grains can really be formed against the strong UV-optical radiation field around the SN, we can compare the derived dust shell size estimate with a dust sublimation radius $r_{\text{sub}}$, within which the dust grains are completely sublimated due to intense UV-optical radiation from the SN.
The dust sublimation radius is given as \citep{wax00,vel16}
\begin{equation}
r_{\text{sub}} = \left( \frac{L_{\text{abs}}}{16 \pi \sigma_{\text{sb}}T_{\text{sub}}^4} \frac{\langle Q_{\text{UV}}\rangle}{\langle Q_{\text{IR}}\rangle} \right)^{1/2},
\end{equation}
where $\langle Q_{\text{UV}}\rangle$ and $\langle Q_{\text{IR}}\rangle$ are the Planck-averaged UV-optical absorption and IR emission efficiencies of the dust grains, respectively, and $L_{\text{abs}}$ is the SN UV-optical luminosity absorbed by the dust.
Because graphite grains have a higher sublimation temperature than silicate grains, here we consider the case of the graphite sublimation radius defined by $T_{\text{sub}} = 1,800$~K.
As described above, we can approximate $\langle Q_{\text{UV}}\rangle \sim \langle Q_{\text{IR}}\rangle \sim 1$ on the assumption of a large dust grain size \citep[e.g.,][]{fox10}.
We can expect that the UV-optical luminosity of KISS15s is always, at most, on the order of $L_{\text{abs}} \sim L_{\text{BB, opt}} \sim 10^{43}~\text{erg}~\text{s}^{-1}$ (Figure~\ref{fig:wise_sed}).
Thus,
\begin{equation}
r_{\text{sub}} = 7.1~\text{lt-days}~\left( \frac{L_{\text{BB, opt}}}{10^{43}~\text{erg}~\text{s}^{-1}} \right)^{1/2} \left( \frac{T_{\text{sub}}}{1,800~\text{K}} \right)^{-2}.
\label{eqn:sub}
\end{equation}
The dust sublimation radius of $7.1~\text{light days}$ is compared to the black-body radius $r_{\text{BB, IR}}$ in the middle panel of Figure~\ref{fig:wise_luminosity}.
The dust shell radius is larger than the dust sublimation radius ($r_{\text{BB, IR}} > r_{\text{sub}}$), which indicates that the dust grains can indeed exist at the inferred location of $r_{\text{BB, IR}}$.

On the assumption of single size dust grains and that the dust grains are mostly optically thin to IR emission, the total dust mass $M_{d}$ required to produce the observed IR emission can be estimated to be \citep[e.g.,][]{smi_geh05}
\begin{equation}
M_{d} = \left( \frac{a\rho}{3\sigma_{\text{sb}} \langle Q_{\text{IR}}\rangle T_{d}^4} \right)L_{\text{BB, IR}},
\label{eqn:dustmass}
\end{equation}
where $\rho$ is the mass density of the dust grain.
Substituting the mass density of a graphite grain ($\rho = 2.25$ g~cm${}^{-3}$), a grain size of $a=0.5$~$\mu$m, a dust temperature of $T_{d} = 1,190$~K, $L_{\text{BB, IR}} = 2 \times 10^{42}$~erg~s${}^{-1}$ (Figure~\ref{fig:wise_luminosity}), and $\langle Q_{\text{IR}}\rangle \sim 1$ into Equation~\ref{eqn:dustmass}, we obtain
\begin{equation}
M_{d} \simeq 3 \times 10^{-4} M_{\odot} \left( \frac{a}{0.5~\mu\text{m}} \right) \left( \frac{T_{d}}{1,190~\text{K}} \right)^{-4} \left( \frac{L_{\text{BB, IR}}}{2\times 10^{42}~\text{erg}~\text{s}^{-1}} \right).
\end{equation}
It should be noted that the estimated total mass of the newly formed large dust grains is consistent with that observed in SN~2010jl \citep{gal14}.
The derived value sets a lower limit of the total dust mass if the dust IR emission is self-shielded.
The dust grains radiating $\sim 3~\mu$m IR emission will undergo rapid radiative cooling.
Therefore, KISS15s may currently harbor a cool dust component of a mass of $\gg 3 \times 10^{-4}~M_{\odot}$, which may be detected by future follow-up observations at MIR and longer wavelengths.

\subsubsection{Bolometric luminosity}
\label{sec:bolometric}

\begin{figure}[tbp]
\center{
\includegraphics[clip, width=3.3in]{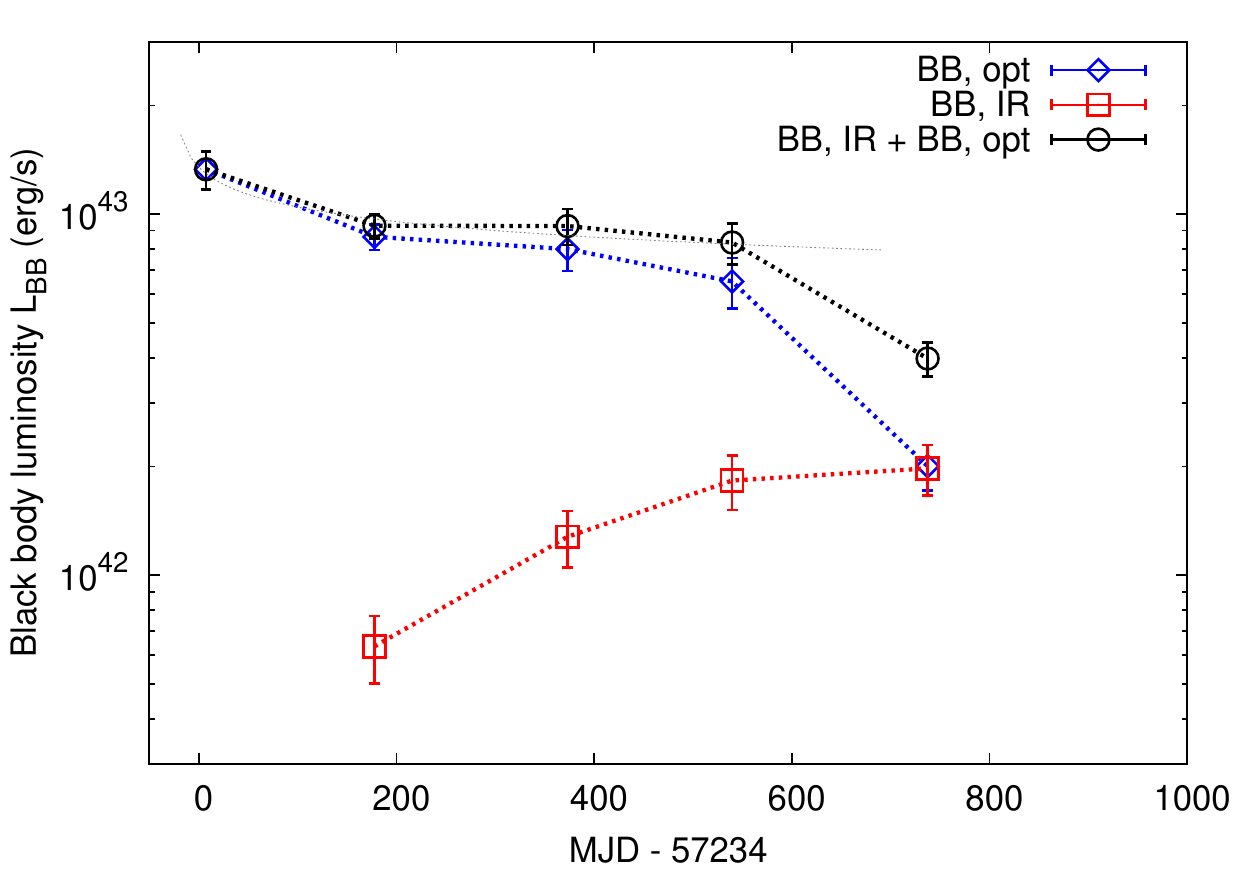}
}
 \caption{Pseudo-bolometric light curves of KISS15s. The optical black-body luminosity $L_{\text{BB,opt}}$ is calculated by assuming the temporal evolution of the $g-i$ color temperature $T_{\text{BB,opt}}$ shown in Figure~\ref{fig:lc_opt_bb_tmp}.
 The IR black-body luminosity is the same as that shown in the top panel of Figure~\ref{fig:wise_luminosity}.
 The host galaxy extinction of $E(B-V)_{\text{host}} = 0.6$~mag is assumed.
 The best-fit power-law model ($L_{\text{bol}} \propto t^{-0.16}$) is also shown.
 }
 \label{fig:lc_opt_bb_lum}
\end{figure}

An estimate of a bolometric luminosity light curve can be obtained from black-body fitting for optical and IR data points, as described in Sections~\ref{sec:powerlaw_fitting} and \ref{sec:ir_lightcurves}.
Because the IR black-body component of KISS15s can naturally be attributed to the newly formed dust in the ejecta-CSM interaction region (rather than IR echoes from preexisting dust), the bolometric luminosity is estimated to be the sum of optical and IR black-body components \citep[][]{fra14}.
Here we define the pseudo-bolometric luminosity as the sum of the optical and IR black-body luminosities, $L_{\text{bol}} = L_{\text{BB,opt}} + L_{\text{BB,IR}}$, which can be calculated from the best-fit black-body models shown in Figure~\ref{fig:wise_sed}.

Figure~\ref{fig:lc_opt_bb_lum} shows the temporal evolution of the optical and IR black-body luminosities and the pseudo-bolometric luminosity of KISS15s.
The power-law index of the optical luminosity is $L_{\text{BB,opt}} \propto t^{-0.23}$ on the assumption of $\text{MJD}_{0} = 57209.2$ (Section~\ref{sec:powerlaw_fitting}).
The relative luminosity contribution of the IR luminosity relative to the optical luminosity becomes stronger at later epochs; consequently, the optical + IR pseudo-bolometric luminosity shows slower temporal evolution compared to the optical luminosity.
The power-law index of the pseudo-bolometric luminosity light curve is given by
\begin{equation}
L_{\text{bol}} \propto t^{-0.16}
\label{eqn:bolometric_index}
\end{equation}
at $< 600$~days since discovery.

After the rapid decline in the optical luminosity at $>600$~days, the pseudo-bolometric luminosity accordingly decrease suddenly, as the IR luminosity remains nearly constant at these epochs.
This behavior of the bolometric luminosity evolution indicates that the sudden decrease in optical luminosity is not due to dust extinction by newly formed dust; instead, this behavior supports a scenario in which the intrinsic weakening of the ejecta-CSM interaction causes accelerated fading of the optical luminosity, as described in Section~\ref{sec:powerlaw_fitting} \citep[see][]{mae13,mor14b}.

The peak luminosity of KISS15s is estimated to be $L_{\text{peak}} \sim 1.0 \times 10^{43}~\text{erg}~\text{s}^{-1}$, and the total radiated energy at the first 600~days is roughly $\sim 4.0 \times 10^{50}~\text{erg}$.
This estimate indicates that KISS15s is one of the most luminous SNe IIn discovered in the local universe so far, comparable to the luminous ($L_{\text{peak}} \sim 3.0 \times 10^{43}~\text{erg}~\text{s}^{-1}$) type IIn SN~2010jl \citep{fra14}.
In the case of SN~2010jl, if the observed luminosity is assumed to be solely powered by ${}^{56}$Ni $\rightarrow$ ${}^{56}$Co $\rightarrow$ ${}^{56}$Fe radioactive decay, it requires an unreasonably large ${}^{56}$Ni mass of $M_{{}^{56}\text{Ni}} \sim 3.4~M_{\odot}$ \citep[see][for details]{mor13}.
From the same discussion, the required ${}^{56}$Ni mass to account for the luminosity observed in KISS15s is $M_{{}^{56}\text{Ni}} \sim 1.1~M_{\odot}$.
Although the ${}^{56}$Ni mass production over $1~M_{\odot}$ may be possible with the SNe of massive stars \citep[e.g.,][]{ume08}, a more natural explanation for the high luminosity, as well as the persistent light curves even after the decay time of ${}^{56}$Co \citep[$\tau_{\text{Co}} = 111.3$~days;][]{nad94}, is that KISS15s is powered by the strong ejecta-CSM interaction as is the case for other luminous SNe IIn.

\subsection{Emission line properties}
\label{line_property}

\subsubsection{Spectral profile of the H$\alpha$ emission line}
\label{balmerfit}

\begin{figure}[tbp]
\center{
\includegraphics[clip, width=3.2in]{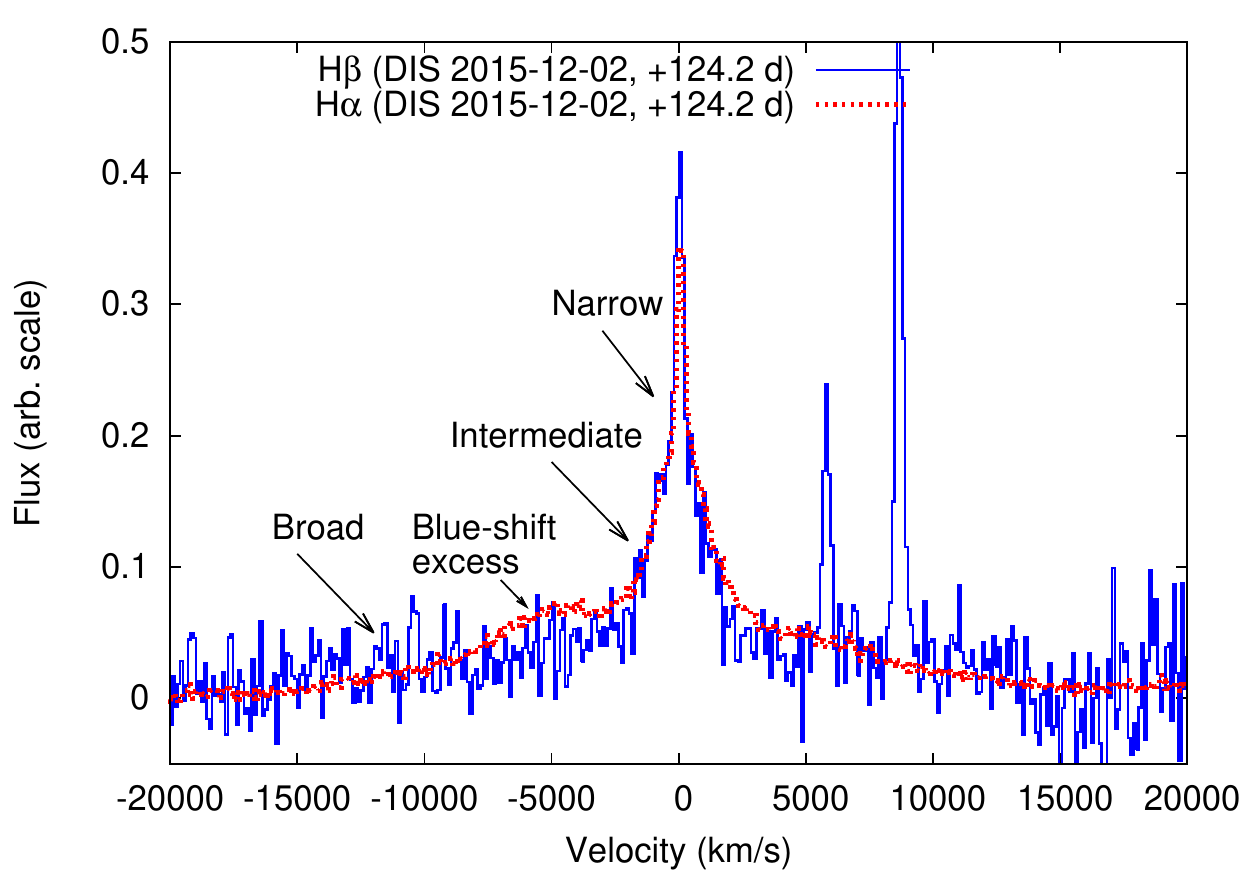}
\includegraphics[clip, width=3.2in]{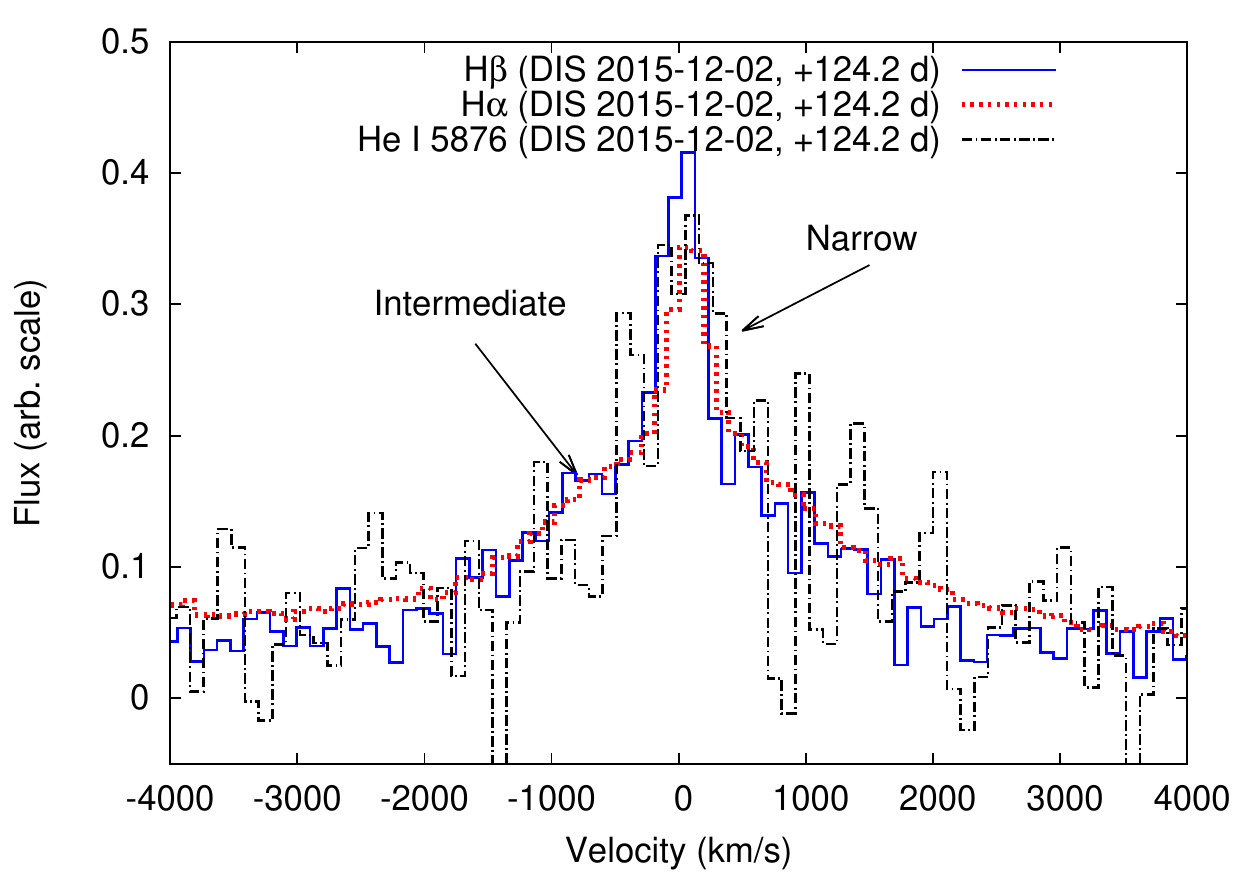}
}
 \caption{
Top panel: comparison between the spectral profiles of H$\alpha$ and H$\beta$ emission lines (DIS low-resolution spectrum obtained on 2015 December 2).
The systemic redshift is defined as $z=0.03782$, and the lower velocities correspond with shorter wavelengths. The continuum-subtracted flux is scaled to roughly match the line flux of the intermediate-width components.
Note that the two narrow lines at $v \sim 5,000-9,000$ are [\ion{O}{3}]$\lambda\lambda 4959,5007$ emission lines from the host galaxy.
Bottom panel: a close-up of the same spectra as the top panel.
In the bottom panel, the spectral profile of the \ion{He}{1} $\lambda$5876 emission line is also shown.
 }
 \label{fig:spec_halphahbeta}
\end{figure}

\begin{figure}[tbp]
\center{
\includegraphics[clip, width=3.2in]{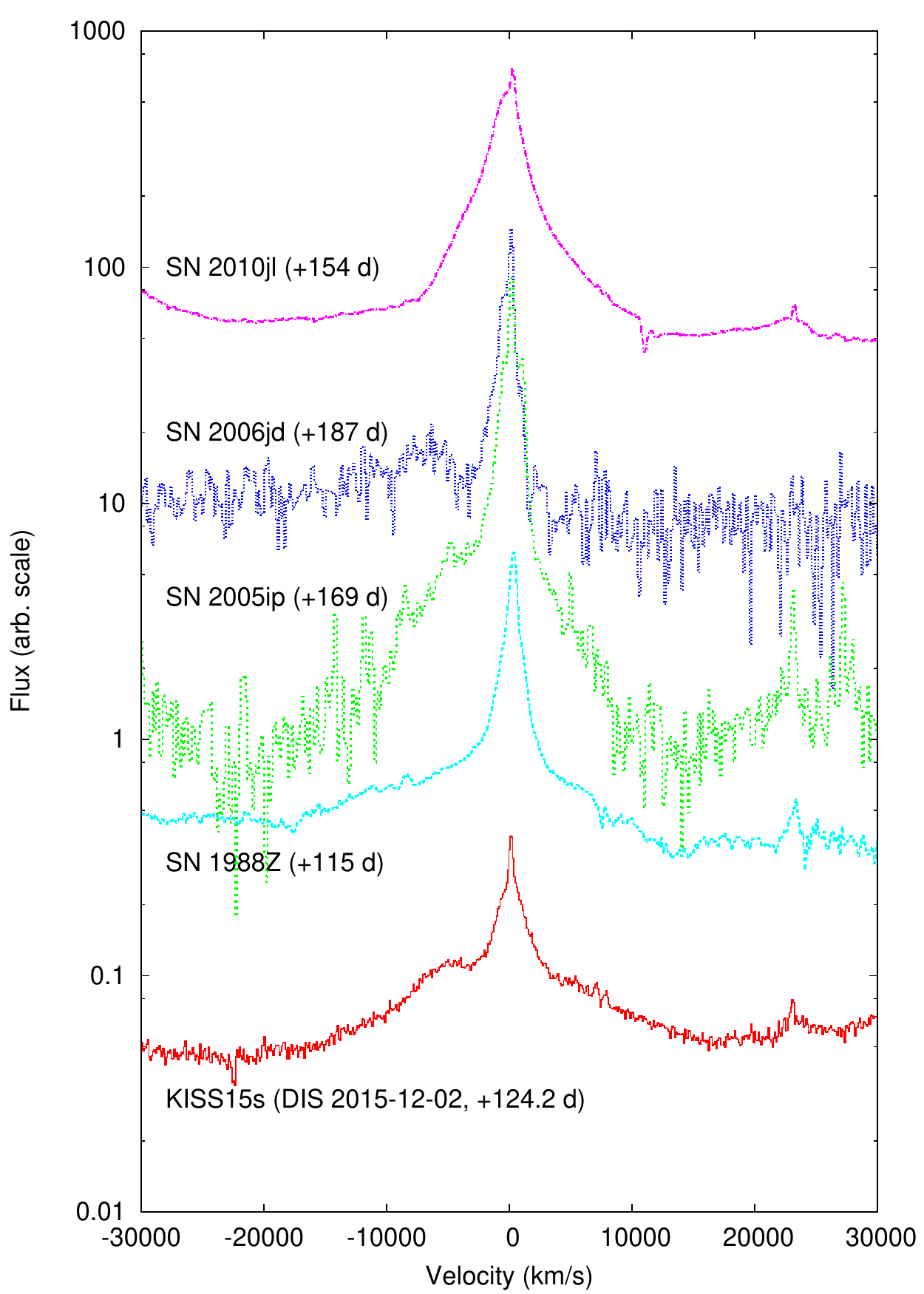}
}
 \caption{Comparisons between the H$\alpha$ profiles of KISS15s and other SN~1988Z-like SNe IIn during similar phases, SN~1988Z \citep{tur93}, SN~2005ip, SN~2006jd \citep{str12}, and SN~2010jl \citep{jen16}. The spectral data of the SNe IIn were taken from WISeREP.
 }
 \label{fig:spec_halpha_comparison}
\end{figure}

\begin{figure*}[tbp]
\center{
\includegraphics[clip, width=3.2in]{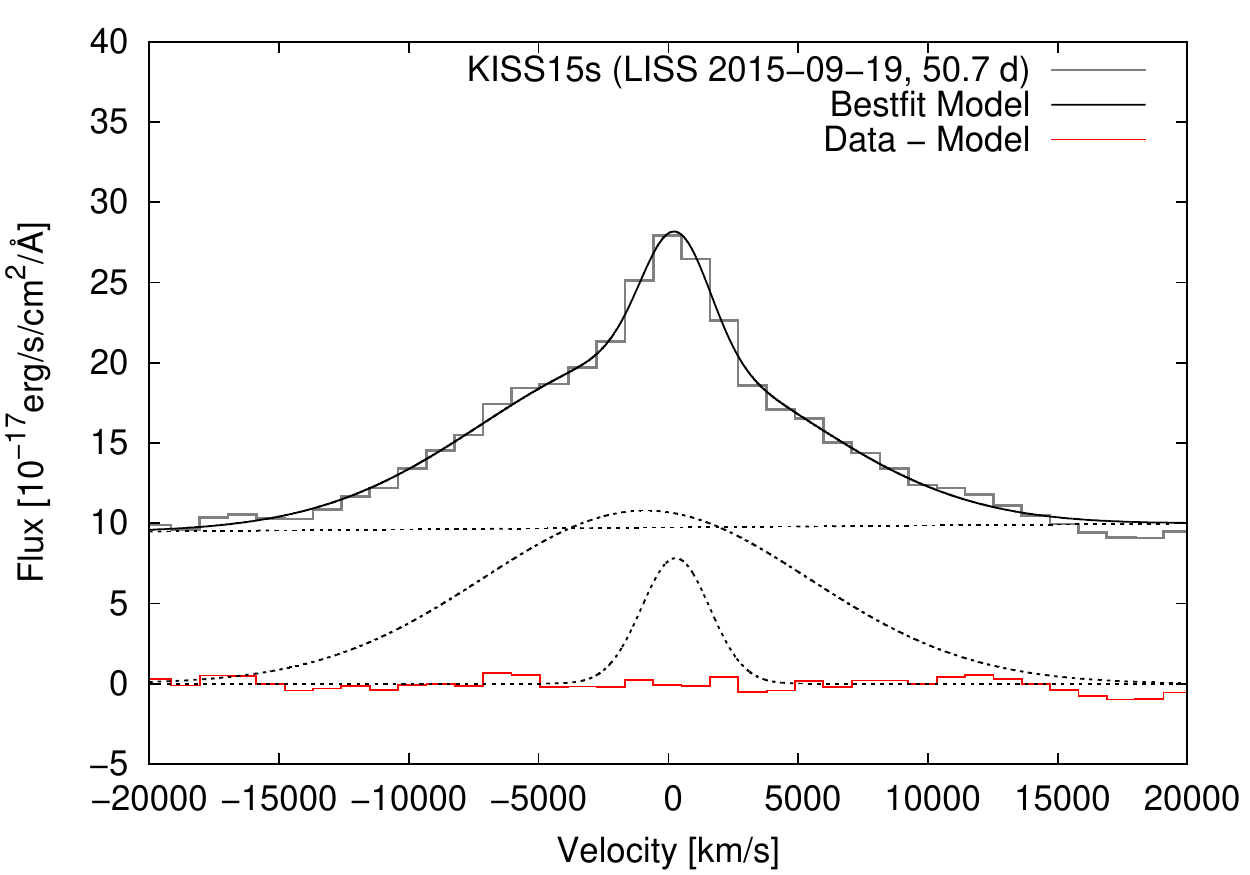}
\includegraphics[clip, width=3.2in]{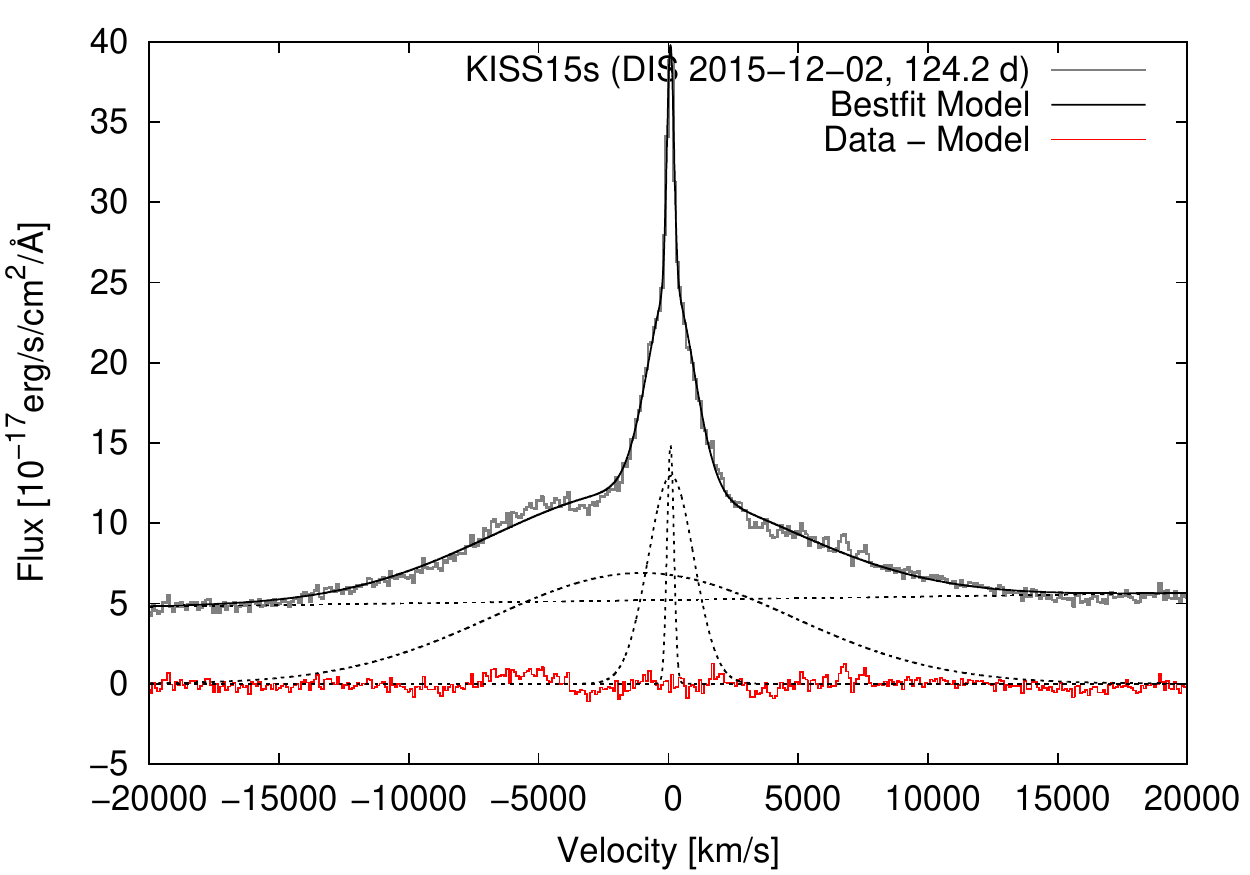}
\includegraphics[clip, width=3.2in]{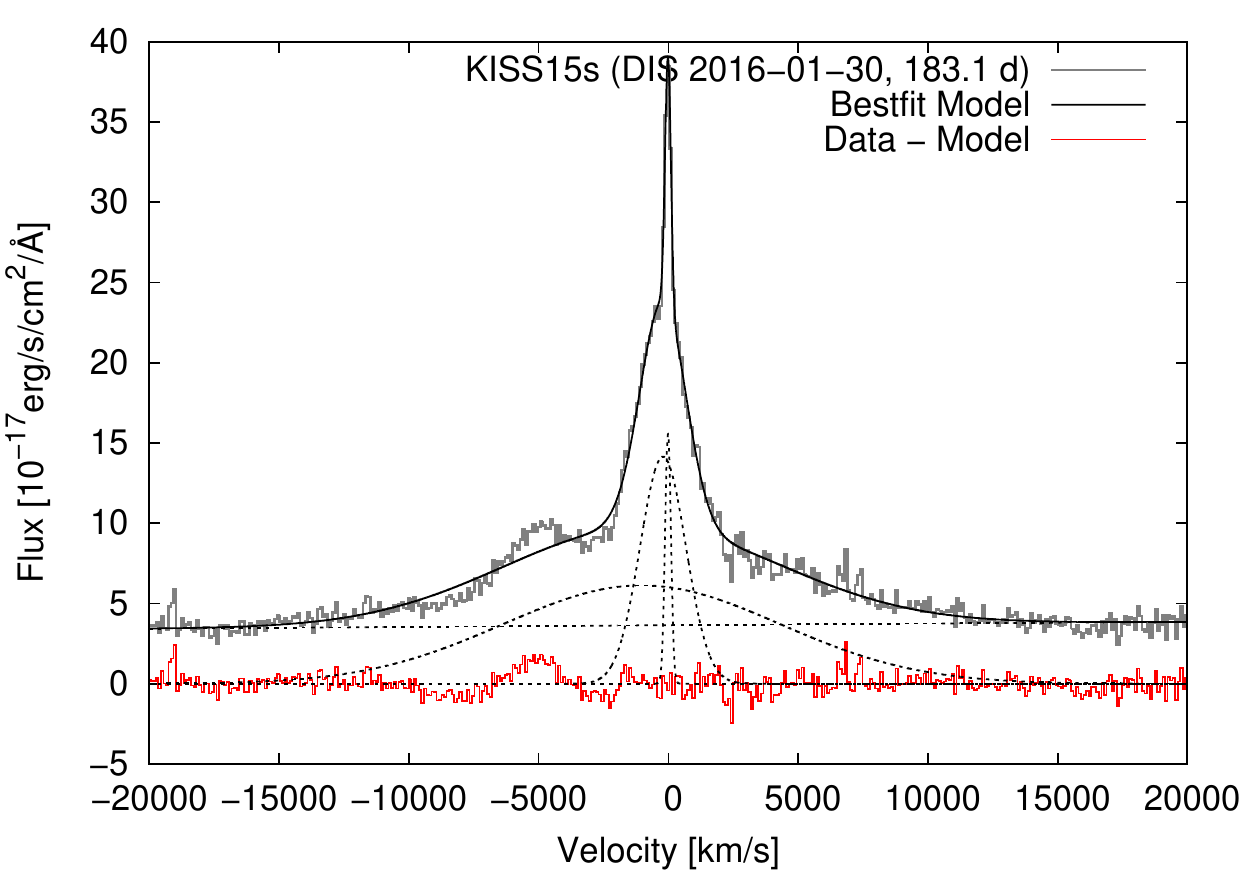}
\includegraphics[clip, width=3.2in]{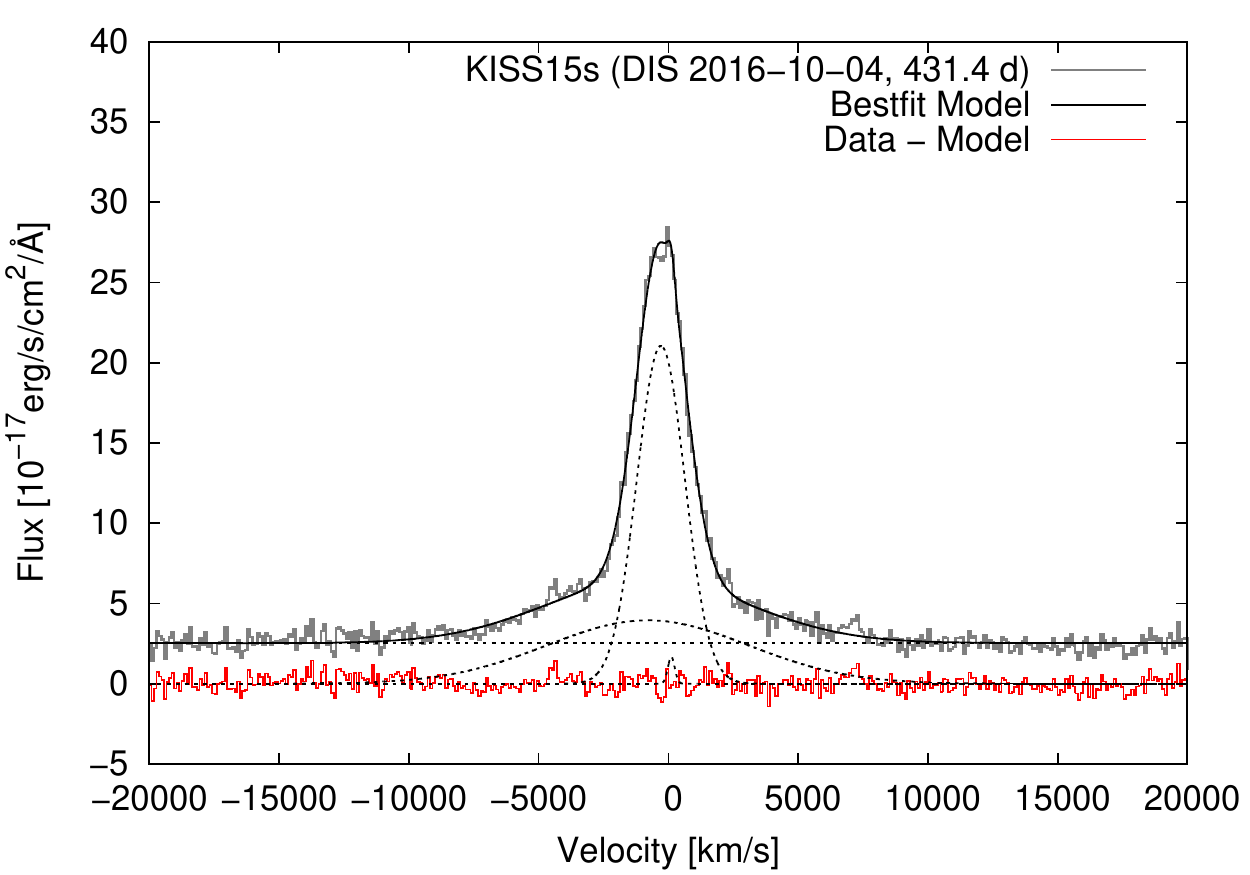}
\includegraphics[clip, width=3.2in]{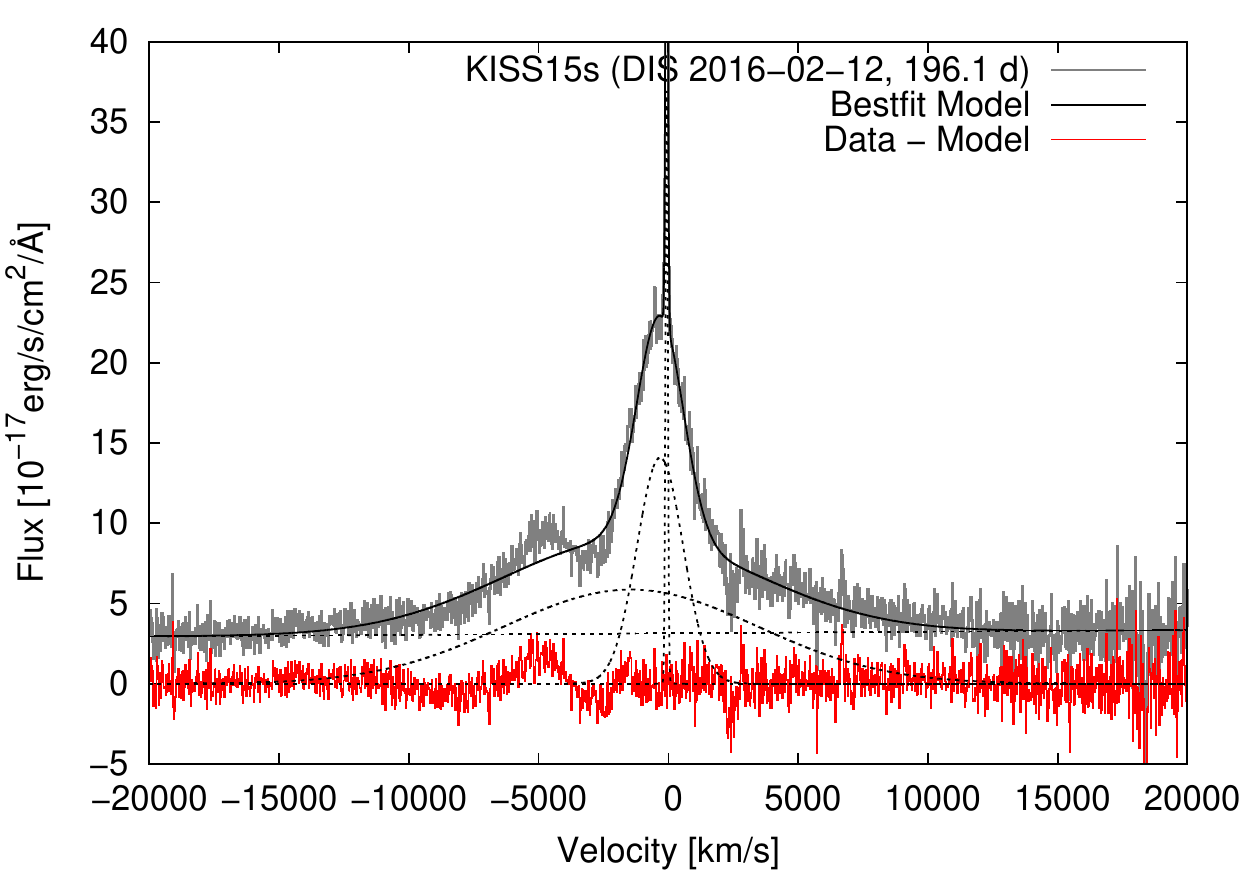}
\includegraphics[clip, width=3.2in]{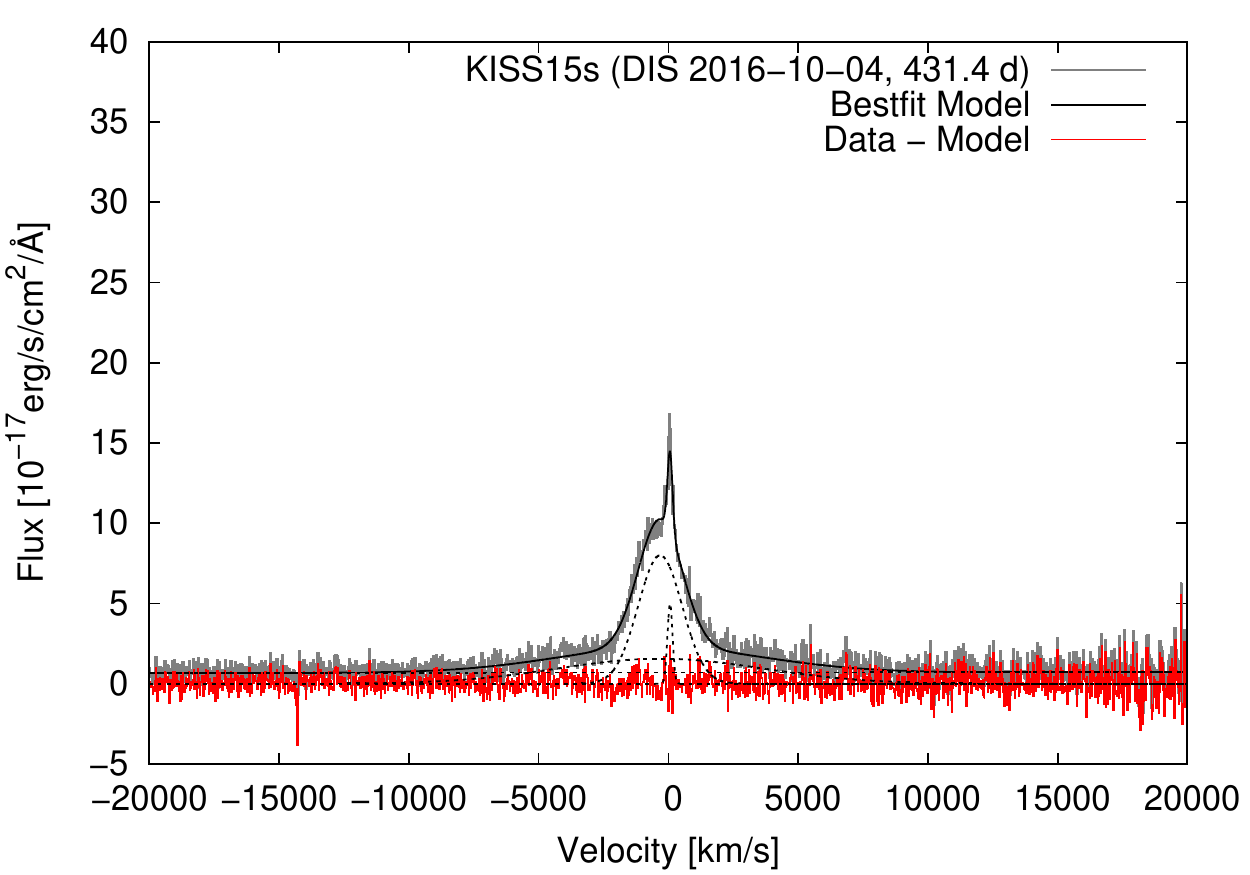}
}
 \caption{
 H$\alpha$ emission line profiles from 2015 September 19 (Nayuta/LISS), 2015 December 2, 2016 January 30, and 2016 October 4 (ARC3.5-m/DIS red-arm low-resolution).
 The horizontal axis is the rest-frame velocity relative to the systemic redshift of $z=0.03782$.
 Best-fit model spectra (Gaussian components, single power-law, and the sum of them) are also plotted (dotted lines); the LISS spectrum is fitted by two Gaussians and the DIS spectra are fitted by three Gaussians.
 Narrow- and intermediate-velocity width components are unresolved in the LISS spectrum, and the narrow component is unresolved in the DIS spectra.
The bottom two panels are for high-resolution DIS spectra.
 }
 \label{fig:spec_ana}
\end{figure*}

\begin{figure*}[tbp]
\center{
\includegraphics[clip, width=3.2in]{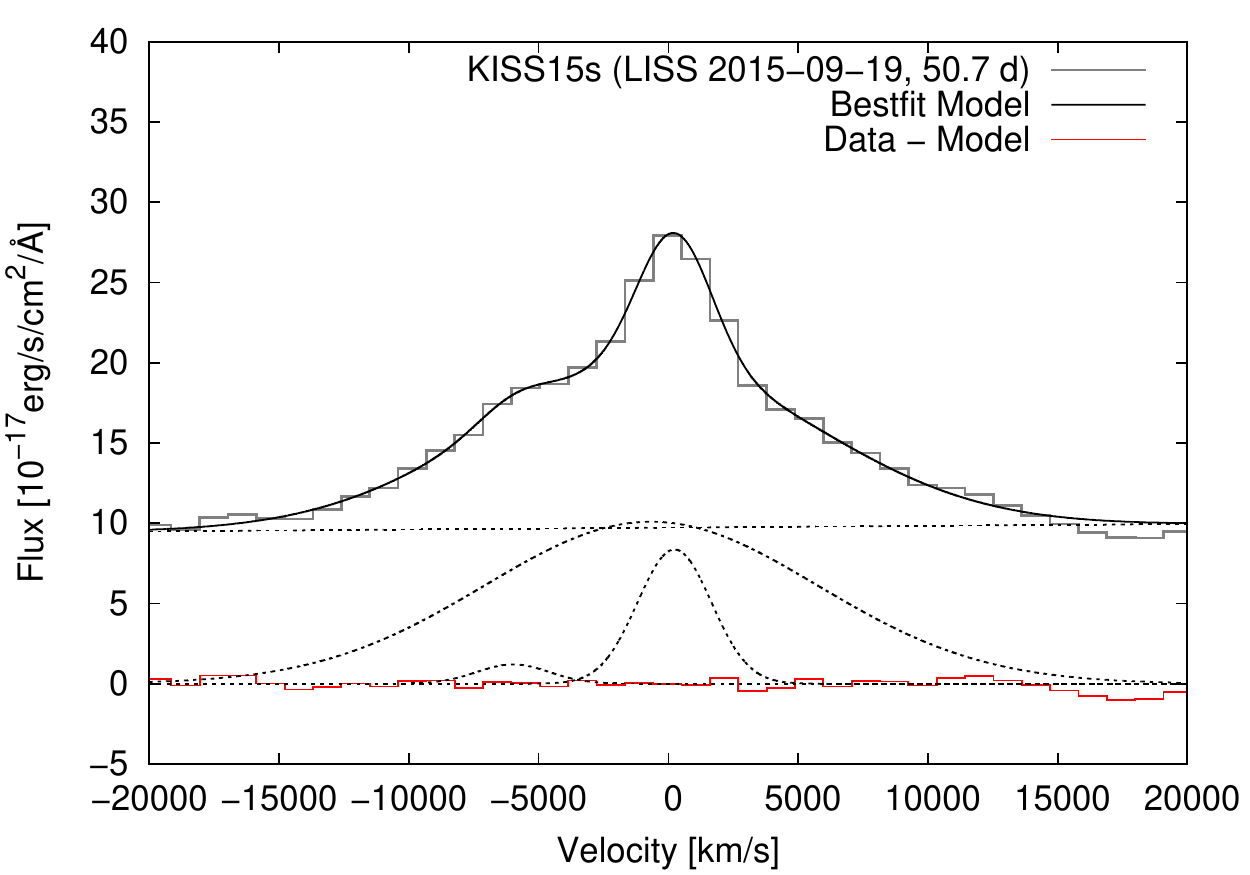}
\includegraphics[clip, width=3.2in]{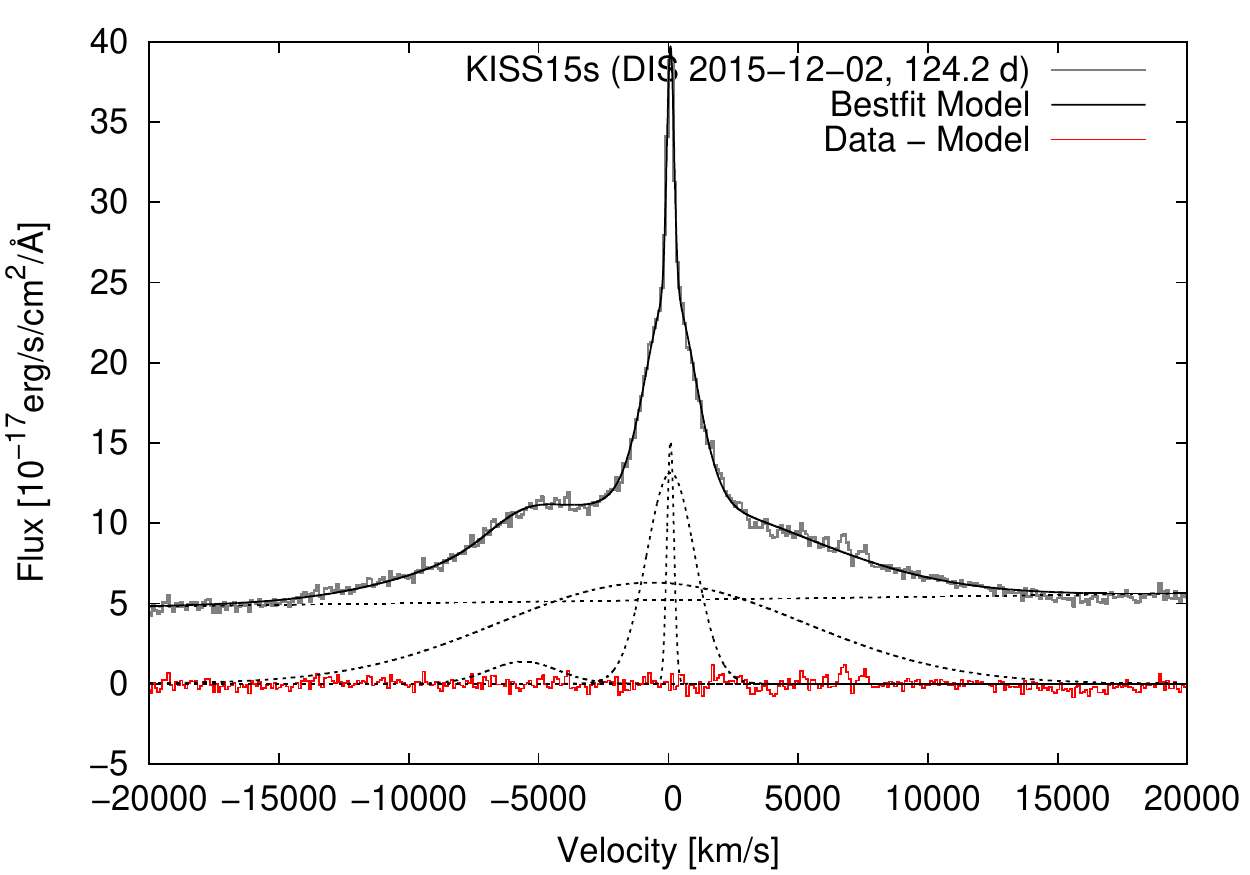}
\includegraphics[clip, width=3.2in]{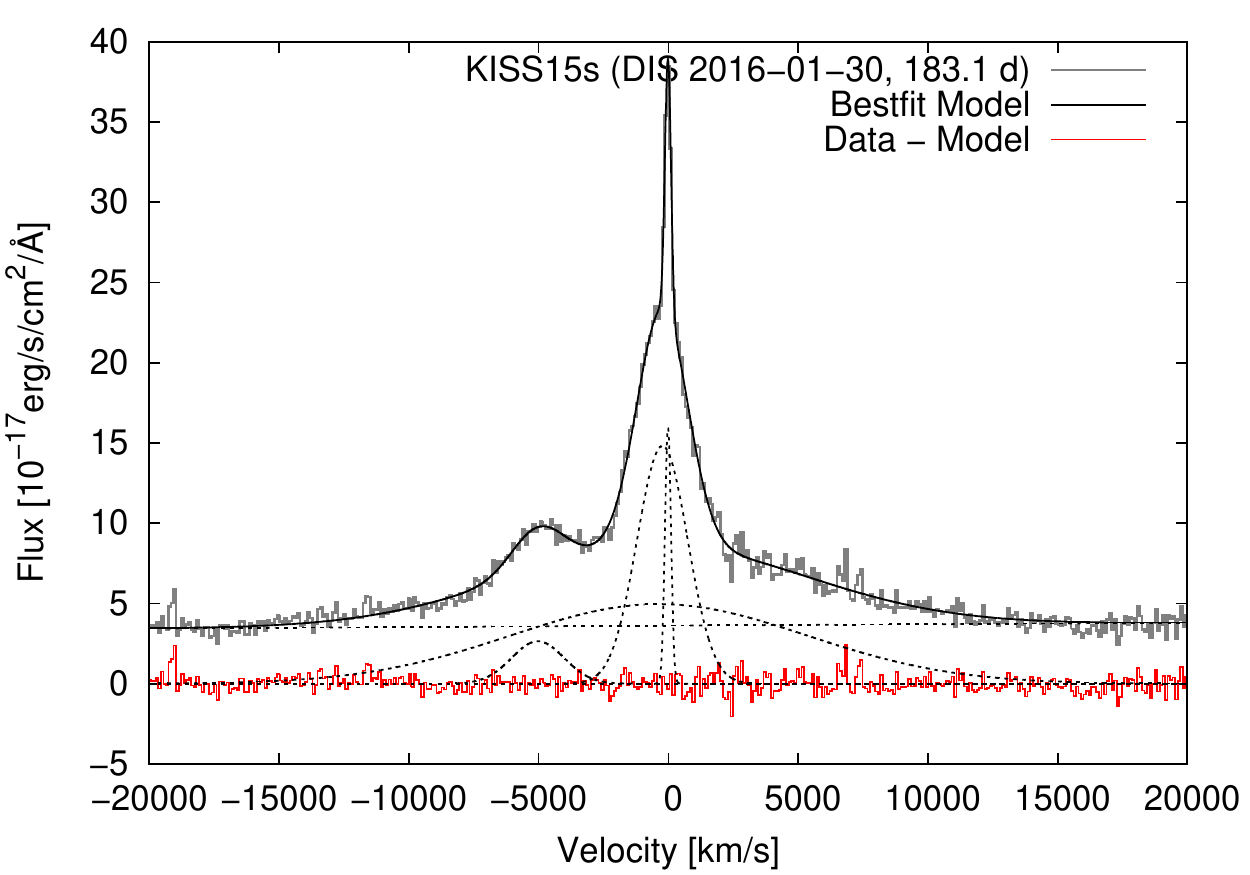}
\includegraphics[clip, width=3.2in]{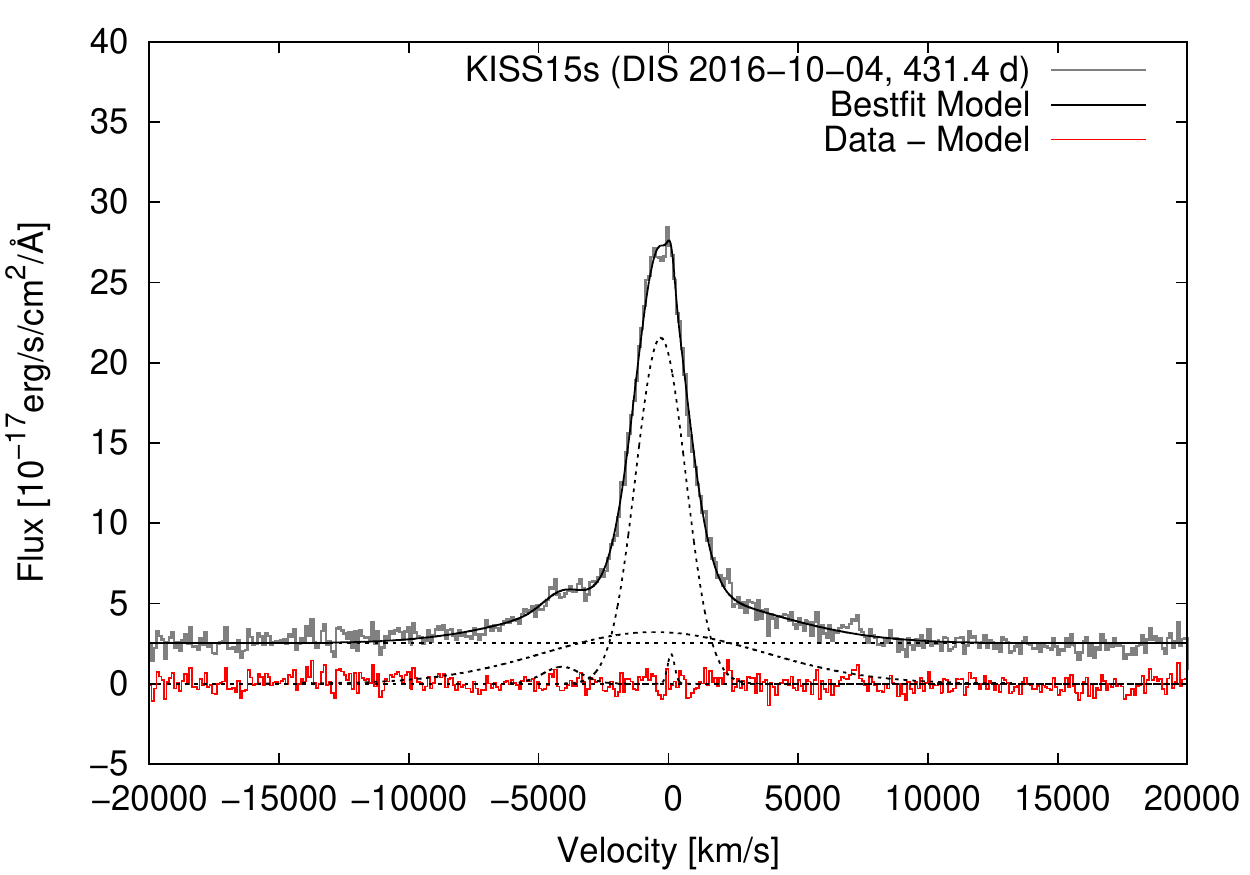}
\includegraphics[clip, width=3.2in]{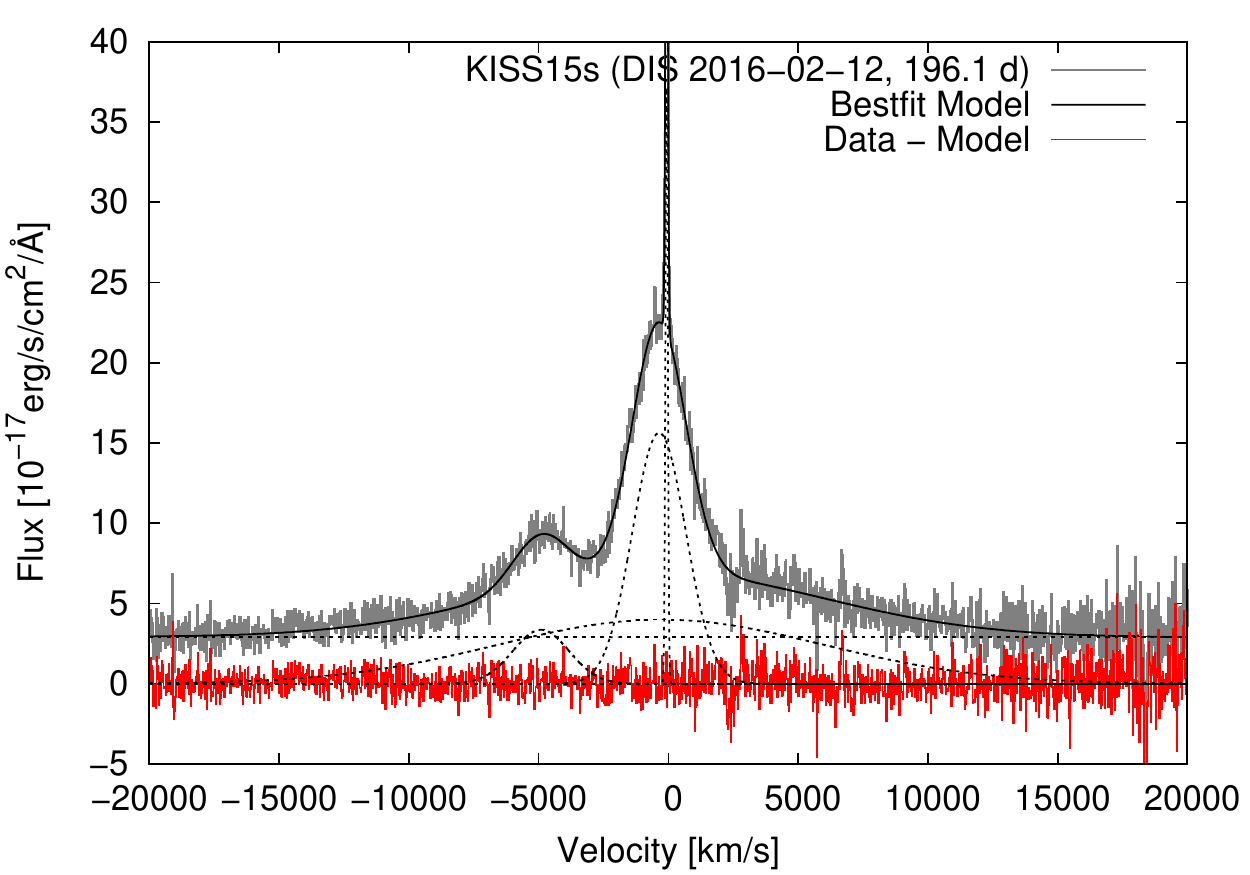}
\includegraphics[clip, width=3.2in]{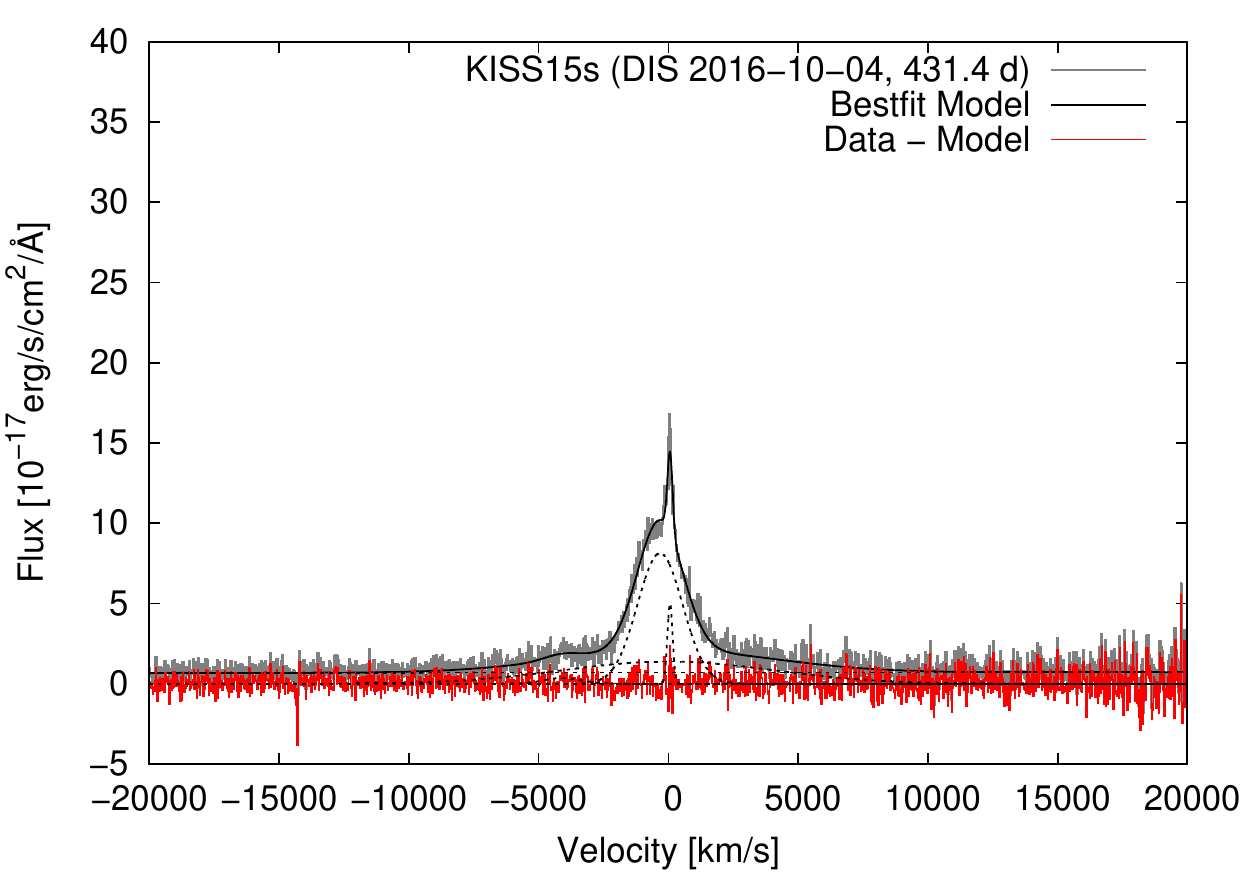}
}
 \caption{
Same as Figure~\ref{fig:spec_ana}, but one additional Gaussian is included to account for the blue-shifted excess emission component.
For the LISS spectrum, the additional Gaussian component is forced to have the same velocity width as the intermediate-width line (assuming they both are unresolved).
 }
 \label{fig:spec_ana3}
\end{figure*}

\begin{deluxetable*}{lllll|ll}
\tablewidth{700pt}
\tabletypesize{\scriptsize}
\tablecaption{Best-fit Gaussian model for the H$\alpha$ line profile without the blue-shift excess component. \label{spectral_fit_1}}
\tablehead{
\colhead{Parameter name} & \colhead{LISS} & \colhead{DIS low 1}  & \colhead{DIS low 2} & \colhead{DIS low 3} & \colhead{DIS high 1} & \colhead{DIS high 2}\\
\colhead{Date} & \colhead{2015-09-19} & \colhead{2015-12-02}  & \colhead{2016-01-30} & \colhead{2016-10-04} & \colhead{2016-02-12} & \colhead{2016-10-04}\\
\colhead{Days since discovery} & \colhead{50.7 d} & \colhead{124.2 d}  & \colhead{183.1 d} & \colhead{431.4 d} & \colhead{196.1 d} & \colhead{431.4 d}
}
\startdata
Norm. ($10^{-17}$ erg s${}^{-1}$ cm${}^{-2}$ \AA${}^{-1}$)   &    9.73 $\pm$     0.07 &     5.21 $\pm$      0.01 &     3.64 $\pm$      0.02 &     2.53 $\pm$      0.02   & 	    3.13 $\pm$     0.04 &     0.70 $\pm$      0.02 \\
Index                       &     0.38 $\pm$     0.07 &     1.18 $\pm$      0.03 &     0.85 $\pm$      0.07 &     0.05 $\pm$      0.11   & 	    0.92 $\pm$     0.17 &     0.61 $\pm$      0.52 \\
$\sigma_{\text{obs}, b}$ (km s$^{-1}$) & 6313.04 $\pm$   119.98 &  5985.78 $\pm$     35.73 &  5551.69 $\pm$     69.37 &  4038.60 $\pm$    118.35   & 	 5111.18 $\pm$    84.04 &  4104.14 $\pm$    184.58 \\
${\lambda}_{\text{obs}, b}$ (\AA)  & 6791.15 $\pm$     2.27 &  6787.18 $\pm$      0.75 &  6786.82 $\pm$      1.32 &  6792.28 $\pm$      1.84   & 	 6776.98 $\pm$     1.45 &  6803.70 $\pm$      2.93 \\
Flux$_{b}$ ($10^{-17}$ erg s${}^{-1}$ cm${}^{-2}$)                  & 3866.77 $\pm$    75.84 &  2343.72 $\pm$     15.08 &  1929.56 $\pm$     23.79 &   905.28 $\pm$     21.53   & 	 1697.40 $\pm$    30.34 &   363.60 $\pm$     15.10 \\
$\sigma_{\text{obs}, i}$ (km s$^{-1}$)    & 1311.16 $\pm$    63.24 &   912.79 $\pm$      9.96 &   914.67 $\pm$     11.88 &   960.62 $\pm$     10.94   & 	  927.25 $\pm$     9.91 &   881.11 $\pm$     15.37 \\
${\lambda}_{\text{obs}, i}$ (\AA) &  6817.42 $\pm$     1.46 &  6813.48 $\pm$      0.19 &  6806.17 $\pm$      0.28 &  6804.27 $\pm$      0.23   & 	 6803.52 $\pm$     0.25 &  6803.59 $\pm$      0.34 \\
Flux$_{i}$ ($10^{-17}$ erg s${}^{-1}$ cm${}^{-2}$) &  584.51 $\pm$    39.49 &   673.38 $\pm$      7.44 &   736.09 $\pm$     12.11 &  1150.25 $\pm$     18.41   & 	  742.90 $\pm$    10.66 &   400.67 $\pm$      9.95 \\
$\sigma_{\text{obs}, n}$ (km s$^{-1}$)    &    \nodata &   136.72 $\pm$      3.91 &   117.25 $\pm$      4.22 &   135.73 $\pm$     55.20   & 	   46.43 $\pm$     1.50 &   101.61 $\pm$     14.95 \\
${\lambda}_{\text{obs}, n}$ (\AA) &    \nodata &  6812.97 $\pm$      0.08 &  6810.74 $\pm$      0.10 &  6813.73 $\pm$      1.80   & 	 6809.51 $\pm$     0.03 &  6812.58 $\pm$      0.17 \\
Flux$_{n}$ ($10^{-17}$ erg s${}^{-1}$ cm${}^{-2}$)  &    \nodata &   115.63 $\pm$      3.66 &   103.86 $\pm$      4.26 &    12.89 $\pm$      4.92   & 	   97.73 $\pm$     2.36 &    28.87 $\pm$      3.26 \\
\enddata
\tablecomments{
The fitting parameters are a normalization and a spectral index of the power-law continuum, observed-frame velocity widths, observed-frame central wavelengths, and integrated fluxes for the broad, intermediate, and narrow components (denoted by subscripts $b$, $i$, and $n$, respectively).
The Galactic extinction is corrected; however, the host galaxy extinction of $E(B-V)_{\text{host}} = 0.6$~mag is not corrected.
The DIS high-resolution spectra are not spectro-photometrically calibrated; thus, the flux values include absolute flux calibration uncertainties. The discovery date is set to MJD=57234.}
\end{deluxetable*}

\begin{deluxetable*}{lllll|ll}
\tablewidth{700pt}
\tabletypesize{\scriptsize}
\tablecaption{Best-fit Gaussian model for the H$\alpha$ line profile with a blue-shift excess component. \label{spectral_fit_2}}
\tablehead{
\colhead{Parameter name} & \colhead{LISS} & \colhead{DIS low 1}  & \colhead{DIS low 2} & \colhead{DIS low 3} & \colhead{DIS high 1} & \colhead{DIS high 2} \\
\colhead{Date} & \colhead{2015-09-19} & \colhead{2015-12-02}  & \colhead{2016-01-30} & \colhead{2016-10-04} & \colhead{2016-02-12} & \colhead{2016-10-04}\\
\colhead{Days since discovery} & \colhead{50.7 d} & \colhead{124.2 d}  & \colhead{183.1 d} & \colhead{431.4 d} & \colhead{196.1 d} & \colhead{431.4 d}
}
\startdata
Norm. ($10^{-17}$ erg s${}^{-1}$ cm${}^{-2}$ \AA${}^{-1}$) &     9.72 $\pm$     0.07 &     5.20 $\pm$      0.01 &     3.61 $\pm$      0.02 &     2.52 $\pm$      0.02   & 	    2.90 $\pm$     0.06 &     0.70 $\pm$      0.03 \\
Index &      0.36 $\pm$     0.07 &     1.13 $\pm$      0.03 &     0.72 $\pm$      0.08 &     0.00 $\pm$      0.11   & 	   -0.06 $\pm$     0.24 &     0.47 $\pm$      0.54 \\
$\sigma_{\text{obs}, b}$ (km s$^{-1}$) & 6427.16 $\pm$   158.91 &  6076.60 $\pm$     44.98 &  5922.66 $\pm$    100.59 &  4355.02 $\pm$    191.32   & 	 6526.60 $\pm$   198.63 &  4201.36 $\pm$    283.85 \\
${\lambda}_{\text{obs}, b}$ (\AA) & 6795.94 $\pm$     3.11 &  6795.87 $\pm$      1.17 &  6802.42 $\pm$      2.18 &  6798.58 $\pm$      2.94   & 	 6804.27 $\pm$     3.14 &  6811.12 $\pm$      5.70 \\
Flux$_{b}$ ($10^{-17}$ erg s${}^{-1}$ cm${}^{-2}$) & 3685.78 $\pm$   107.11 &  2174.53 $\pm$     22.13 &  1669.75 $\pm$     33.36 &   796.34 $\pm$     34.03   & 	 1482.86 $\pm$    44.84 &   330.64 $\pm$     24.83 \\
$\sigma_{\text{obs}, i}$ (km s$^{-1}$) & 1420.75 $\pm$    84.34 &   971.30 $\pm$     12.54 &  1012.55 $\pm$     15.51 &   998.11 $\pm$     14.29   & 	 1070.06 $\pm$    13.36 &   904.97 $\pm$     18.84 \\
${\lambda}_{\text{obs}, i}$ (\AA) & 6816.41 $\pm$     1.48 &  6813.15 $\pm$      0.19 &  6805.61 $\pm$      0.28 &  6804.06 $\pm$      0.23   & 	 6802.67 $\pm$     0.24 &  6803.46 $\pm$      0.39 \\
Flux$_{i}$ ($10^{-17}$ erg s${}^{-1}$ cm${}^{-2}$) &  677.10 $\pm$    62.69 &   726.85 $\pm$      9.45 &   853.24 $\pm$     15.79 &  1223.74 $\pm$     26.65   & 	  951.19 $\pm$    16.34 &   417.99 $\pm$     12.97 \\
$\sigma_{\text{obs}, n}$ (km s$^{-1}$) &    \nodata &   140.43 $\pm$      4.06 &   121.09 $\pm$      4.59 &   140.88 $\pm$     49.27   & 	   47.20 $\pm$     1.30 &   102.97 $\pm$     14.33 \\
${\lambda}_{\text{obs}, n}$ (\AA) &     \nodata &  6812.98 $\pm$      0.08 &  6810.72 $\pm$      0.10 &  6813.57 $\pm$      1.56   & 	 6809.51 $\pm$     0.03 &  6812.56 $\pm$      0.17 \\
Flux$_{n}$ ($10^{-17}$ erg s${}^{-1}$ cm${}^{-2}$) &    \nodata &   120.79 $\pm$      3.83 &   109.74 $\pm$      4.45 &    14.78 $\pm$      5.01   & 	   99.93 $\pm$     2.36 &    29.34 $\pm$      3.20 \\
$\sigma_{\text{obs}, \text{add}}$ (km s$^{-1}$) &     \nodata &  1405.69 $\pm$     79.96 &  1069.12 $\pm$     57.61 &   762.90 $\pm$    123.95   & 	 1145.75 $\pm$    49.37 &   749.03 $\pm$    399.18 \\
${\lambda}_{\text{obs}, \text{add}}$ (\AA) & 6675.12 $\pm$     9.46 &  6679.37 $\pm$      1.90 &  6692.49 $\pm$      1.30 &  6714.40 $\pm$      2.89   & 	 6695.31 $\pm$     0.96 &  6714.74 $\pm$      5.04 \\
Flux$_{\text{add}}$ ($10^{-17}$ erg s${}^{-1}$ cm${}^{-2}$) &   95.13 $\pm$    35.07 &   108.08 $\pm$      9.93 &   158.10 $\pm$     12.25 &    44.83 $\pm$     10.84   & 	  215.18 $\pm$    12.51 &    14.89 $\pm$      8.61 \\
\enddata
\tablecomments{Same as Table~\ref{spectral_fit_1}. Subscript $``\text{add}''$ denotes the additional blue-shifted component. The intermediate and blue-shifted components in the LISS spectrum are unresolved and assumed to have the same velocity width.}
\end{deluxetable*}

Figure~\ref{fig:spec_halphahbeta} compares the spectral profiles of H$\alpha$ and H$\beta$ emission lines of the DIS low-resolution spectrum obtained on 2015 December 2.
It is clear that the H$\beta$ and H$\alpha$ emission lines have nearly identical spectral profiles, and both of them have broad ($v_{\text{FWHM}}\sim$ 14,000 km~s${}^{-1}$), intermediate ($v_{\text{FWHM}}\sim$ 2,000 km~s${}^{-1}$), and narrow ($v_{\text{FWHM}} < 100$~km~s${}^{-1}$) components.
It should be noted that the \ion{He}{1} emission lines and the H$\gamma$ emission line also have similar line profiles (Figure~\ref{fig:spec_comparison}), although the \ion{He}{1} emission line luminosities are much weaker than the H$\beta$ and H$\alpha$ emission lines.
As shown in Figure~\ref{fig:spec_halpha_comparison} (see also Figure~\ref{fig:spec_comparison}), the optical spectral properties of KISS15s are similar to type~IIn SN~1988Z and other SN~1988Z-like subclasses of SN IIn \citep[see e.g.,][]{sta91,pas02,mil10,str12}.

To quantitatively evaluate the line widths and line intensities of the H$\alpha$ emission, we fitted multi-component Gaussian models to Nayuta/LISS and ARC3.5-m/DIS spectra in the spectral region around the H$\alpha$ emission line. 
$\chi^2$ minimization was performed using the Python version of {\tt MPFIT} \citep{mar09}\footnote{{\tt MPFIT.py} was developed by Mark Rivers and Sergey Koposov: \href{https://github.com/segasai/astrolibpy}{https://github.com/segasai/astrolibpy}}.
Nayuta/LISS and ARC3.5-m/DIS low-resolution spectra were fitted in an observed-frame wavelength range of $\lambda_{\text{obs}}=5,860-7,440$\AA, and the ARC3.5-m/DIS high-resolution spectra were fitted at $\lambda_{\text{obs}}=6,240-7,310$\AA.

First, the DIS low- and high-resolution spectra were fitted by three Gaussians (broad, intermediate, and narrow) and a single power-law model.
The Nayuta/LISS very low-resolution spectrum was fitted by two Gaussians and a single power-law model, because the spectral resolution of Nayuta/LISS is lower than the velocity width of the intermediate component; thus, the narrow component is mixed with the intermediate component.
The measurement uncertainties of the model parameters were evaluated by 10,000 trials of Monte Carlo resampling, where 10,000 mock spectra were generated by adding Gaussian flux noise to the original spectrum using the calculated flux density errors.
Figure~\ref{fig:spec_ana} shows the profiles of the H$\alpha$ emission of KISS15s.
The best-fit model spectra and the fitting residuals are also shown in the same figure, and the best-fit model parameters are tabulated in Table~\ref{spectral_fit_1}.
The fitted models mostly explain the observed emission line profiles, except for a blue-shift excess component at $v \sim -5,000$ km~s${}^{-1}$, as indicated in Figure~\ref{fig:spec_halphahbeta}.

Next, we performed another model fitting to the H$\alpha$ profiles by adding an additional Gaussian component to account for the blue-shift excess component.
Figure~\ref{fig:spec_ana3} shows the same H$\alpha$ emission line profiles as Figure~\ref{fig:spec_ana}, but fitted by the refined model with the additional Gaussian blue-shift excess component.
The best-fit parameters of the four Gaussians model are tabulated in Table~\ref{spectral_fit_2}.
The best-fit model shown in Figure~\ref{fig:spec_ana3} reasonably reproduces the observed line profiles, and the fitting residuals no longer show any signature of remaining spectral components.
We also tried to model the line profiles with additional Gaussian absorption of the broad component, instead of the additional Gaussian emission component, but no reasonable fitting was achieved.
This analysis suggests that the apparent asymmetry of the entire H$\alpha$ line profile of KISS15s is not due to absorption of broad or intermediate emission lines by opaque gases or dusts located in the ejecta-CSM interaction region \citep[as suggested for other SNe IIn by, e.g.,][]{smi09,fox11,smi12,tad13,fra14,che17}.
Instead, the observed line profile of KISS15s is most likely due to the presence of an additional blue-shifted symmetric intermediate Gaussian emission component with respect to the symmetric narrow, intermediate, and broad Gaussians, which may be related to the inhomogeneity of the CSM distributions \citep[][]{smi15,and16}.
There is no evidence of Lorenzian-like broad wing components in the observed line profiles of KISS15s, suggesting that the electron scattering in optically-thick ejecta-CSM regions is not relevant in KISS15s \citep[see e.g.,][]{smi10,fra14,bor15,hua18}.
Considering the better fitting for the observed spectra compared to the three Gaussians model, below we focus on the fitting results from the ``additional Gaussian model'' or the four Gaussians model (Table~\ref{spectral_fit_2} and Figure~\ref{fig:spec_ana3}).

The narrow emission line component is partially resolved in the DIS high-resolution spectrum obtained at MJD=57430.1 (``DIS high 1'' in Tables~\ref{spectral_fit_1} and \ref{spectral_fit_2}).
The rest-frame velocity FWHM ($\text{FWHM}_{\text{rest}}$) can be estimated from the observed-frame line width ($\sigma_{\text{obs}}$), the instrumental broadening width ($\sigma_{\text{inst}}$), and the source redshift $z$ as
\begin{equation}
\text{FWHM}_{\text{rest}} = 2\sqrt{2\ln 2}\frac{\sqrt{\sigma_{\text{obs}}^2 - \sigma_{\text{inst}}^2}}{1+z}.
\label{fwhm_rest}
\end{equation}
Substituting $\sigma_{\text{obs}, n} = 47.2$~km~s${}^{-1}$ and $\sigma_{\text{inst}} = 20.8$~km~s${}^{-1}$ taken from Tables~\ref{spectral_fit_2} and \ref{obslog_dis}, respectively, to Equation~\ref{fwhm_rest}, the rest-frame velocity FWHM of the narrow component of KISS15s is evaluated as $\text{FWHM}_{\text{rest}, n} = 114.8$~km~s${}^{-1}$.
It should also be noted that the H$\alpha$ line flux of the narrow component is $\sim 10^{-15}$~erg~s${}^{-1}$~cm${}^{2}$, which is roughly in agreement with the host galaxy H$\alpha$ line flux measured from the SDSS spectrum (see Section~\ref{sec:hostproperties} for details).
Therefore, although it is still possible that there is a flux contribution to the narrow line from the unshocked CSM photoionized by strong radiation from KISS15s, we can assume that the observed narrow emission line component predominantly comes from the foreground/background \ion{H}{2} regions in the host galaxy that may not directly be related to the SN explosion event of KISS15s.
We put a rough upper limit on the velocity width of the undetected narrow emission line component from the CSM formed by the progenitor's stellar wind as $\text{FWHM}_{w} \lesssim 100$~km~s${}^{-1}$, which means that $v_w \lesssim 100$~km~s${}^{-1}$ on the assumption of $v_{w} \simeq \text{FWHM}_{w}$ \citep[e.g.,][]{tad15} (see Section~\ref{sec:massloss} for further discussion).

\begin{figure}[tbp]
\center{
\includegraphics[clip, width=3.4in]{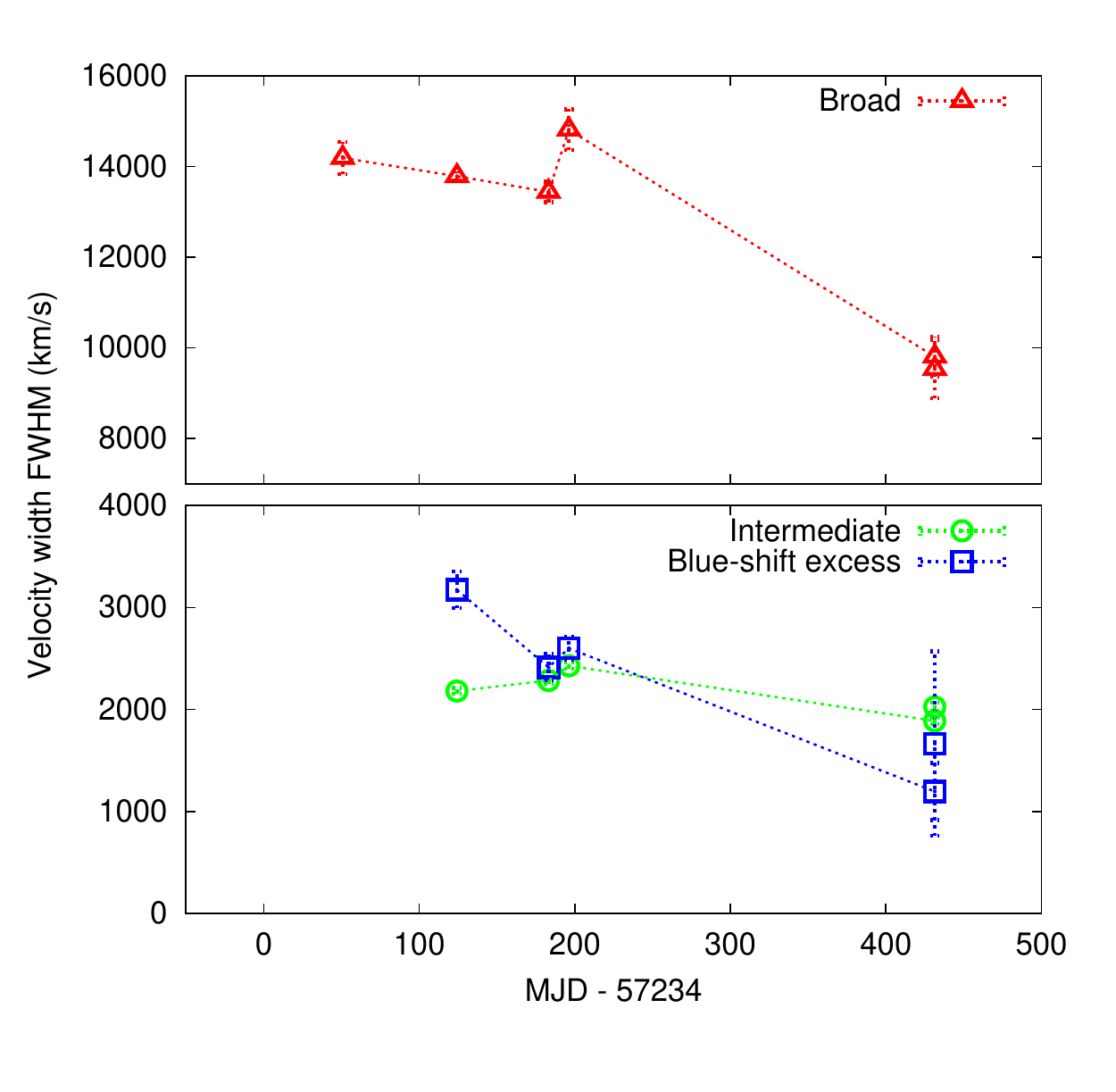}
\includegraphics[clip, width=3.4in]{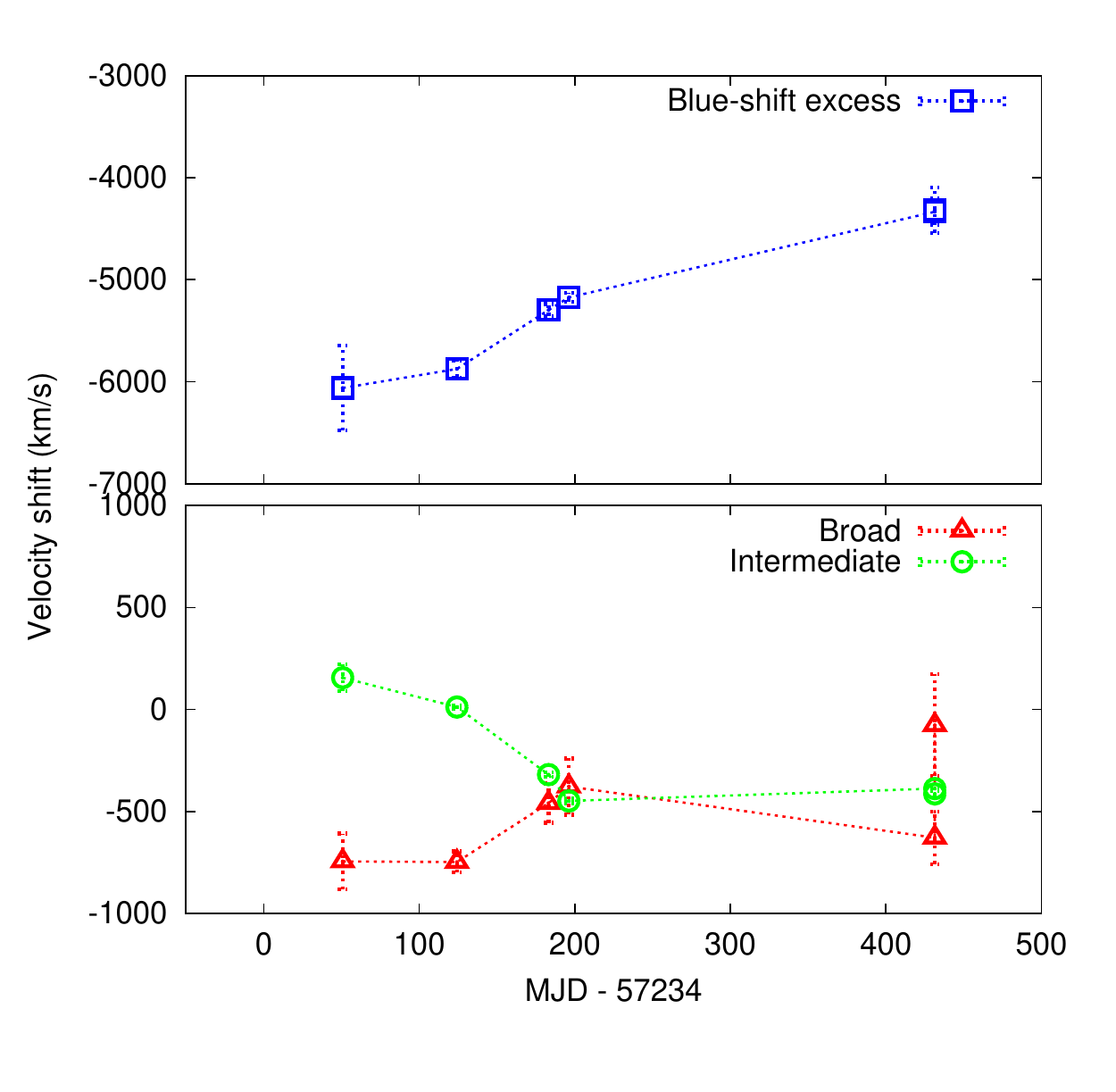}
}
 \caption{
Rest-frame intrinsic velocity width and the velocity shift of the decomposed broad, intermediate, and blue-shift excess Gaussian components of the H$\alpha$ line profile of KISS15s, as a function of observed time. The velocity widths of the intermediate and blue-shift components are unresolved in the Nayuta/LISS and are omitted from the plot.
 }
 \label{fig:spec_lightcurves_1}
\end{figure}

\begin{figure}[tbp]
\center{
\includegraphics[clip, width=3.2in]{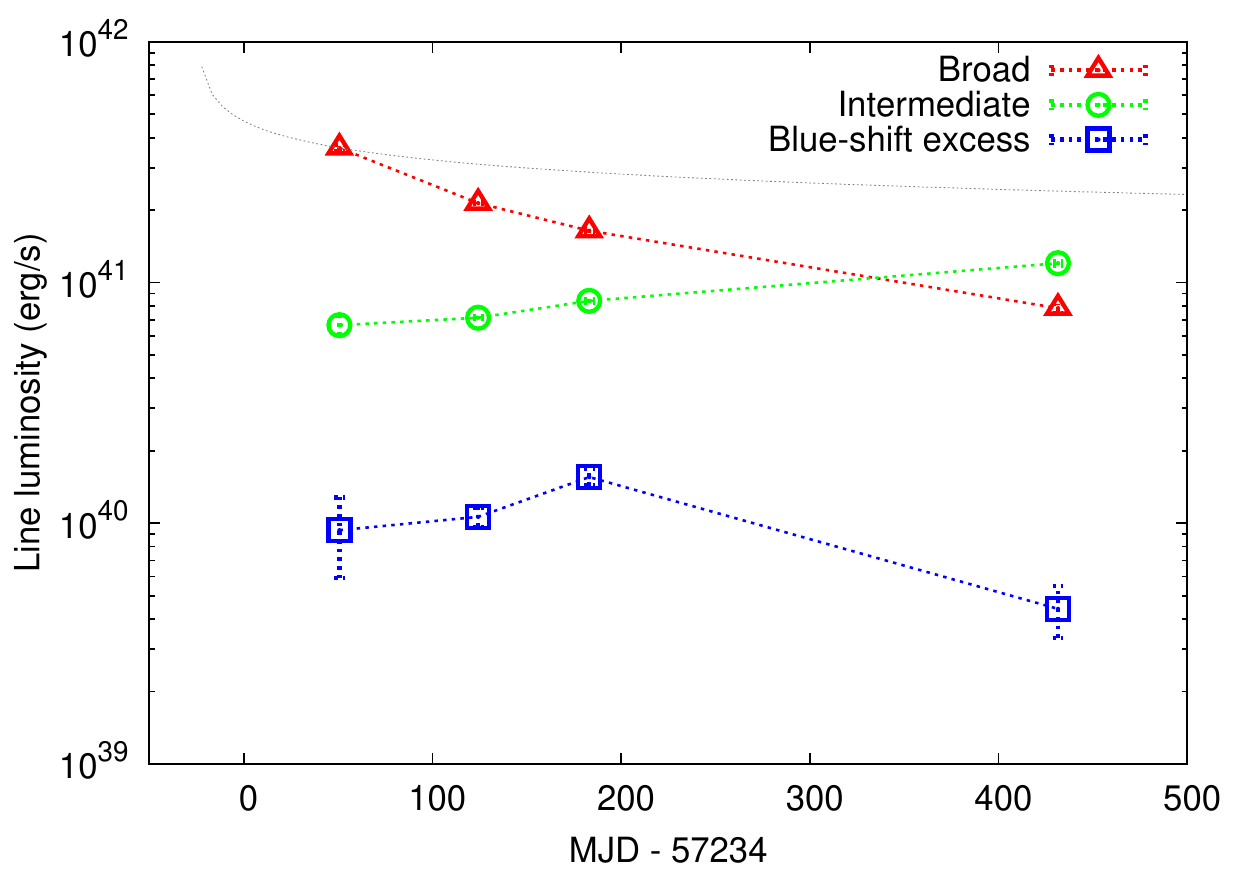}
\includegraphics[clip, width=3.2in]{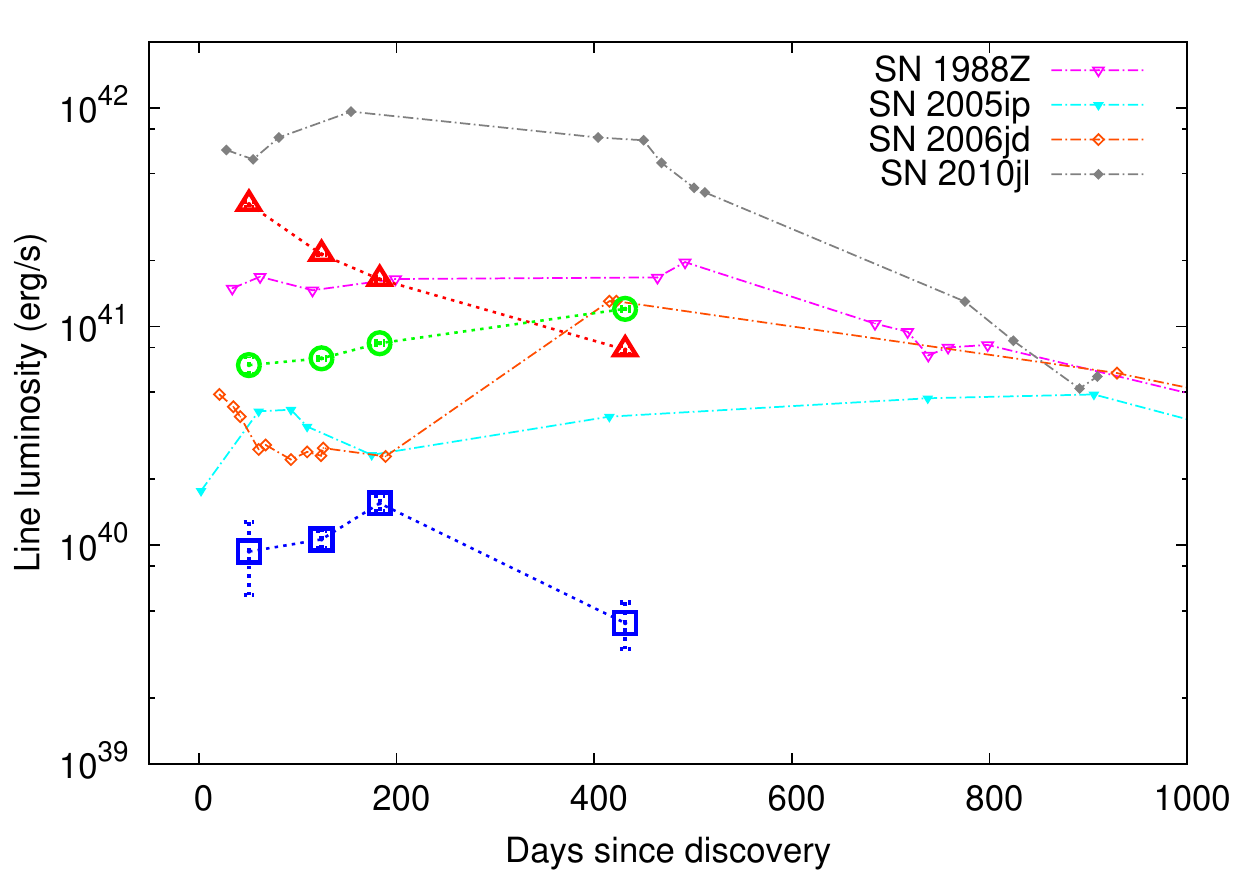}
}
 \caption{
Light curves of the line luminosity of the decomposed broad, intermediate, and blue-shift excess Gaussian components of the H$\alpha$ line profile of KISS15s. 
The line luminosities evaluated from the spectro-photometrically calibrated low-resolution spectra are shown.
The host galaxy extinction of $E(B-V)_{\text{host}} = 0.6$~mag is corrected.
For comparison, an arbitrary-scaled power-law of $L_{\text{BB, opt}} \propto t^{-0.23}$ (Equation~\ref{eqn:bolometric_index}) is also shown.
Bottom panel compares the time evolution of the H$\alpha$ line luminosity of KISS15s with those of other SN~1988Z-like SNe, SN~1988Z \citep{are99}, SN~2005ip, SN~2006jd \citep{str12}, and SN2010jl \citep{jen16}.
 }
 \label{fig:spec_lightcurves_2}
\end{figure}

Figure~\ref{fig:spec_lightcurves_1} shows the temporal evolution of the rest-frame velocity width FWHM ($\text{FWHM}_{\text{rest}}$; Equation~\ref{fwhm_rest}) and the velocity shift relative to the systemic redshift of $z=0.03782$ of the decomposed H$\alpha$ emission line components of KISS15s.
The velocity width of the broad component gradually decreases as a function of time from $\text{FWHM} \sim 14,000$~km~s${}^{-1}$ to $\sim 10,000$~km~s${}^{-1}$, and the velocity shift only shows weak variability.
It is interesting to note that the velocity widths of the intermediate and blue-shift excess components are consistent with each other, suggesting that the two components are produced by the same emission mechanism.
The line of sight velocity of the two intermediate-width components are, however, very different; the velocity shift of the intermediate component decreases from $\sim 0$~km~s${}^{-1}$ to about $-500$~km~s${}^{-1}$, on the other hand that of the blue-shift component increases from about $-6,000$~km~s${}^{-1}$ to about $-4,500$~km~s${}^{-1}$.
The coexistence of these two emission line components implies that the emission region has a spherically asymmetric geometry, which may be related to the inhomogeneous structure of the CSM produced by the intrinsic non-spherically-symmetric progenitor's stellar wind \citep[see Section~\ref{sec:spec_lightcurve} for further discussion;][and references therein]{bor15,smi17b}.

Figure~\ref{fig:spec_lightcurves_2} presents the light curves of the decomposed H$\alpha$ emission line luminosities.
The broad emission line luminosity monotonically decreases, as with the power-law luminosity evolution of the optical continuum light curve.
The similarity of the luminosity evolution between the broad H$\alpha$ line and optical continuum implies that the broad emission line component and optical continuum emission are powered by the same energy source, which may be the ingoing and/or outgoing intense radiation from the ejecta-CSM interaction region \citep[e.g.,][]{chu94,che17}.
The intermediate and blue-shifted components show different temporal evolutions with each other, as well as with the broad component.
The luminosity of the intermediate component increases during the observations, which may imply that the volume of the ejecta-CSM interaction region is increasing.
On the other hand, the blue-shift component becomes weaker at the last epoch of the spectroscopic observation, suggesting that this component is produced from a separate region from which the intermediate component emerges.

\subsubsection{Other spectral features}
\label{sec:otherfeatures}

In addition to \ion{He}{1}$\lambda$5876, the \ion{He}{1}$\lambda$7065 emission line is clearly visible in the KISS15s spectra (Figure~\ref{fig:spec_comparison}), and it seems to have at least two spectral components with intermediate ($\sim$ 2,000 km~s${}^{-1}$) and narrow ($<$300 km~s${}^{-1}$) line widths.
The broad emission bump at wavelengths of $\lambda_{\text{rest}} \sim 8,600$~\AA\ is probably due to a blend of broad \ion{O}{1}$\lambda$8446.

KISS15s shows prominent broad emission line features of the \ion{Ca}{2} IR triplet ($\lambda\lambda$8498, 8542, 8662), as is also observed in several other SNe IIn such as SN~1988Z \citep[e.g.,][]{tur93}.
Blended \ion{Ca}{2} IR triplet emission shows a temporal evolution similar to that of the H$\alpha$ emission line, although the detailed spectral decomposition is difficult due to insufficient spectral coverages of our data and blending of multiple broad/intermediate lines of \ion{Ca}{2} and probably \ion{O}{1}$\lambda$8446 \citep[e.g.,][]{jen16}.
The strong, broad/intermediate \ion{Ca}{2} triplet emission with no associated P Cygni absorption feature must be emitted by the same region as that with the strong H$\alpha$ emission line, i.e., the CDS and/or photoionized ejecta \citep{chu94,des16}.

The origin of the broad bump at the wavelengths of $\lambda_{\text{rest}} \sim 7,300$~\AA\ in KISS15s spectra is unclear, but it may be due to a blend of broad lines of \ion{He}{1} and \ion{Ca}{2} and several forbidden lines \citep[e.g.,][]{gra14,smi17}.
The flux excess in the wavelength range of $\lambda_{\text{rest}}<5,700$\AA\ generally observed in spectra of SNe IIn, which is composed of a blend of \ion{Fe}{2} and several other emission lines \citep{sta91,str12}, is not observed or is significantly weaker in KISS15s compared to SN~1988Z (Figure~\ref{fig:spec_comparison}).

\section{Discussion}
\label{sec:discussion}

\subsection{Host galaxy properties}
\label{sec:hostproperties}

\begin{figure}[tbp]
\center{
\includegraphics[clip, width=3.6in]{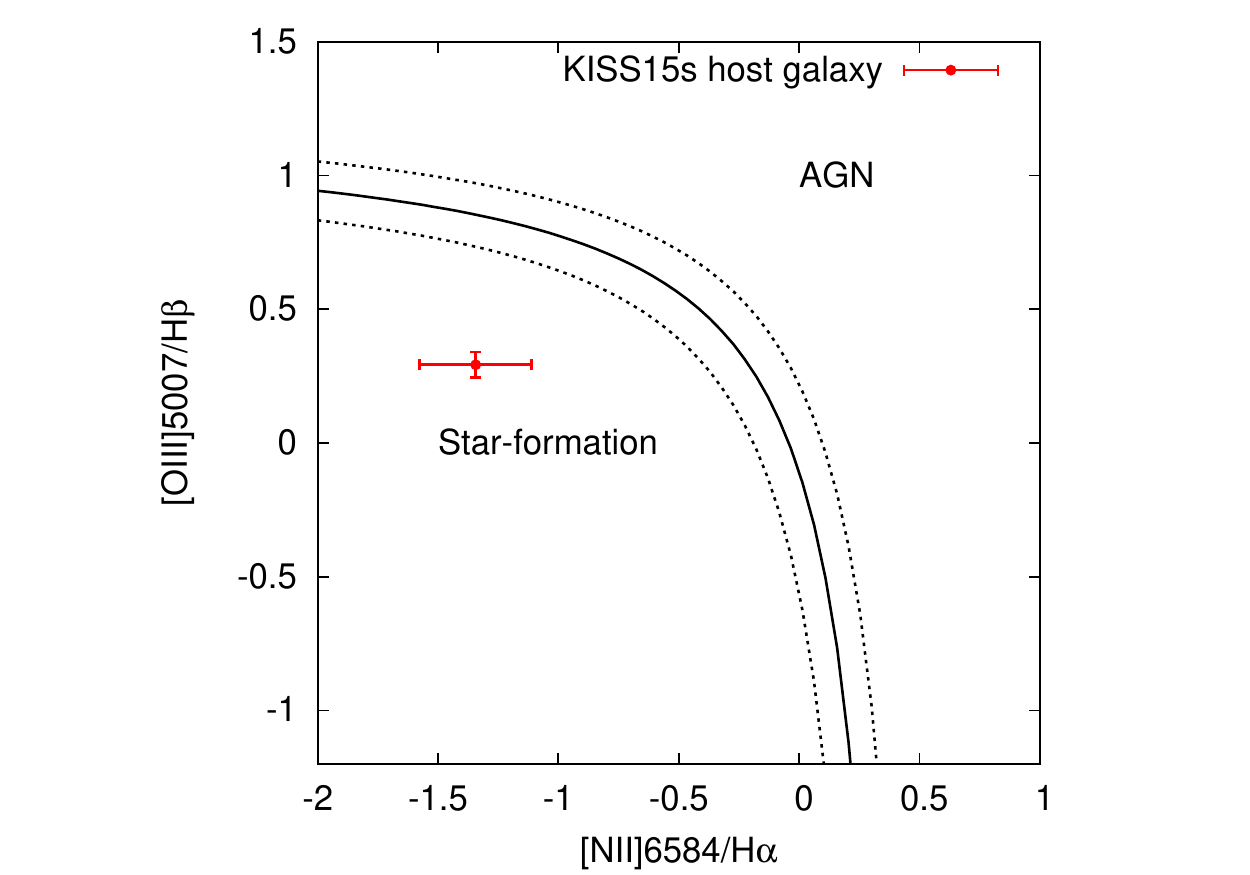}
}
 \caption{BPT diagram of the host galaxy SDSS~J030831.67-005008.6. The location of the line (and $\pm 0.1$~dex of the line) dividing star-forming galaxies and active galactic nuclei is taken from \cite{kew01}.}
 \label{fig:bpt_diagram}
\end{figure}

The host galaxy of KISS15s, SDSS~J030831.67-005008.6, was imaged by the SDSS Legacy Survey, and the central region was observed spectroscopically through the SDSS 3 arcsecond fiber spectrograph \citep{yor00}.
Three spectra obtained at different epochs are available in the SDSS database.
We used the spectrum taken at MJD = 52264 (which has the highest S/N ratio among the three spectra) in the analyses below.
Note that the emission lines detected in the SDSS spectrum of SDSS~J030831.67-005008.6 are unresolved, and the velocity width is constrained to $\sigma_{v}\lesssim 70$ km~s${}^{-1}$.

According to the outputs of the SDSS {\tt photo} pipeline\footnote{\href{http://www.sdss.org/dr12/algorithms/magnitudes/}{http://www.sdss.org/dr12/algorithms/magnitudes/}}, the light profile of SDSS J030831.67-005008.6 is best-fit by a pure exponential model.
The host galaxy $g$- and $r$-band absolute magnitude is $M_g = -17.01\pm 0.02$ and $M_r = -17.18\pm0.02$ mag, and the $u-r$ color is $1.46\pm 0.13$~mag (corrected for Galactic extinction), which are typical values of a ``blue cloud'' star forming galaxy \citep[e.g.,][]{sch14}.
The effective radius of the exponential profile model $r_e$ is 
$4.3''\pm0.2''$ (3.1 $\pm$ 0.1~kpc) in the $r$-band.
Figure~\ref{fig:bpt_diagram} shows a BPT diagram \citep{bal81} of SDSS J030831.67-005008.6, where the Galactic extinction corrected line intensities are calculated from SDSS spectroscopy pipeline outputs ({\tt spZline}).
Based on the BPT classification, this galaxy is a low-metallicity star-forming galaxy.
The SDSS spectroscopy pipeline outputs indicate that the stellar mass of this galaxy is $\log(M_*[M_{\odot}])=8-9$ \citep{kau03,tre04}.
The luminosity-metallicity relation \citep[Equation~1 of][]{lee06} suggests that this galaxy is metal-poor \citep[12+log O/H $\sim$ $8.1 \pm 0.4$; see also][]{aar15}.
Alternatively, the Galactic extinction corrected line intensity ratio analyses using {\tt HII-CHI-mistry}\footnote{HII-CHI-mistry Version 3.0: \href{http://www.iaa.es/~epm/HII-CHI-mistry.html}{http://www.iaa.es/~epm/HII-CHI-mistry.html}.} Python code \citep{per14} and line ratios ([\ion{O}{2}]3727, [\ion{O}{3}]4363, [\ion{O}{3}]5007, [\ion{N}{2}]6584, and [\ion{S}{2}]6717+6731 relative to H$\beta$, taken from the SDSS pipeline outputs) provide a constraint on the gas-phase metallicity of the host galaxy of 
\begin{equation}
12+\log \text{O/H} = 7.873 \pm 0.385.
\end{equation}
Thus, the host galaxy SDSS J030831.67-005008.6 can be considered a low-mass, low-metallicity star forming galaxy, which is the low end of the metallicity distribution of explosion sites of SNe IIn, particularly of 1988Z-like SNe IIn \citep[e.g.,][]{sto11,hab14,tad15}.

The SDSS spectrum of the host galaxy shows strong recombination lines, such as H$\alpha$, H$\beta$, and [\ion{O}{3}] emission lines, from the \ion{H}{2} regions in the galaxy.
The SDSS spectroscopy pipeline indicates that the narrow H$\alpha$ emission line flux from the central 3-arcsec region of the galaxy is $f_{\text{H}\alpha, \text{host}} = 7.39$ ($\pm$0.15) $\times 10^{-16}$ erg~s${}^{-1}$~cm${}^{-2}$, and its equivalent is EW = $116.1 \pm 2.4$~\AA.
The star-formation rate (SFR) estimated from the H$\alpha$ emission line luminosity of the host galaxy, $L_{\text{H}\alpha, \text{host}} = 4\pi d_L^2 f_{\text{H}\alpha, \text{host}} = 2.17 (\pm 0.04) \times 10^{39}~\text{erg}~\text{s}^{-1}$, is as follows \citep{ken94,ken98}:
\begin{eqnarray}
\text{SFR(H}\alpha) &=& 7.9~M_{\odot} \text{yr}^{-1} \times \left( \frac{L_{\text{H}\alpha, \text{host}}}{10^{42}~\text{erg}~\text{s}^{-1}} \right) \nonumber \\
&\simeq& 0.02~M_{\odot}~\text{yr}^{-1}. \nonumber
\end{eqnarray}

As already pointed out in Section~\ref{balmerfit}, it should be noted that this H$\alpha$ flux is comparable to that observed in the DIS spectra of KISS15s as the narrow H$\alpha$ component, namely $\sim 1 \times 10^{-15}$ erg~s${}^{-1}$~cm${}^{-2}$.
Considering the local variation in H$\alpha$ luminosity within the galaxy, it is possible that the narrow H$\alpha$ flux in the KISS15s spectrum is mostly from star-forming (\ion{H}{2}) regions in the host galaxy and is not directly related to the SN event.
The other narrow emission lines of [\ion{O}{3}] and [\ion{S}{2}] observed in the KISS15s spectrum are probably also due to flux contributions from the \ion{H}{2} regions in the host galaxy \citep[see the cases of SN 1987F and SN 2006qq;][]{fil89,tad13}.

Generally speaking, the low-metallicity environment suggests that the dust content of the host galaxy is small.
The Balmer decrement of the SDSS spectrum of SDSS J030831.67-005008.6 is H$\alpha$/H$\beta \sim$ 3.24 $\pm$ 0.31 (Galactic extinction corrected).
This value is only slightly larger than the Case B H$\alpha$/H$\beta$ ratio of 2.86 for an electron density of $n=100$ cm${}^{-3}$ and electron temperature $T_e=10^4$ K \citep[e.g.,][]{ost89,bri04,dom13}.
The Balmer decrement of $\sim 3.24$ implies that the host galaxy extinction, if any, is $E(B-V)_{\text{host}} \sim 0.12$~mag (assuming SMC-like extinction and $R_{V} = 2.93$), which is much smaller than the host galaxy extinction inferred from the KISS15s spectrum ($E(B-V)_{\text{host}} \sim 0.6$~mag; Section~\ref{note_on_host_extinction}).
However, the explosion site of KISS15s, which is located at the edge of the host galaxy, could be much deeper inside the dust lane of the galactic disk or embedded in a dusty nebula and thus more heavily obscured, as inferred in Section~\ref{note_on_host_extinction}.
It is also possible that KISS15s is affected by CSM dust.
A promising candidate progenitor of SNe IIn would be the luminous blue variables \citep[LBVs;][]{gal09,dwa11}; observationally circumstellar environments of LBVs are preferentially dusty \citep{uma12}.

The comoving volume within the redshift of SDSS~J030831.67-005008.6 is $0.0143$~Gpc${}^{3}$.
The volumetric rate of the core-collapse SNe at the local universe is estimated to be $\sim 0.7 \times 10^{-4}$~SN~Mpc${}^{-3}$~yr${}^{-1}$ \cite[e.g.,][]{li11b}.
SNe IIn comprise roughly 7~\% of the entire core-collapse SNe population \citep{li11,smi11b,shi17}.
From these values, we can estimate that SNe IIn are occurring at a rate of $\sim 70$~year${}^{-1}$ out to the redshift of SDSS~J030831.67-005008.6 (at the comoving distance of $150$~Mpc) \citep[see also][]{fox10}.
Considering the limiting magnitudes and the sky coverage of the KISS survey ($\sim$300~deg${}^{2}$), the event rate of $\sim 70$~yr${}^{-1}$ is consistent with the fact that a handful of SNe IIn within 150~Mpc (including the 1988Z-like KISS15s) had been discovered in 4 years' worth of survey data \citep[see][]{morokuma14}.
It also indicates that 1988Z-like SNe IIn are common among the total population of SNe IIn \citep[e.g.,][]{tad15}.

\subsection{Energy source of the optical continuum emission}
\label{sec:energy_source}

Eruptions of LBVs produce $M>-16$ mag optical transients with similar optical spectra to SNe IIn, referred to as SN impostors \citep[e.g.,][]{van00,mau06,smi11,smi14,jae15}.
If the host galaxy extinction is assumed to be zero, KISS15s may possibly be identified as an SN impostor, in terms of its faintness of the optical continuum emission ($M_g \sim -16.6$ mag at the peak epoch; Figure~\ref{fig:lightcurve_5}).
However, as discussed in Sections~\ref{note_on_host_extinction} and \ref{sec:ir_lightcurves}, there is strong evidence of the host galaxy extinction of $E(B-V)_{\text{host}} \sim 0.6$ mag; thus, the intrinsic peak absolute magnitude of KISS15s is estimated to be $M_g\sim -18.8$ mag (Figure~\ref{fig:lc_comparisons}).
According to the luminosity function of SNe IIn compiled by \cite{ric14}, the inferred intrinsic absolute magnitude of KISS15s is consistent with the typical magnitude range of SNe IIn ($M_B=-15.81$ to $-21.25$ mag in the 2$\sigma$ range; the mean is $M_B=-18.53$ mag).
Moreover, KISS15s shows strong broad and intermediate emission lines in the optical spectra and late-time hot dust thermal emission, both of which are difficult to explain without the strong shock interaction between SN ejecta and dense CSM.
These observational properties suggest that KISS15s is not an SN impostor but rather is a true SN event.

Since its discovery, the optical continuum emission of KISS15s has shown a slow temporal evolution up to $\sim$~600 days followed by a sudden decrease, with a faster rate of decline rate in late epochs (Sections~\ref{sec:powerlaw_fitting} and \ref{sec:bolometric}).
The long-duration optical continuum emission of SNe IIn is undoubtedly the result of strong ejecta-CSM interactions \citep{che81,sch90,smi17b}.
If we assume that the event duration is solely determined by the outer extent of the dense CSM \citep[e.g.,][]{smi14}, the long duration of $t_{\text{duration}} \sim $ 600~days of the KISS15s light curve may imply that the progenitor star of KISS15s had experienced continuous mass-loss by extreme stellar winds blowing for 
\begin{eqnarray}
t_\text{wind} & \sim & \frac{v_{\text{s}}}{v_{w}} \times t_{\text{duration}}\nonumber\\ 
&\approx& 82~\text{yrs}~\left( \frac{v_{s}}{2,000~\text{km}~\text{s}^{-1}} \right) \times \left( \frac{v_{w}}{40~\text{km}~\text{s}^{-1}} \right)^{-1} \left( \frac{t_{\text{duration}}}{600~\text{d}} \right) \nonumber
\end{eqnarray}
before the SN explosion, where $v_{w}$ and $v_{s}$ denote the stellar wind velocity and the ejecta-CSM shock velocity, respectively \citep{smi17b,smi17}.
Here we adopt typical values for SNe IIn.
The long duration light curves, corresponding to the centuries of massive stellar winds before the SN explosion, are commonly observed in 1988Z-like SNe IIn (Figure~\ref{fig:lc_comparisons}); however, the physical mechanism driving such stellar winds in the final stage of stellar evolution is unclear \citep[][and references therein]{shi14,smi14,smi17,yar17}.
As already mentioned in Sections~\ref{sec:powerlaw_fitting} and \ref{sec:bolometric}, the sudden luminosity drop observed in KISS15s at $\gtrsim 600$~days after first detection may possibly be interpreted as a signature of the shock wave exit from the dense CSM \citep[as suggested for the luminosity evolution of SN~2010jl;][]{fra14,mor14b,des15}.

As discussed in Section~\ref{sec:bolometric}, the power-law index of the time evolution of the bolometric light curve of KISS15s is $\alpha=-0.16$ (Equation~\ref{eqn:bolometric_index}).
Comparing this value to a simple interaction model that relates the power-law index of the time evolution of the bolometric light curve $\alpha$ to that of the density radial profile of the expanding SN ejecta $\rho_{\text{ej}}\propto r^{-n}$ and the stellar wind CSM of $\rho_{CSM}\propto r^{-s}$ as \cite[e.g.,][]{mor13,mor14,ofe14b,fra14}
\begin{equation}
\alpha=\frac{6s-15+2n-ns}{n-s}, 
\end{equation}
we obtain $n=21$ in the case of a steady stellar mass-loss $s=2$.
This power-law index is steeper than the values expected for the SN ejecta density structure of red supergiants (RSGs; $n\simeq12$) and Wolf-Rayet stars ($n \simeq 10$) \citep[e.g.,][]{che94,mat99,mor13,mor14,ofe14b}.
Instead, if we assume $n$ to be in the range of $n=10-12$, the CSM density slope $s$ can be constrained to be $s=1.6-1.8$, suggesting that the mass-loss rate was nearly steady but gradually decreased as the progenitor got closer to the time of explosion \citep{mor14}.
In conclusion, the photometric light curve of the continuum emission of KISS15s is consistent with the scenario that KISS15s is an SN IIn in which the continuum luminosity (at least after the time of discovery) is powered by the strong interaction between SN II ejecta and the CSM formed by centuries of nearly steady mass-loss episodes of the progenitor star before the SN explosion.

\subsection{CSM properties of KISS15s: temporal variation in the emission line profile}
\label{sec:spec_lightcurve}

The line-of-sight velocity of KISS15s (the bottom panel of \ref{fig:spec_lightcurves_1}) shows complex behaviors.
Focusing on the intermediate component, the velocity shift decreases from $0$ to $-500$~km~s${}^{-1}$.
An increase in the blue-shift velocity of an intermediate-width emission line component in SNe IIn is interpreted either as increasing extinction of the red wing by newly formed dust \citep{mae13,gal14} or as radiative acceleration of the line emission regions due to the intense radiation from inner regions \citep{fra14}.
In the case of KISS15s, the emission line profile of the intermediate component remains consistent with a symmetric Gaussian while the velocity shift changes.
The persistently symmetric line profile disagrees with the expectation from the dust extinction model and supports the scenario that the time evolution of the velocity shift is probably due to the radiative acceleration of the line-emitting region.
Contrary to the intermediate emission line, the blue-shift excess component of KISS15s shows deceleration since discovery, which may be due to a braking effect by the swept-up CSM \citep{fra14}.

The emission line profile of KISS15s largely resembles those of 1988Z-like SNe IIn  (Figure~\ref{fig:spec_halpha_comparison}).
It has been suggested that the broad and intermediate velocity components in the SN~1988Z-like SNe IIn can be interpreted as two distinct line-emitting regions, where the former is related to the unshocked expanding ejecta and the latter to the shocked ejecta/CSM (or CDS) regions.
The line velocity widths are determined by a combination of bulk motion of the line emitting region, electron scattering broadening, and so forth, and it is uncertain what physical mechanisms are dominantly responsible for producing the observed emission line velocity widths.
The intermediate and blue-shifted excess emission line components have similar velocity widths but have a relative velocity shift of about $-5,000$~km~s${}^{-1}$ (Section~\ref{balmerfit}).
This suggests that two separate shocked regions may have propagated through the non-spherically symmetric CSM, in which the large velocity shift component corresponds to emission from a high velocity ejecta-CSM shock propagating through a more rarified CSM region toward the observer \citep[e.g.,][]{chu94,fox10,smi12b}.

Interestingly, similar blue-shifted excess emission with a velocity shift of about $-5,000~\text{km}~\text{s}^{-1}$ (decelerated from $-5,500~\text{km}~\text{s}^{-1}$ to $-4,000~\text{km}~\text{s}^{-1}$) was detected in the helium lines of SN~2010jl; however, in the case of SN~2010jl, blue-shifted excess emission is not detected in the hydrogen lines \citep{fra14,bor15}.
\cite{bor15} attributed this blue-shifted emission component to photoionized ejecta in which hydrogen is underabundant, expanding into a CSM region of relatively small density toward the observer.
In the case of KISS15s, the blue-shifted component is clearly detected in the hydrogen lines and has a consistent Gaussian profile with the central component (Section~\ref{balmerfit}).
Thus, the blue-shifted excess in KISS15s is interpreted as photoionized hydrogen-abundant CSM material located in the high velocity ejecta-CSM interaction region.

In the literature, it is generally assumed that the shock front velocity $v_{s}$ is on the order of the velocity width of the intermediate component \citep[][]{kie12,tad13}, i.e., $v_{s} \sim v_{i}\sim 2,000$~km~s${}^{-1}$ for KISS15s.
In addition, an estimate of an upper limit on $v_{s}$ can be derived from the line-of-sight velocity shift of the blue-shift excess emission component,  $v_{s} \lesssim 5,000$~km~s${}^{-1}$ (see Section~\ref{balmerfit} for details).
Adopting $t_{\text{duration}}=600$~days, the expanding shock reaches $r_{s} = v_{s} \times t_{\text{duration}} \sim 4$~light-days.
The inferred values of the shock front velocity $v_{s}$ and the radial extent $r_{s}$ are comparable to the IR black-body expansion velocity $v_{\text{dust}}$ and the IR black-body radius $r_{\text{BB, IR}}$ (Equations~\ref{eqn:diff_velocity} and \ref{eqn:bb_radius}, respectively), which further supports the scenario that the hot dust thermal emission probed by the W1- and W2-band observations is produced by newly formed dust in the ejecta-CSM interaction region or CDS.

The velocity width of the broad emission line decreases from $\sim 14,000$~km~s${}^{-1}$ to $\sim 10,000$~km~s${}^{-1}$ (Figure~\ref{fig:spec_lightcurves_1}).
Although the line-broadening mechanism, and consequently the origin, of the broad emission lines are unclear, the extreme broadness suggests that the broad line emission region is related to the inner part of the ejecta-CSM system of KISS15s, in which a high gas temperature and a large bulk motion of the gas can be expected.
Assuming that the velocity width reflects the thermal motion of free electrons, the decreasing velocity width indicates that the temperature of the broad emission line region rapidly decreases by a factor of $2$ during our observations, due to radiative cooling or adiabatic expansion.

\subsection{Progenitor mass-loss rate}
\label{sec:massloss}

Assuming a steady stellar wind CSM ($s=2$; Section~\ref{sec:energy_source}), the progenitor mass-loss rate $\dot{M}$ just before the SN explosion can be related to the bolometric luminosity via a kinetic-to-radiation conversion efficiency factor $\epsilon < 1$, as
\begin{equation}
L_{\text{bol}} = \epsilon \frac{dE_{\text{kin}}}{dt} = \frac{1}{2}\epsilon \frac{\dot{M}}{v_w}v_s^3,
\label{eq:massloss_lbol}
\end{equation}
where $E_{\text{kin}}$ denotes the kinetic energy of the thin shocked shell, and $\epsilon$ is usually assumed to be in a range of $0.1-0.5$ \citep{mor13,mor14,mor14c,ofe14b}.
As reference values, we adopt $\epsilon = 0.3$ and $v_s \sim v_{i} \sim 2,000$~km~s${}^{-1}$ (Section~\ref{sec:spec_lightcurve}).
The absence of the narrow P Cygni absorption in the APO/DIS high-resolution spectra ($\sigma_\text{inst} = 20.8~\text{km}~\text{s}^{-1}$; Section~\ref{data:dis}) may provide a constraint on the stellar wind velocity of KISS15s as $v_w \lesssim \text{FWHM}_\text{inst}/(1+z) = 47.2~\text{km}~\text{s}^{-1}$; here, we adopt $v_w = 40$~km~s${}^{-1}$ \citep[for example, $v_w = 40$~km~s${}^{-1}$ is assumed in the case of SN~2005ip by][]{smi17}.

As shown in Section~\ref{sec:bolometric}, the pseudo-bolometric luminosity of KISS15s at $\sim 600$~days since discovery is roughly $L_{\text{bol}} \sim 0.8 \times 10^{43}~\text{erg}~\text{s}^{-1}$ (Figure~\ref{fig:lc_opt_bb_lum}), giving a mass-loss rate estimate of
\begin{eqnarray}
\dot{M} & \simeq & 0.4~\text{M}_{\odot}~\text{yr}^{-1}~ \left(\frac{L_{\text{bol}}}{0.8\times 10^{43}~\text{erg}~\text{s}^{-1}}\right)\nonumber\\
        &\times& \left(\frac{v_w}{40~\text{ km}~\text{s}^{-1}}\right)\left(\frac{\epsilon}{0.3}\right)^{-1} \left(\frac{v_s}{2,000~\text{km}~\text{s}^{-1}}\right)^{-3}.
\label{eqn_massloss_lbol}
\end{eqnarray}
The observed duration of $t_{\text{duration}} \sim 600$~days for KISS15s indicates that the amount of shocked CSM around KISS15s at the epoch of our last spectroscopic observation on 2016 October 4 is \citep[][]{fra14}
\begin{eqnarray}
M_{\text{shocked~CSM}} &=& \frac{v_s}{v_w}\dot{M} \times t_{\text{duration}} \nonumber\\
&\simeq& 35~\text{M}_{\odot}~\left(\frac{L_{\text{bol}}}{0.8\times 10^{43}~\text{erg}~\text{s}^{-1}}\right)\nonumber\\
&\times&\left(\frac{\epsilon}{0.3}\right)^{-1} \left(\frac{v_s}{2,000~\text{km}~\text{s}^{-1}}\right)^{-2} \times \left( \frac{t_{\text{duration}}}{600~\text{days}} \right).
\end{eqnarray}

Instead, in the literature, it is usually assumed that the luminosity of the intermediate H$\alpha$ line (denoted as $L_{\text{H}\alpha}$) is proportional to the bolometric luminosity $L_{\text{bol}}$ and thus is a good indicator of $L_{\text{bol}}$ \citep[e.g.,][]{sal98,kie12,tad13,jae15}.
$L_{\text{H}\alpha}$ can be written as 
\begin{equation}
L_{\text{H}\alpha} = \frac{1}{2}\epsilon_{\text{H}\alpha} \frac{\dot{M}}{v_w}v_s^3,
\label{eq:massloss}
\end{equation}
where $\epsilon_{H\alpha}$ represents the efficiency of the conversion of the dissipated kinetic energy into the H$\alpha$ luminosity in the shock wave.
Note that because at most only half of the total radiation can contribute to the ionization of the shell, $\epsilon_{H\alpha}$ is less than 0.5 \citep{sal98}.
For KISS15s, the intermediate H$\alpha$ emission line luminosity in late epochs, corrected for the host galaxy extinction of $E(B-V)_{\text{host}} = 0.6$~mag, is $L_{H\alpha} \simeq 1 \times 10^{41}$ erg~s${}^{-1}$ (Figure~\ref{fig:spec_lightcurves_2}).
To make the mass-loss rate values between Equations~\ref{eq:massloss_lbol} and \ref{eq:massloss} consistent with each other, $\epsilon_{H\alpha}$ must be much smaller than $\epsilon$, i.e.,
\begin{eqnarray}
\frac{\epsilon_{H\alpha}}{\epsilon} &=& \left( \frac{L_{H\alpha}}{L_{\text{bol}}} \right) \nonumber\\
&=& 0.013  \left( \frac{L_{H\alpha}}{1.0 \times 10^{41}~\text{erg}~\text{s}^{-1}} \right)\left( \frac{L_{\text{bol}}}{0.8 \times 10^{43}~\text{erg}~\text{s}^{-1}} \right)^{-1}.
\end{eqnarray}
The low value of $\epsilon_{H\alpha}$ probably reflects a low conversion efficiency from the ionizing continuum to the H$\alpha$ emission line due to thermalization of the H$\alpha$ line inside the dense CSM \citep[e.g.,][]{fra14}.
In the case of $\epsilon = 0.3$, $\epsilon_{H\alpha}$ becomes 0.004, which is an order of magnitude smaller than the canonical value of $\epsilon_{H\alpha} = 0.05$ usually assumed in the literature \citep[e.g.,][]{sal98,tad13}.

It should be kept in mind that these estimates strongly depend on the assumed parameters.
The mass-loss rate estimate based on the above equations is uncertain in that the wind velocity $v_{w}$ is not observationally constrained for KISS15s, and continuum and emission line luminosities $L_{\text{bol}}$ and $L_{\text{H}\alpha}$ are time-variable.
The assumption that the velocity width of the intermediate H$\alpha$ emission line represents the shock velocity $v_{s}$ may be an oversimplification, and the mass-loss rate estimate of KISS15s becomes much smaller ($\dot{M} \simeq 0.1~M_{\odot}~\text{yr}^{-1}$) if we instead adopt a higher value of $v_{s} = 3,000~\text{km}~\text{s}^{-1}$, as inferred for SN~2010jl \citep{ofe14b}.
Moreover, the above equations assume spherically-symmetric CSM distributions; thus, the mass-loss rate estimate is subjected to uncertainties depending on the inhomogeneities of the CSM \citep[e.g.,][]{fra14}.

The mass-loss rate of KISS15s estimated with Equation~\ref{eqn_massloss_lbol} is largely on the upper bound of those measured in other SNe IIn \citep[e.g.,][]{kie12,tad13,mor14}.
For example, using Equation~\ref{eqn_massloss_lbol} and adopting $\epsilon=0.3$, the mass-loss rate of the highly luminous SN~2010jl is estimated to be $\dot{M}\sim 0.4~M_{\odot}~\text{yr}^{-1}$ \citep[][]{fra14}\footnote{\cite{fra14} implicitly assumed $\epsilon=1$ when deriving $\dot{M} = 0.11~M_{\odot}~\text{yr}^{-1}$ as a mass-loss rate estimate for SN~2010jl \citep[see Equations~9 and 11 of][]{fra14}.}.
Such a high progenitor mass-loss rate is three to four orders of magnitude higher than typical values for RSGs, and is consistent with those expected for eruptive mass-loss of LBVs \citep[e.g.,][]{kie12,tad13,smi14,gol17}.
However, as noted by \cite{tad13}, we should not take the rough mass-loss rate estimate described above as conclusive evidence for the LBV progenitor channel for SNe IIn.
According to studies of local environments of SNe IIn in their host galaxies, SNe IIn generally do not trace star-forming regions in the host galaxies, suggesting that the majority of SN IIn events are not related to the most massive stars such as LBVs \citep[e.g.,][]{hab14,kun18}.
Moreover, as mentioned in Section~\ref{sec:energy_source}, the long duration of several hundreds of days of the light curves of KISS15s implies that the high progenitor mass-loss rates are sustained at least for decades before the SN explosions.
Similar enhanced mass-loss for $\sim$ 1,000 years before core collapse (i.e., well before Ne, O, or Si burning phases) may also explain the continuous strong CSM interaction in other SN~1988Z-like SNe IIn, such as SNe 2005ip and 1988Z \citep{smi17}.
These observations suggest that there must be unresolved mechanisms that trigger the large persistent mass-loss at the final stage of the massive star evolution, other than the eruptive mass-loss observed in LBVs and RSGs \citep[e.g.,][]{mor14,smi17}.

Physical mechanisms enhancing the mass-loss from RSGs several decades before the explosion are currently not uniquely identified \citep[e.g.,][]{shi14,smi14c,woo17}.
Further long-term observations of SNe IIn will provide important information about the mass-loss mechanisms in the late stage of massive star evolution.

\subsection{Model of the emission regions of KISS15s}
\label{sec:geometry_model}

\begin{figure*}[tbp]
\center{
\includegraphics[clip, width=5.3in]{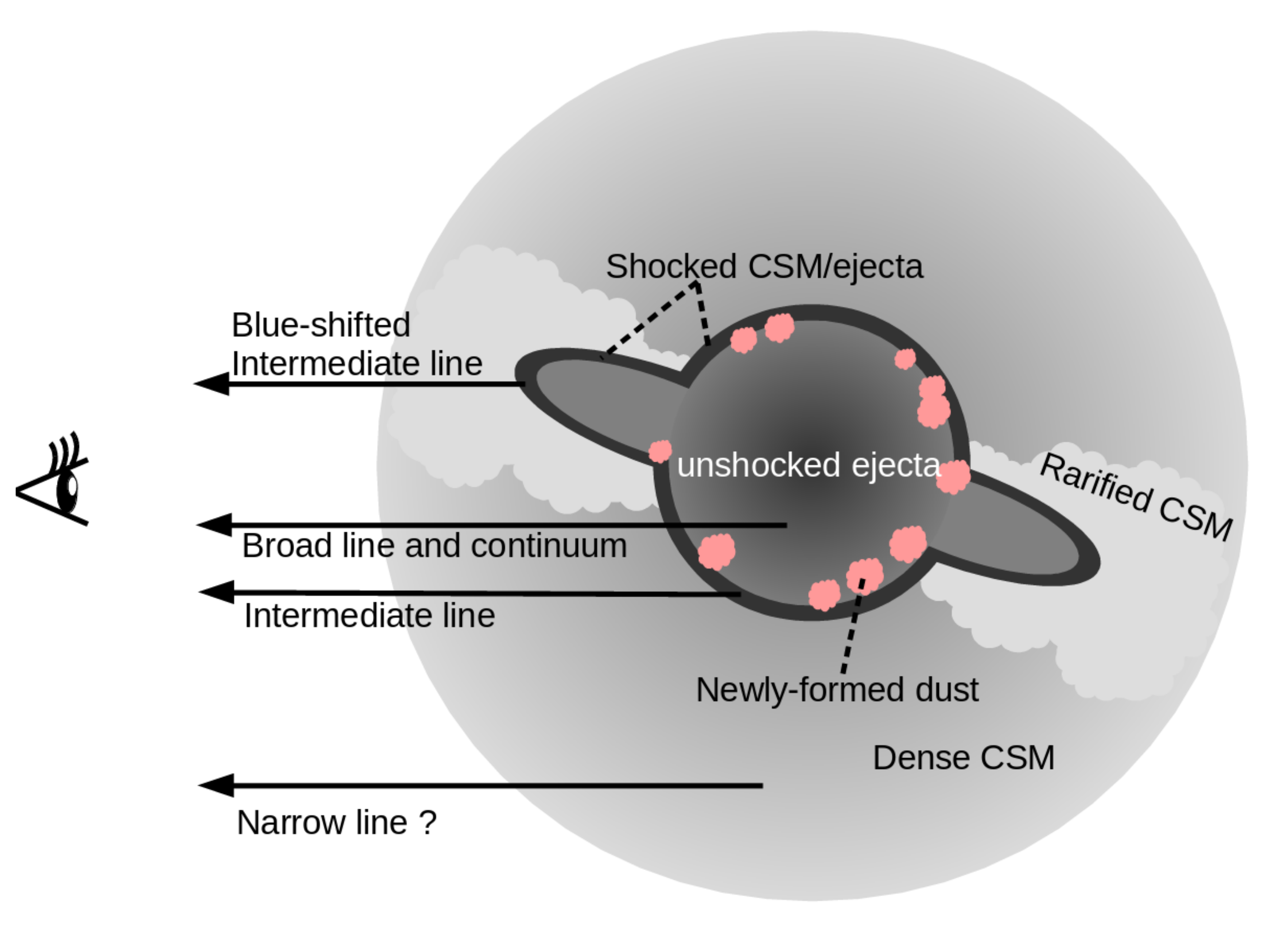}
}
 \caption{
Schematic drawing of the proposed ejecta-CSM interaction region of KISS15s (see text in Section~\ref{sec:geometry_model}). 
The broad component (FWHM: $\sim$~14,000~km~s${}^{-1}$; velocity shift: about $-700$~km~s${}^{-1}$) and the optical continuum are produced by unshocked ejecta, ionized by intense ingoing radiation from the reverse shock.
The intermediate component (FWHM: $\sim$~2,000~km~s${}^{-1}$; velocity shift: about $-500$~km~s${}^{-1}$) is attributed to the emission from the shocked ejecta-CSM region or CDS.
The ejecta expanding through a less dense CSM region have a higher velocity, in which a blue-shifted intermediate-width emission line (FWHM: $\sim$~2,000~km~s${}^{-1}$; velocity shift: about $-5,000$~km~s${}^{-1}$) is generated.
The observed narrow emission lines (FWHM $\lesssim$~100~km~s${}^{-1}$) are predominantly from the \ion{H}{2} regions in the host galaxy, but there may be flux contributions from the emission of unshocked CSM, photoionized by SN emission.
New dust can be formed in the CDS and produce the hot dust thermal emission, as indicated by red blobs in the illustration. 
 }
 \label{fig:schematic}
\end{figure*}

The observations of the emission line profile and the light curves of KISS15s provide constraints on the geometry of its emission regions.
Figure~\ref{fig:schematic} summarizes the proposed picture of the ejecta-CSM interaction region of KISS15s, based on a toroidal CSM geometry model \citep{chu94,mau14,smi15,smi17b}.
As mentioned in Sections~\ref{line_property} and \ref{sec:spec_lightcurve}, such a CSM density distribution formed by the wide-angle bipolar stellar wind can result in the inhomogeneous expanding ejecta distribution and produce multiple velocity components in the H$\alpha$ emission line (see below).
It should be noted that although the model illustrated in Figure~\ref{fig:schematic} employs a bipolar outflow-like geometry, we do not exclude the possibility of one-sided asymmetry \citep[see e.g.,][]{fra14}.

The broad emission lines (FWHM $\sim$ 14,000 km~s${}^{-1}$) are probably produced from unshocked ejecta, which is ionized by high-energy ingoing radiation from the reverse shock \citep[e.g.,][]{chu94,che17}.
As shown in Figure~\ref{fig:spec_lightcurves_2}, the continuum luminosity shows a similar declining rate with the broad H$\alpha$ emission line, suggesting that the continuum emission may possibly be powered by the same energy source with a broad emission line, and the emission regions of the broad emission line and optical continuum are co-spatial \citep[e.g,][]{smi17b}.

The intermediate emission line component can be produced from the shocked CSM/ejecta or CDS ionized by the intense radiation from the forward or reverse shocks \citep[e.g.,][and references therein]{smi17}.
The blue-shifted intermediate emission line component has a consistent velocity width of $\sim 2,000$~km~s${}^{-1}$ with the intermediate emission line at the systemic redshift, suggesting that the physical conditions between the two emission regions should be similar to each other; nevertheless, the line-of-sight velocity is different by $\sim 5,000$~km~s${}^{-1}$ (Section~\ref{sec:spec_lightcurve}).
As discussed in Section~\ref{sec:spec_lightcurve}, and as shown in Figure~\ref{fig:schematic}, one possibility of the origin of these two intermediate-width components is inhomogeneities in the CSM.
If there is a less dense region within the CSM, the shock velocity in the less dense direction can be larger compared to the other directions; thus, it can produce the blue-shifted emission line component \citep[e.g.,][]{chu94,smi09,bor15}.
The geometry shown in Figure~\ref{fig:schematic} is the simplest case, and actual SNe probably have more complex inhomogeneities of the CSM/ejecta distribution.
The absence of the red-shifted counterpart of the blue-shifted intermediate component in the observed spectra can be explained by the absorption due to the opaque continuum photosphere in front of the far-side rarified CSM region \citep[e.g.,][]{des15}; however, it is also possible that the rarified CSM has an intrinsically one-sided structure pointing towards us.

The IR emission can naturally be interpreted as $T_{\text{BB, IR}} \sim 1,200$~K hot dust thermal emission (Section~\ref{sec:ir_lightcurves}).
The constantly high dust temperature during the observations (see Figure~\ref{fig:wise_color_temperature}) suggests that the dust emission is from newly formed dust, not from IR echo or radiatively heated preexisting dust in the CSM.
The newly formed dust responsible for the IR excess emission can be located in the CDS region, where the gas is radiatively cooled to temperatures in which it shifts to supersaturated states, at which point dust grain condensation can be initiated \citep[e.g.,][]{smi09,che17}.
The strong H$\alpha$ emission with the symmetric line profile observed even in late epochs implies that the dust distribution is probably clumpy, which enables the optical emission to leak from the inner part of the ejecta-CSM interaction region \citep{mae13,chu18}.

The observed narrow emission lines are probably dominated by the emission from \ion{H}{2} regions in the host galaxy, which may not be directly related to KISS15s.
Nevertheless, the present observations do not rule out the possibility of a narrow emission line flux contribution from the unshocked CSM photoionized by the SN emission.

Apparently, the proposed toroidal CSM distribution is not a unique solution.
Another possible scenario to explain the blue-shifted H$\alpha$ intermediate-width emission line component is to consider the (one-sided) jet structure of SN ejecta.
To account for the triple-peak H$\alpha$ emission line (composed of central emission line and $\sim$10,000 km~s${}^{-1}$ blue- and red-shifted H$\alpha$ emission lines) observed in SN~2010jp, \cite{smi12b} proposed a model in which a two-sided supernova jet, inclined relative to the observer, passes through initially spherically symmetric CSMs and produces two shocked regions with blue- and red-shifted velocities.
It should be noted that the triple-peak H$\alpha$ emission line profile in SN~2010jp is quite different from the double-peak intermediate line profile observed in KISS15s, and the velocity shift is also more modest in KISS15s ($\sim -5,000$~km~s${}^{-1}$); thus, there is no need to employ the jet structure other than the inhomogeneous expansion due to the aspherical CSM structure in the case of KISS15s.

Further multi-wavelength monitoring observations at optical, IR, X-ray, and radio wavelengths should reveal the details of the ejecta-CSM interaction regions in KISS15s.

\section{Conclusions}
\label{sec:conclusions}

We report the discovery of KISS15s, which is confirmed to be an SN~1988Z-like type IIn supernova.
The host galaxy of KISS15s is a low-mass ($M_{*} = 10^{8-9}$~$M_{\odot}$), low-metallicity ($12+\log \text{O/H} = 7.873 \pm 0.385$), star-forming galaxy at $z=0.038$.
Spectral comparisons with SN~1988Z suggest that it is suffering from host galaxy extinction of $E(B-V)_{\text{host}} = 0.6$~mag on the assumption of an SMC-like extinction curve.

We modeled the H$\alpha$ emission line profile by combining four symmetric Gaussians of narrow, intermediate, blue-shifted intermediate, and broad velocity width components.
While the blue-shift velocities of the narrow, intermediate, and broad components relative to the systemic redshift of $z=0.038$ are smaller than about $-1,000$~km~s${}^{-1}$, the blue-shifted intermediate component shows a large blue-shift of about $-5,000$~km~s${}^{-1}$.
The fitting results reveal the FWHMs of broad, intermediate, and narrow components of $\lesssim 100$, $\sim 2,000$, and $\sim 14,000$ km~s${}^{-1}$, respectively, and the intermediate and blue-shifted intermediate velocity components have consistent velocity widths with respect to the others.
The narrow H$\alpha$ emission line component is most likely dominated by emission from the interstellar medium (\ion{H}{2} regions) in the host galaxy.
However, it is still possible that there is a flux contribution from an unshocked CSM around KISS15s, which is photoionized by strong radiation from KISS15s.
The intermediate H$\alpha$ line luminosity has increased monotonically since the discovery of KISS15s, which implies that the ejecta-CSM interaction is still ongoing, at least during the observations presented in this work.

In the $\lesssim$~600~days since discovery, the optical continuum emission has decreased monotonically.
By contrast, the hot dust IR continuum emission peaking at $\sim 3 \mu$m ($T_{\text{BB, IR}} = 1,190$~K) has increased over time since discovery, suggesting that new dust grains are continuously being formed in the CDS.
The increasing IR luminosity indicates that the IR black-body expansion velocity is $v_{\text{dust}} \sim 1,000-6,000$~km~s${}^{-1}$ (Figure~\ref{fig:wise_luminosity}), which is close to the velocity widths and/or velocity shifts of the H$\alpha$ emission line components, suggesting that the hot dust thermal emission is actually due to newly formed dust in the CDS and is not due to the IR echo by preexisting dust in the CSM.
Unlike the other SNe IIn associated with IR excess emission due to the newly formed dust at the CDS (e.g., SN~2005ip, 2006jd, and 2010jl), the intermediate H$\alpha$ emission line profile of KISS15s shows no evidence of late-time blue-shifted asymmetry, which implies that the geometric configuration of the line-emitting region and dust-forming region in the CDS is different among SNe IIn.

The bolometric luminosity of KISS15s (calculated from the sum of the optical and IR black-body luminosities) roughly follows a power-law luminosity evolution of $L \propto t^{-0.16}$.
The power-law decline of the luminosity can naturally be explained by the interaction between the ejecta and the mostly steady stellar wind (Section~\ref{sec:energy_source}).

The coexistence of the intermediate H$\alpha$ emission line components with large blue-shift and at roughly systemic redshift indicates that the ejecta-CSM interaction region of KISS15s has a non-spherical, inhomogeneous spatial distribution.
We propose the toroidal CSM geometry model where the ejecta expanding through the rarified CSM region can have a higher velocity than that through the dense CSM, producing the blue-shifted intermediate component.
The broad emission line may be produced from unshocked ejecta ionized by the intense radiation from the ejecta-CSM interaction region or CDS.

The progenitor mass-loss rate inferred from the bolometric luminosity is $\dot{M} \sim 0.4$ $M_{\odot}$ yr${}^{-1}$ $(v_{w}/40$ km s${}^{-1}$), where $v_w$ is the observationally unconstrained wind speed of the CSM, suggesting that the progenitor of KISS15s was a red supergiant star or an LBV that had experienced a large mass-loss in the centuries before the explosion.
The comparison between the bolometric continuum luminosity and intermediate-width H$\alpha$ emission line luminosity indicates that the kinetic-to-radiation conversion efficiency for the H$\alpha$ emission is roughly 100 times smaller than that for the bolometric luminosity.

Future follow-up observations of KISS15s at optical, IR, radio, and X-ray wavelengths will provide insight into the late-time evolution of dust formation in KISS15s and more detailed constraints regarding the CSM geometry.


\acknowledgments

\section*{Acknowledgements}

This work was supported by JSPS KAKENHI Grant Number 17J01884 and 15H02075.
We are grateful to all the staff in the Kiso Observatory, the Nishi-Harima Astronomical Observatory, and the Apache Point Observatory for their efforts to maintain the observation systems.
We thank Ryo Tazaki and Takaya Nozawa for fruitful discussions.
This research has made use of NASA's Astrophysics Data System Bibliographic Services.

This work is based in part on observations at Kitt Peak National Observatory, National Optical Astronomy Observatory (NOAO Prop. ID 2017A-0172; PI: T.~Morokuma), which is operated by the Association of Universities for Research in Astronomy (AURA) under cooperative agreement with the National Science Foundation. The authors are honored to be permitted to conduct astronomical research on Iolkam Du'ag (Kitt Peak), a mountain with particular significance to the Tohono O'odham.

This publication makes use of data products from the Two Micron All Sky Survey, which is a joint project of the University of Massachusetts and the Infrared Processing and Analysis Center/California Institute of Technology, funded by the National Aeronautics and Space Administration and the National Science Foundation.

This publication makes use of data products from the Near-Earth Object Wide-field Infrared Survey Explorer (NEOWISE), which is a project of the Jet Propulsion Laboratory/California Institute of Technology. NEOWISE is funded by the National Aeronautics and Space Administration.
This research has made use of the NASA/ IPAC Infrared Science Archive, which is operated by the Jet Propulsion Laboratory, California Institute of Technology, under contract with the National Aeronautics and Space Administration.

The Legacy Surveys consist of three individual and complementary projects: the Dark Energy Camera Legacy Survey (DECaLS; NOAO Proposal ID \# 2014B-0404; PIs: David Schlegel and Arjun Dey), the Beijing-Arizona Sky Survey (BASS; NOAO Proposal ID \# 2015A-0801; PIs: Zhou Xu and Xiaohui Fan), and the Mayall z-band Legacy Survey (MzLS; NOAO Proposal ID \# 2016A-0453; PI: Arjun Dey). DECaLS, BASS and MzLS together include data obtained, respectively, at the Blanco telescope, Cerro Tololo Inter-American Observatory, National Optical Astronomy Observatory (NOAO); the Bok telescope, Steward Observatory, University of Arizona; and the Mayall telescope, Kitt Peak National Observatory, NOAO. The Legacy Surveys project is honored to be permitted to conduct astronomical research on Iolkam Du'ag (Kitt Peak), a mountain with particular significance to the Tohono O'odham Nation.
NOAO is operated by the Association of Universities for Research in Astronomy (AURA) under a cooperative agreement with the National Science Foundation.
This project used data obtained with the Dark Energy Camera (DECam), which was constructed by the Dark Energy Survey (DES) collaboration. Funding for the DES Projects has been provided by the U.S. Department of Energy, the U.S. National Science Foundation, the Ministry of Science and Education of Spain, the Science and Technology Facilities Council of the United Kingdom, the Higher Education Funding Council for England, the National Center for Supercomputing Applications at the University of Illinois at Urbana-Champaign, the Kavli Institute of Cosmological Physics at the University of Chicago, Center for Cosmology and Astro-Particle Physics at the Ohio State University, the Mitchell Institute for Fundamental Physics and Astronomy at Texas A\&M University, Financiadora de Estudos e Projetos, Fundacao Carlos Chagas Filho de Amparo, Financiadora de Estudos e Projetos, Fundacao Carlos Chagas Filho de Amparo a Pesquisa do Estado do Rio de Janeiro, Conselho Nacional de Desenvolvimento Cientifico e Tecnologico and the Ministerio da Ciencia, Tecnologia e Inovacao, the Deutsche Forschungsgemeinschaft and the Collaborating Institutions in the Dark Energy Survey. The Collaborating Institutions are Argonne National Laboratory, the University of California at Santa Cruz, the University of Cambridge, Centro de Investigaciones Energeticas, Medioambientales y Tecnologicas-Madrid, the University of Chicago, University College London, the DES-Brazil Consortium, the University of Edinburgh, the Eidgenossische Technische Hochschule (ETH) Zurich, Fermi National Accelerator Laboratory, the University of Illinois at Urbana-Champaign, the Institut de Ciencies de l'Espai (IEEC/CSIC), the Institut de Fisica d'Altes Energies, Lawrence Berkeley National Laboratory, the Ludwig-Maximilians Universitat Munchen and the associated Excellence Cluster Universe, the University of Michigan, the National Optical Astronomy Observatory, the University of Nottingham, the Ohio State University, the University of Pennsylvania, the University of Portsmouth, SLAC National Accelerator Laboratory, Stanford University, the University of Sussex, and Texas A\&M University.
The Legacy Surveys imaging of the DESI footprint is supported by the Director, Office of Science, Office of High Energy Physics of the U.S. Department of Energy under Contract No. DE-AC02-05CH1123, by the National Energy Research Scientific Computing Center, a DOE Office of Science User Facility under the same contract; and by the U.S. National Science Foundation, Division of Astronomical Sciences under Contract No. AST-0950945 to NOAO.

The Pan-STARRS1 Surveys (PS1) have been made possible through contributions of the Institute for Astronomy, the University of Hawaii, the Pan-STARRS Project Office, the Max-Planck Society and its participating institutes, the Max Planck Institute for Astronomy, Heidelberg and the Max Planck Institute for Extraterrestrial Physics, Garching, The Johns Hopkins University, Durham University, the University of Edinburgh, Queen's University Belfast, the Harvard-Smithsonian Center for Astrophysics, the Las Cumbres Observatory Global Telescope Network Incorporated, the National Central University of Taiwan, the Space Telescope Science Institute, the National Aeronautics and Space Administration under Grant No. NNX08AR22G issued through the Planetary Science Division of the NASA Science Mission Directorate, the National Science Foundation under Grant No. AST-1238877, the University of Maryland, and Eotvos Lorand University (ELTE).
Operation of the Pan-STARRS1 telescope is supported by the National Aeronautics and Space Administration under Grant No. NNX12AR65G and Grant No. NNX14AM74G issued through the NEO Observation Program.

Funding for the SDSS and SDSS-II has been provided by the Alfred P. Sloan Foundation, the Participating Institutions, the National Science Foundation, the U.S. Department of Energy, the National Aeronautics and Space Administration, the Japanese Monbukagakusho, the Max Planck Society, and the Higher Education Funding Council for England. The SDSS Web Site is http://www.sdss.org/.
The SDSS is managed by the Astrophysical Research Consortium for the Participating Institutions. The Participating Institutions are the American Museum of Natural History, Astrophysical Institute Potsdam, University of Basel, University of Cambridge, Case Western Reserve University, University of Chicago, Drexel University, Fermilab, the Institute for Advanced Study, the Japan Participation Group, Johns Hopkins University, the Joint Institute for Nuclear Astrophysics, the Kavli Institute for Particle Astrophysics and Cosmology, the Korean Scientist Group, the Chinese Academy of Sciences (LAMOST), Los Alamos National Laboratory, the Max-Planck-Institute for Astronomy (MPIA), the Max-Planck-Institute for Astrophysics (MPA), New Mexico State University, Ohio State University, University of Pittsburgh, University of Portsmouth, Princeton University, the United States Naval Observatory, and the University of Washington.

The national facility capability for SkyMapper has been funded through ARC LIEF grant LE130100104 from the Australian Research Council, awarded to the University of Sydney, the Australian National University, Swinburne University of Technology, the University of Queensland, the University of Western Australia, the University of Melbourne, Curtin University of Technology, Monash University and the Australian Astronomical Observatory. SkyMapper is owned and operated by The Australian National University's Research School of Astronomy and Astrophysics. The survey data were processed and provided by the SkyMapper Team at ANU. The SkyMapper node of the All-Sky Virtual Observatory (ASVO) is hosted at the National Computational Infrastructure (NCI). Development and support the SkyMapper node of the ASVO has been funded in part by Astronomy Australia Limited (AAL) and the Australian Government through the Commonwealth's Education Investment Fund (EIF) and National Collaborative Research Infrastructure Strategy (NCRIS), particularly the National eResearch Collaboration Tools and Resources (NeCTAR) and the Australian National Data Service Projects (ANDS).

%


\facilities{ARC (DIS), Nayuta (LISS, NIC), Kiso (KWFC), Mayall (KOSMOS), Blanco (DECam), WISE, PS1}
\software{IRAF, Astropy \citep{ast18}, SExtractor \citep{ber96}, HOTPANTS \citep{bec15}, barycorrpy \citep{kan18}, MPFIT \citep{mar09}, GALFIT \citep{pen02}}


\bibliography{./kiss15s}

\appendix

\section{Instrumental line profiles of ARC3.5-m/DIS red-arm spectra}
\label{append:inst_profile}

\begin{figure}[tbp]
\center{
\includegraphics[clip, width=3.4in]{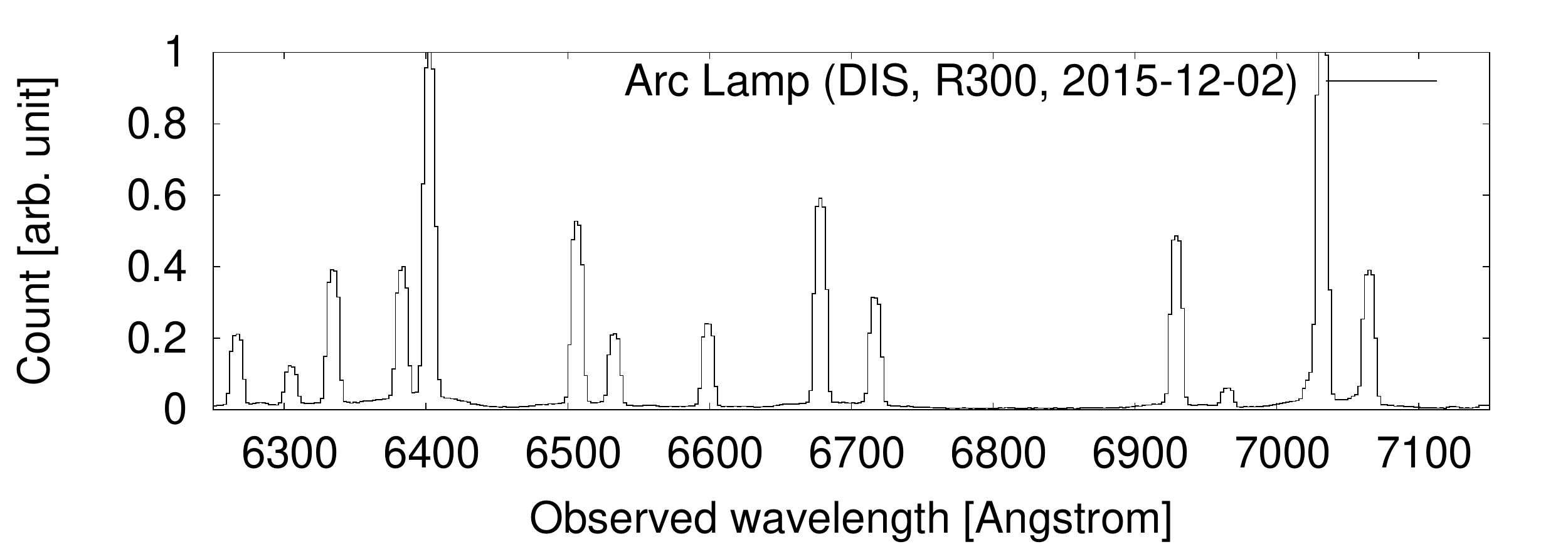}
\includegraphics[clip, width=3.4in]{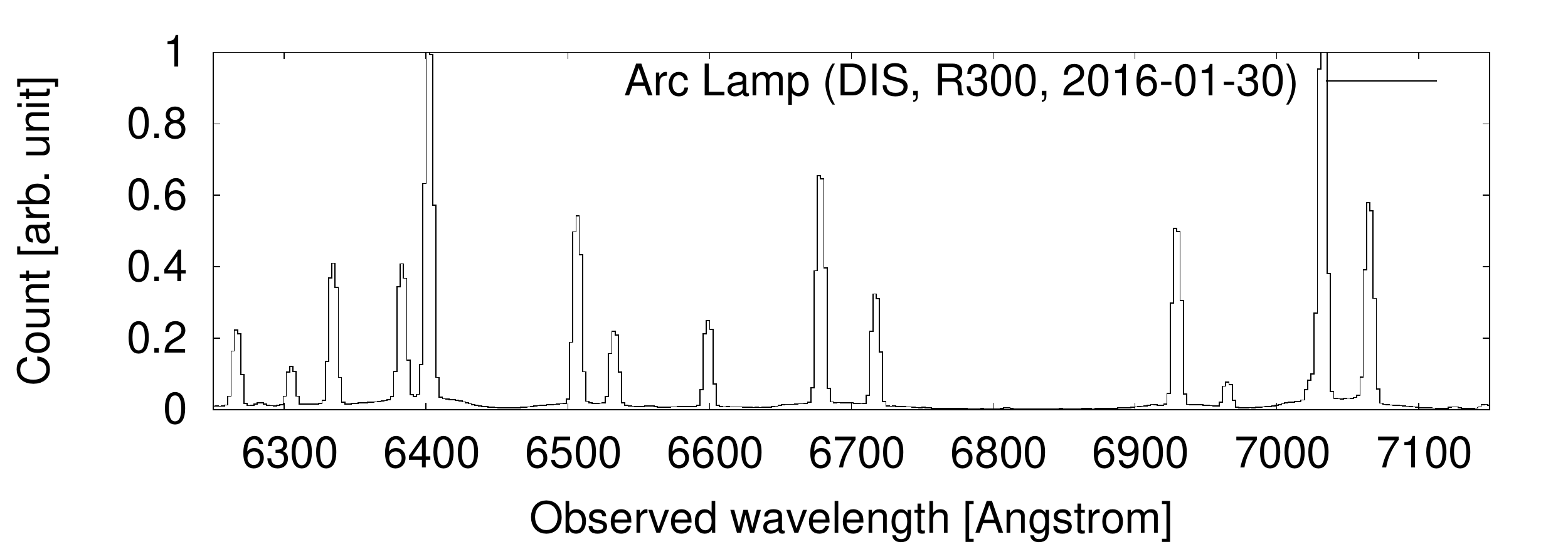}
\includegraphics[clip, width=3.4in]{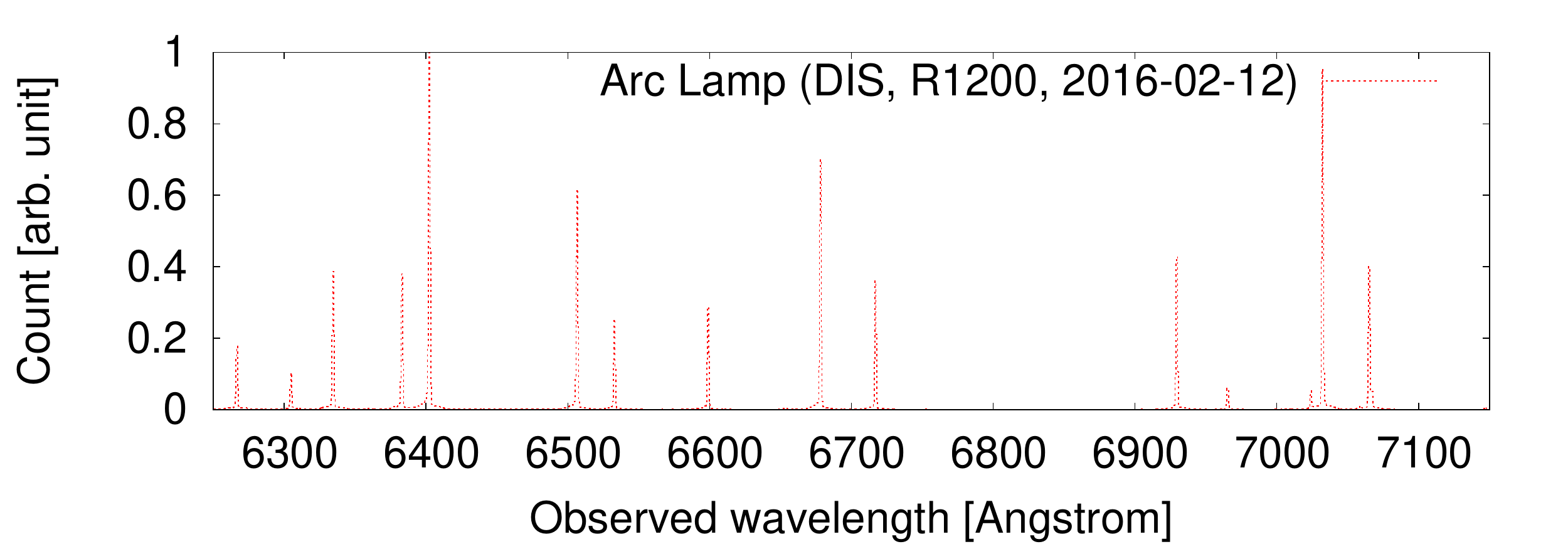}
\includegraphics[clip, width=3.4in]{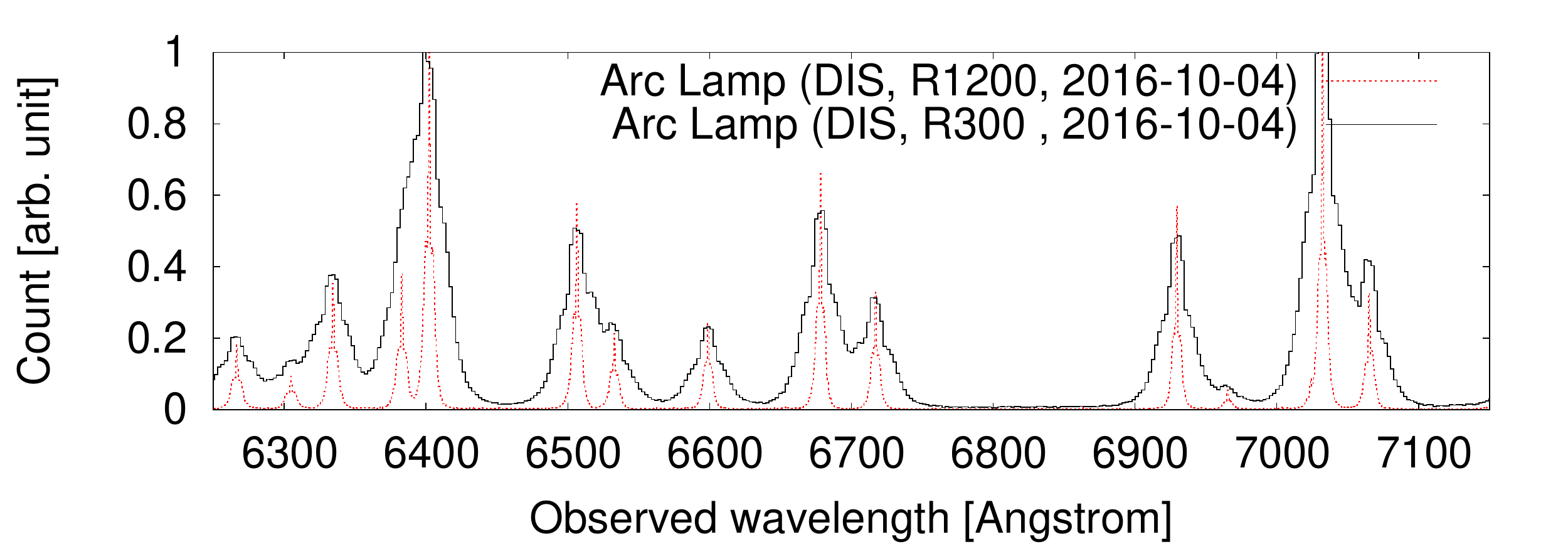}
}
 \caption{ARC3.5-m/DIS red-arm arc lamp spectra at $\lambda=6250-6750$~\AA\ obtained at different epochs.
 On 2016 October 4, the instrumental profile was significantly contaminated by a broad base, which may be due to improper setting of the internal focus of the DIS spectrograph.
 }
 \label{fig:arc_lamp}
\end{figure}

\begin{figure}[tbp]
\center{
\includegraphics[clip, width=3.4in]{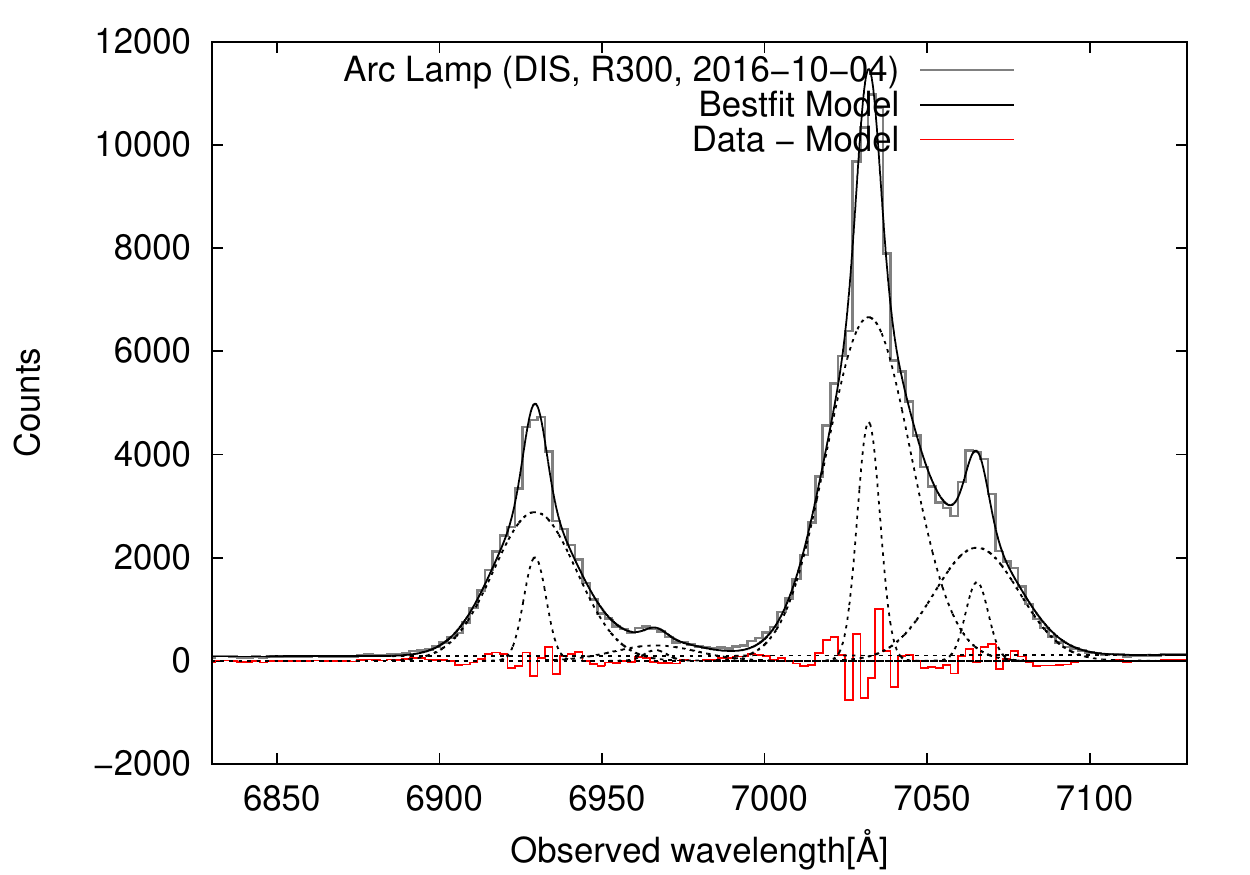}
\includegraphics[clip, width=3.4in]{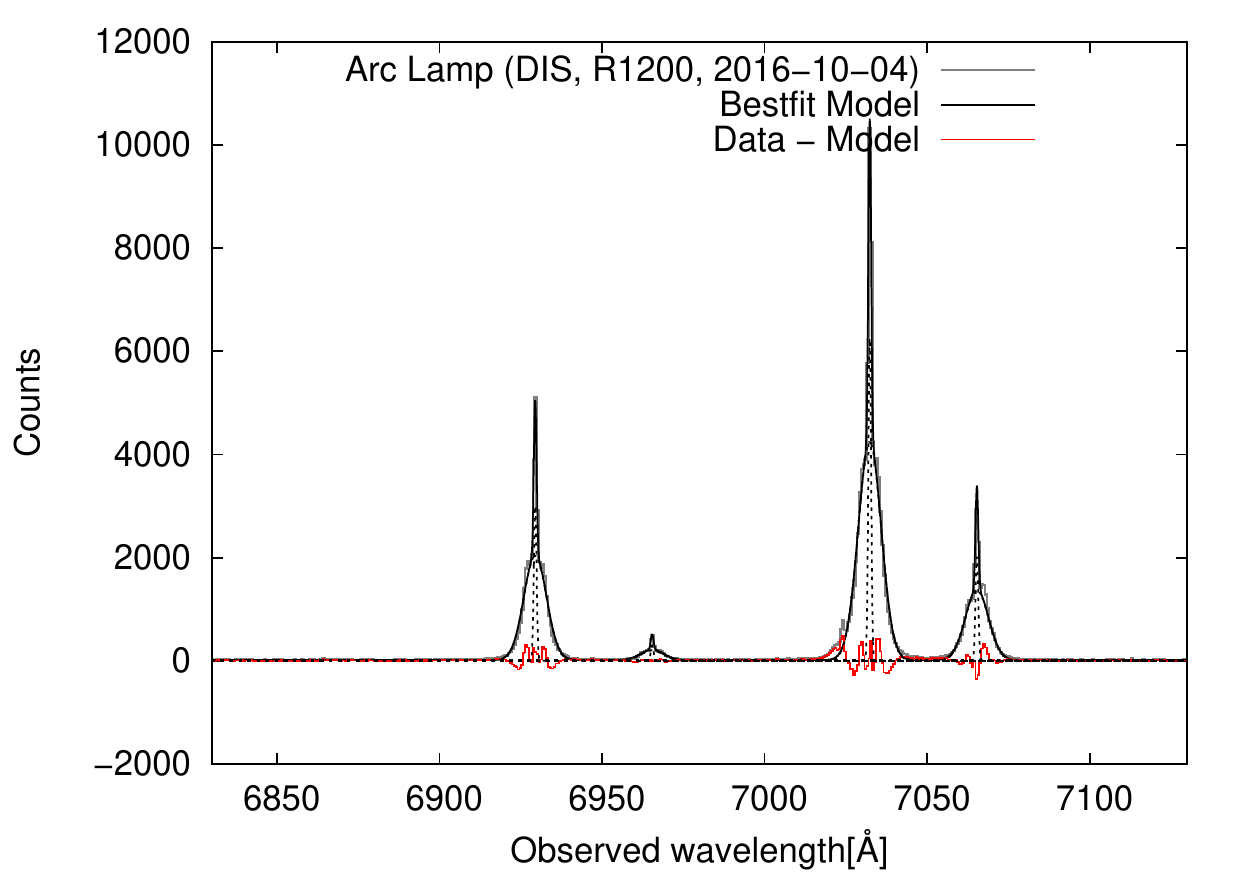}
}
 \caption{Double-Gaussian spectral profile model for ARC3.5-m/DIS red-arm arc lamp R300 and R1200 spectra obtained on 2016 October 4. The line width of the narrow cores is consistent with the instrumental resolution. The additional broad component is probably due to the defocusing of the spectrograph.
 }
 \label{fig:arc_lamp_profile}
\end{figure}

Figure~\ref{fig:arc_lamp} presents the ARC 3.5-m/DIS red-arm R300 and R1200 arc lamp spectra obtained on the same nights with KISS15s observations.
The instrumental line profile at the first three epochs is well reproduced by a single Gaussian function; $\sigma_\text{inst}$ obtained by a single Gaussian fitting for these arc lamp spectra at $\lambda_{\text{obs}} = 6900-7100$~\AA\ are listed in Table~\ref{obslog_dis}.

In the last epoch (2016 October 4), the instrumental profiles of both R300 and R1200 were significantly affected by an additional broad-width wing component, although the instrumental settings were nearly the same as those used in previous observations.
This peculiar instrumental profile is not seen in the DIS blue-arm data (Table~\ref{obslog_dis_hbeta}). 
DIS red arm spectra sometimes exhibit complex instrumental broadening profiles \citep{bal16}.
Figure~\ref{fig:arc_lamp_profile} shows the results of a double Gaussian fitting for R300 and R1200 arc lamp spectra obtained on 2016 October 4.
The line profile is reasonably reproduced by the double Gaussian model, in which the narrow core shows widths that are consistent with previous observations; the widths of the broad wing are several times broader than the narrow core.
This double Gaussian profile may be due to a slight defocusing of the internal focus of the red-arm spectrograph \footnote{\href{https://www.apo.nmsu.edu/arc35m/Instruments/DIS/\#5}{https://www.apo.nmsu.edu/arc35m/Instruments/DIS/\#5}}.
Table~\ref{obslog_dis} shows the widths of both the narrow and broad components.
The broad component dominates the flux of the total line profile; therefore, we assumed $\sigma_\text{inst} = \sigma_\text{inst} (\text{b})$ for the spectra obtained on 2016 October 4.

Although the complex instrumental line profile makes the narrow component fitting uncertain, it does not affect the fitting results for the intermediate and broad line components because the intrinsic velocity widths of these components are at least several times larger than that associated with instrumental broadening.

\section{Profile fitting for the H$\beta$ line}
\label{append:hbeta}

\begin{figure*}[tbp]
\center{
\includegraphics[clip, width=3.2in]{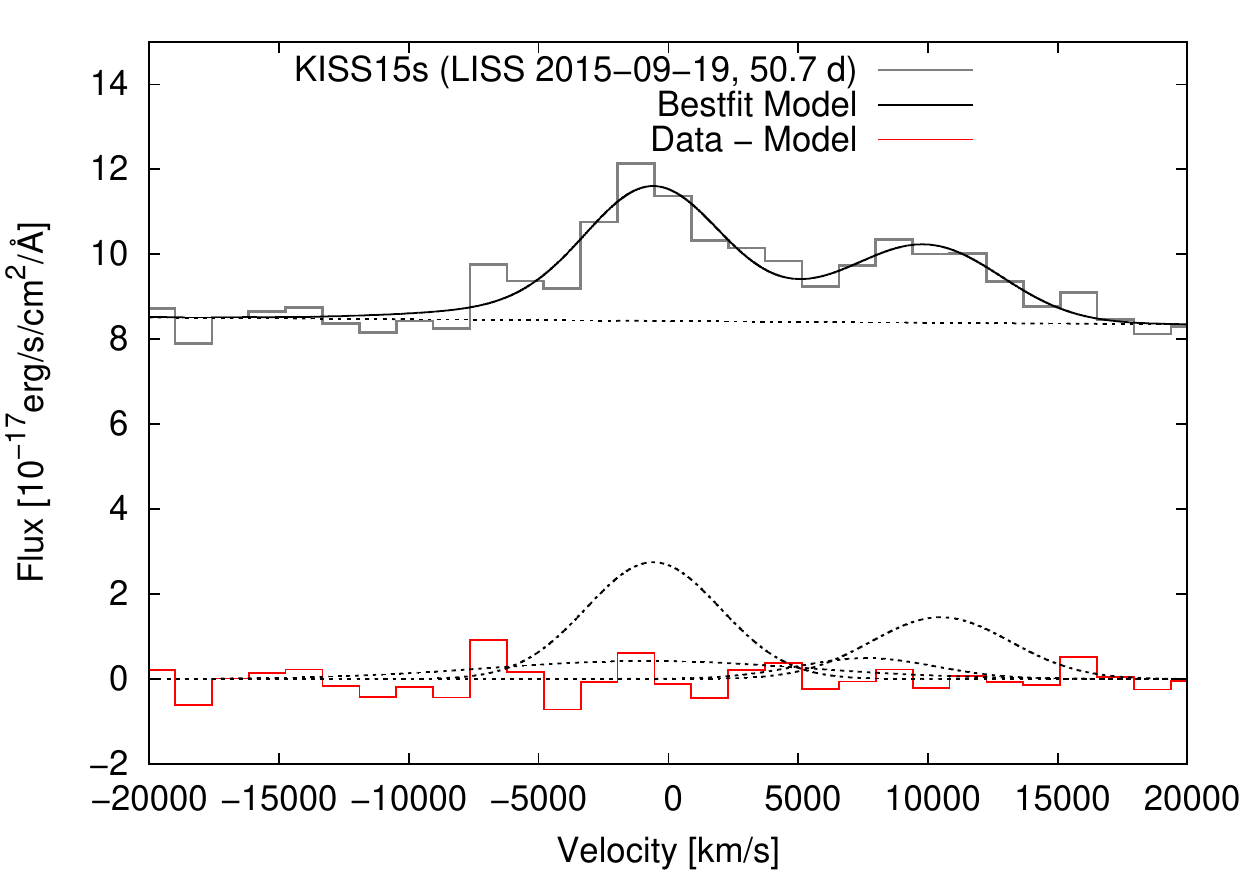}
\includegraphics[clip, width=3.2in]{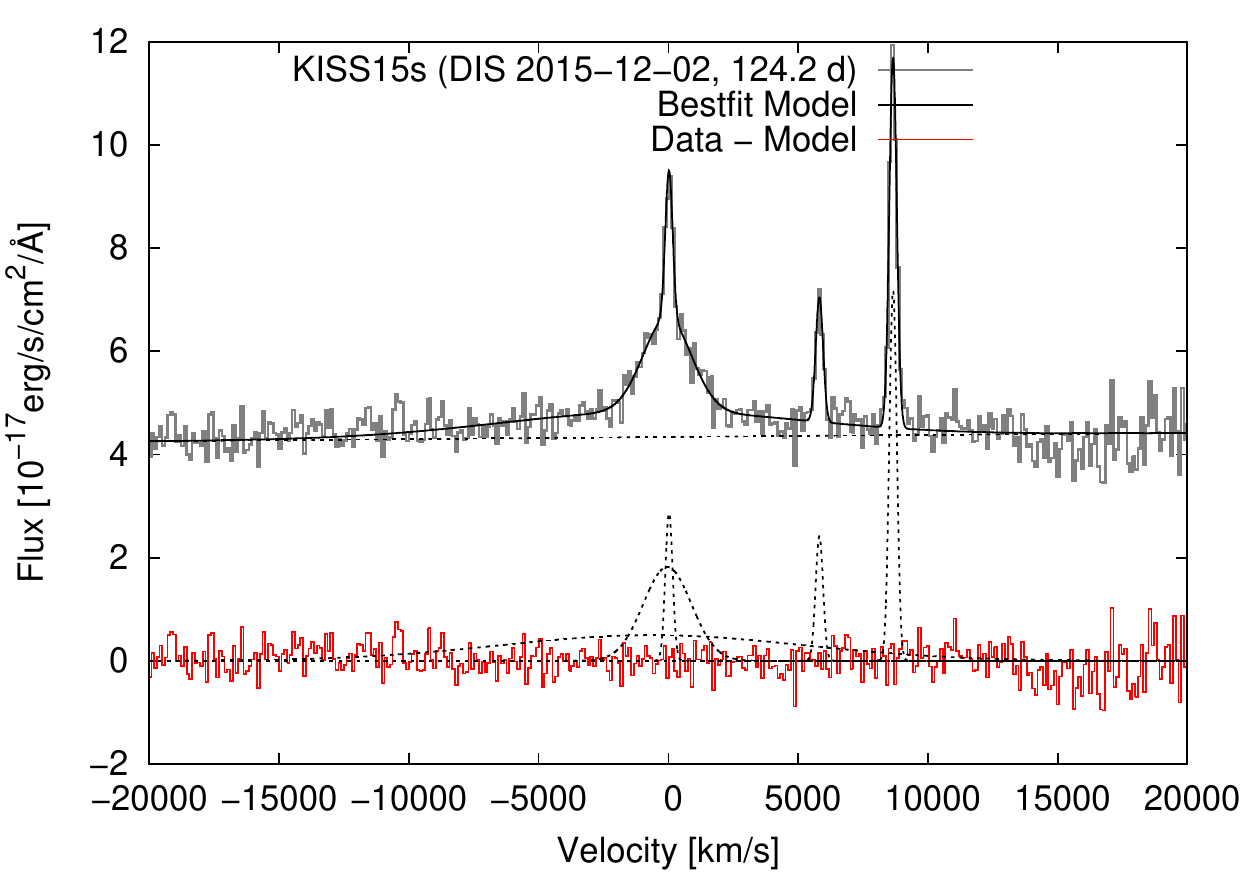}
\includegraphics[clip, width=3.2in]{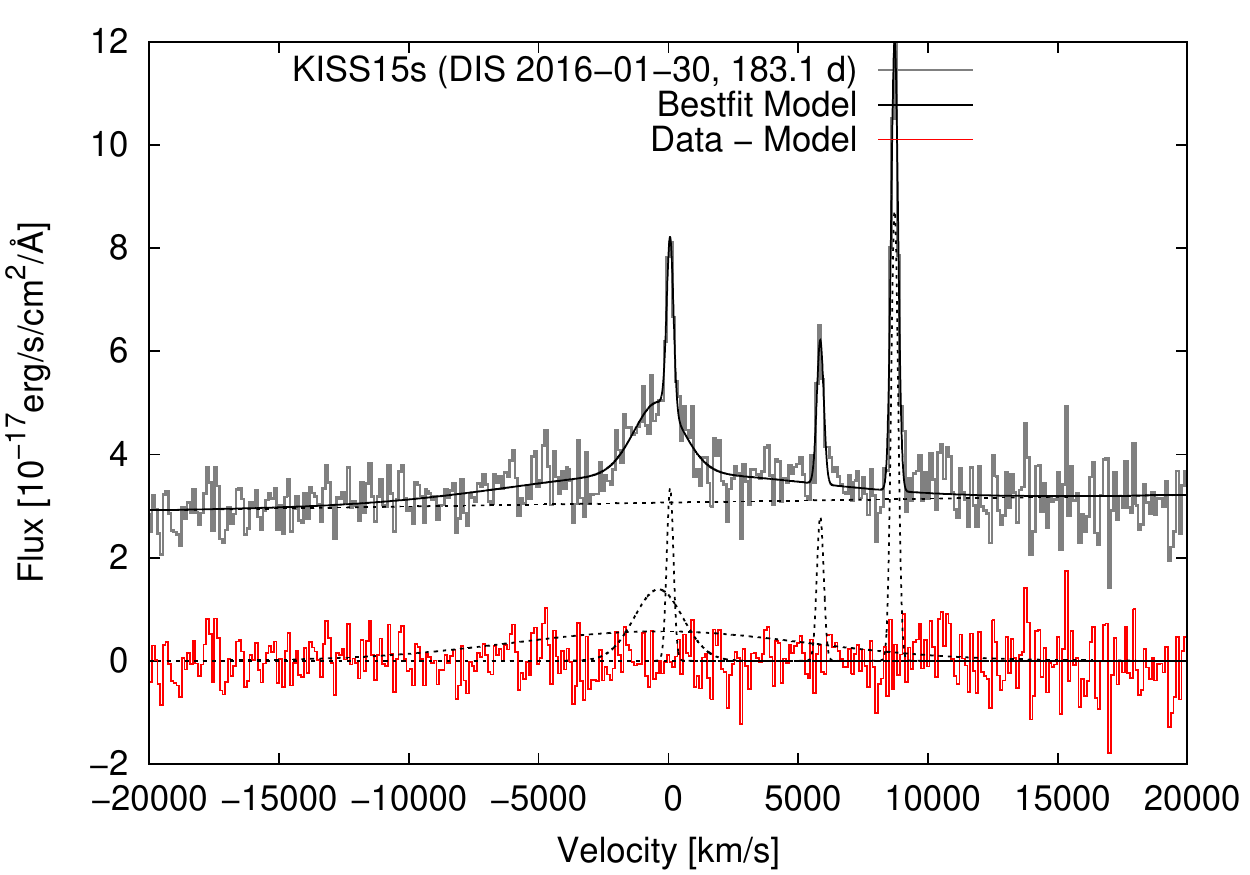}
\includegraphics[clip, width=3.2in]{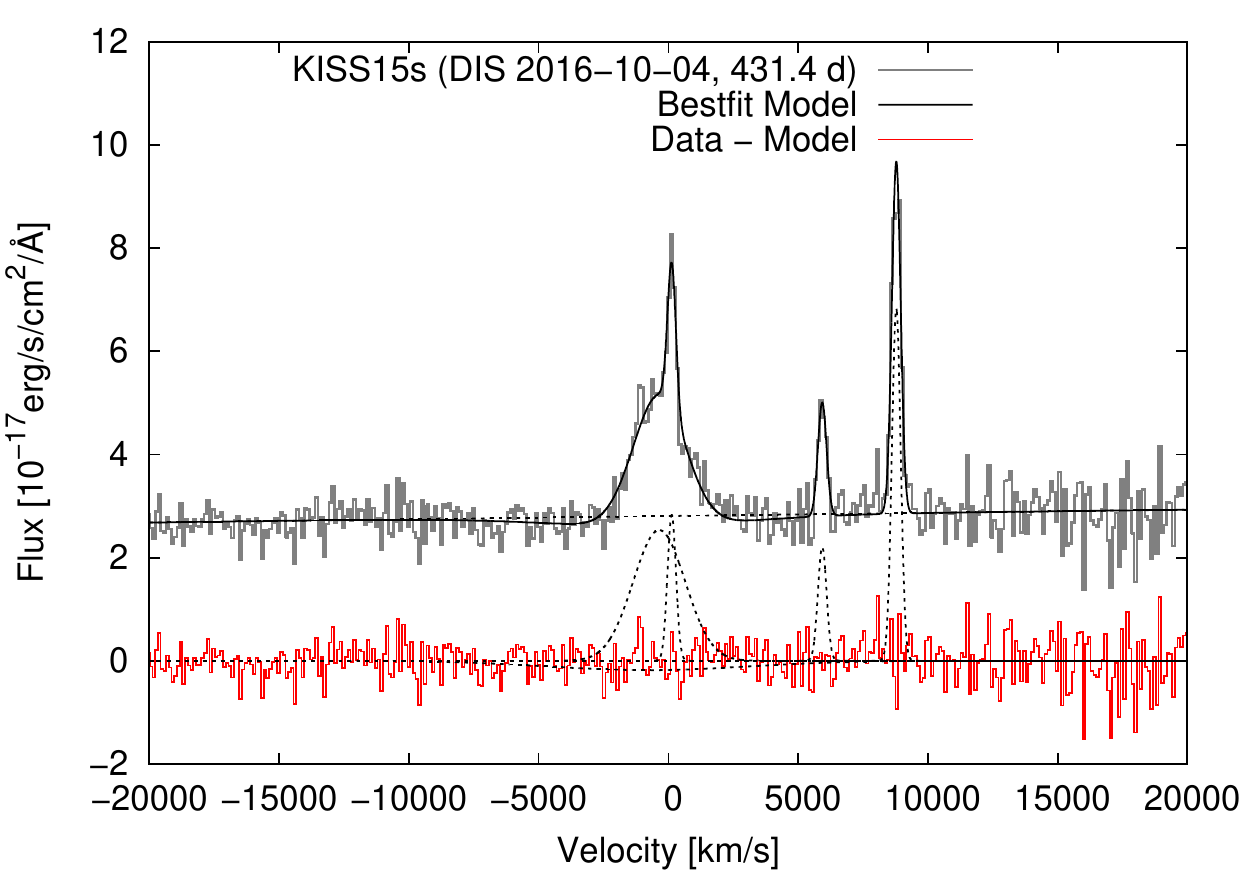}
\includegraphics[clip, width=3.2in]{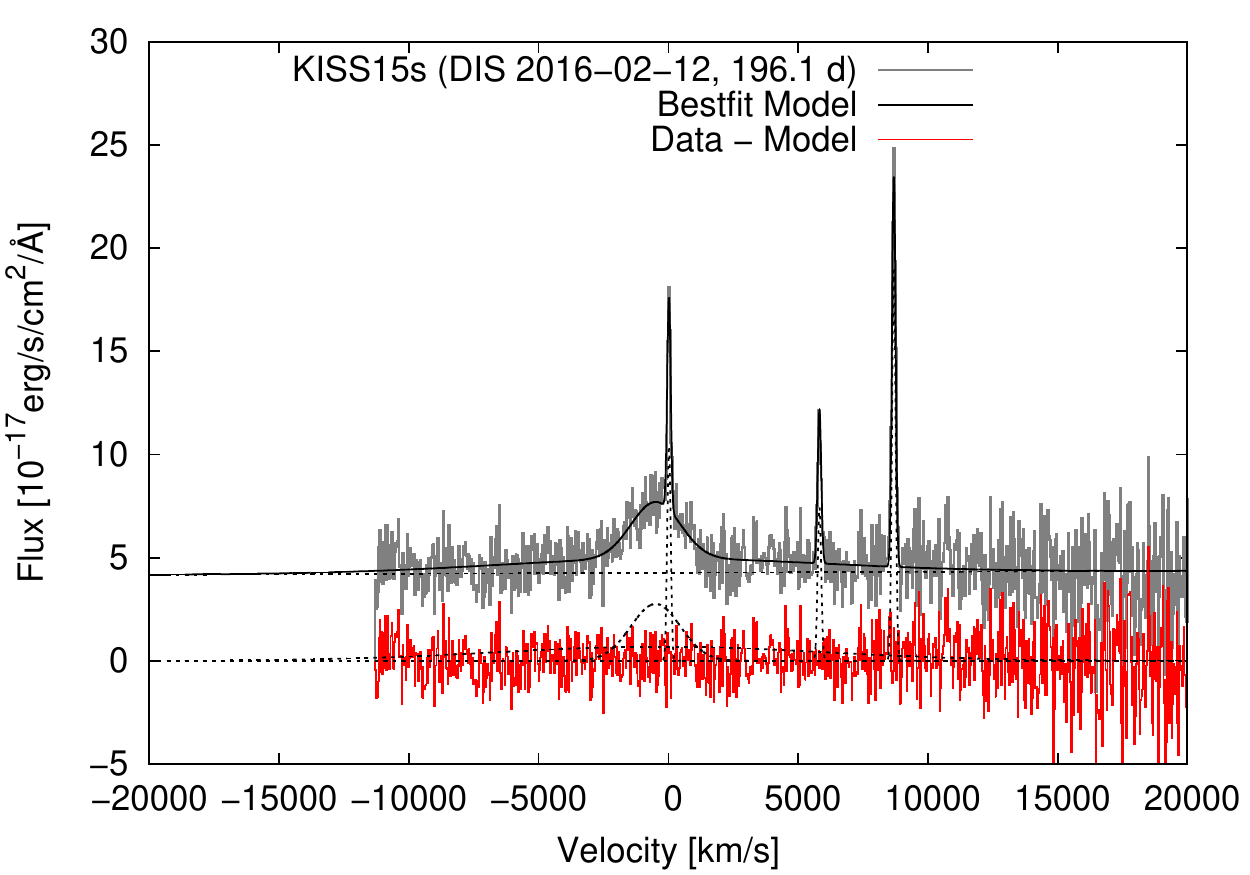}
\includegraphics[clip, width=3.2in]{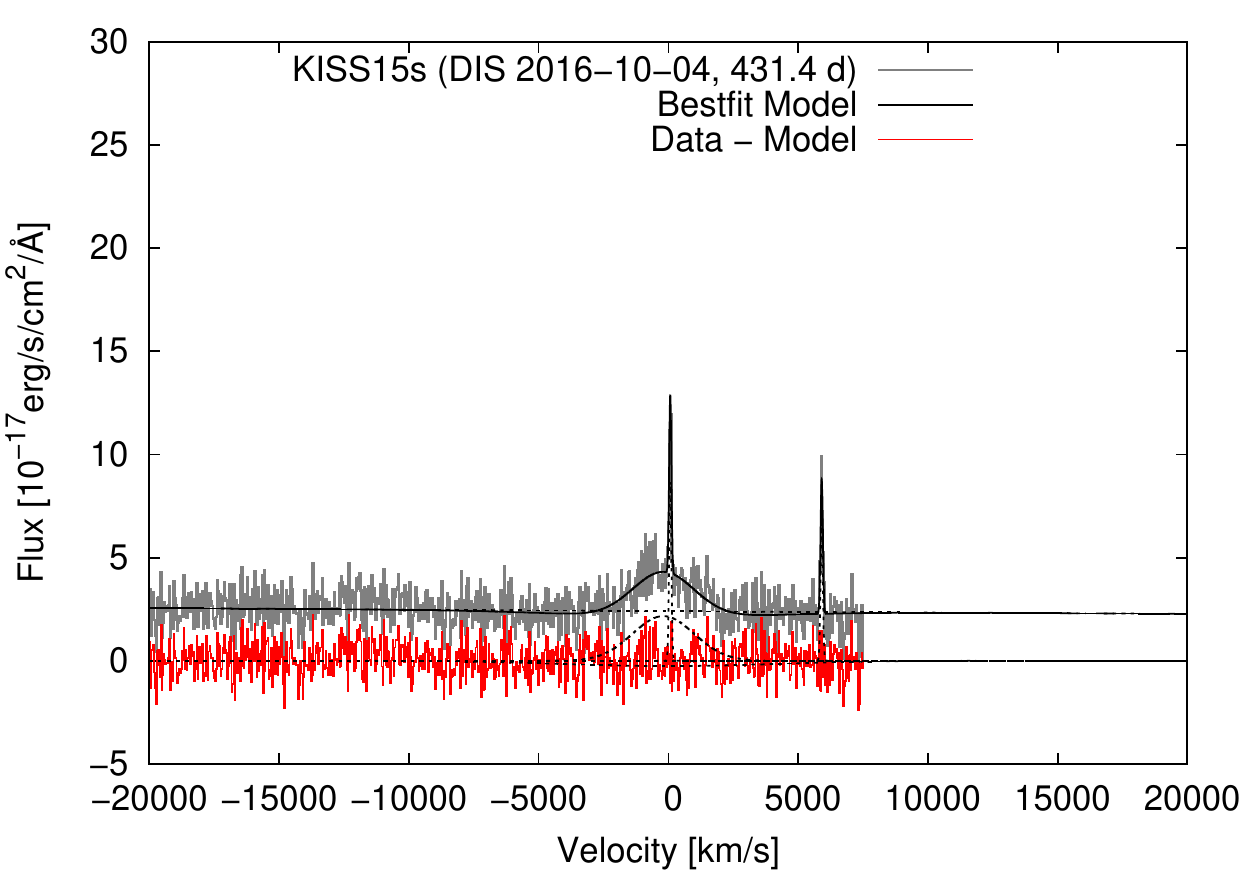}
}
 \caption{Same as Figure~\ref{fig:spec_ana}, but for H$\beta$+[\ion{O}{3}] emission line profiles on 2015 September 19 (Nayuta/LISS), 2015 December 2, 2016 January 30, and 2016 October 4 (ARC3.5-m/DIS blue-arm low-resolution).
The bottom two panels are for high-resolution DIS spectra.
 }
 \label{fig:spec_ana_hbeta}
\end{figure*}

\begin{figure}[tbp]
\center{
\includegraphics[clip, width=3.4in]{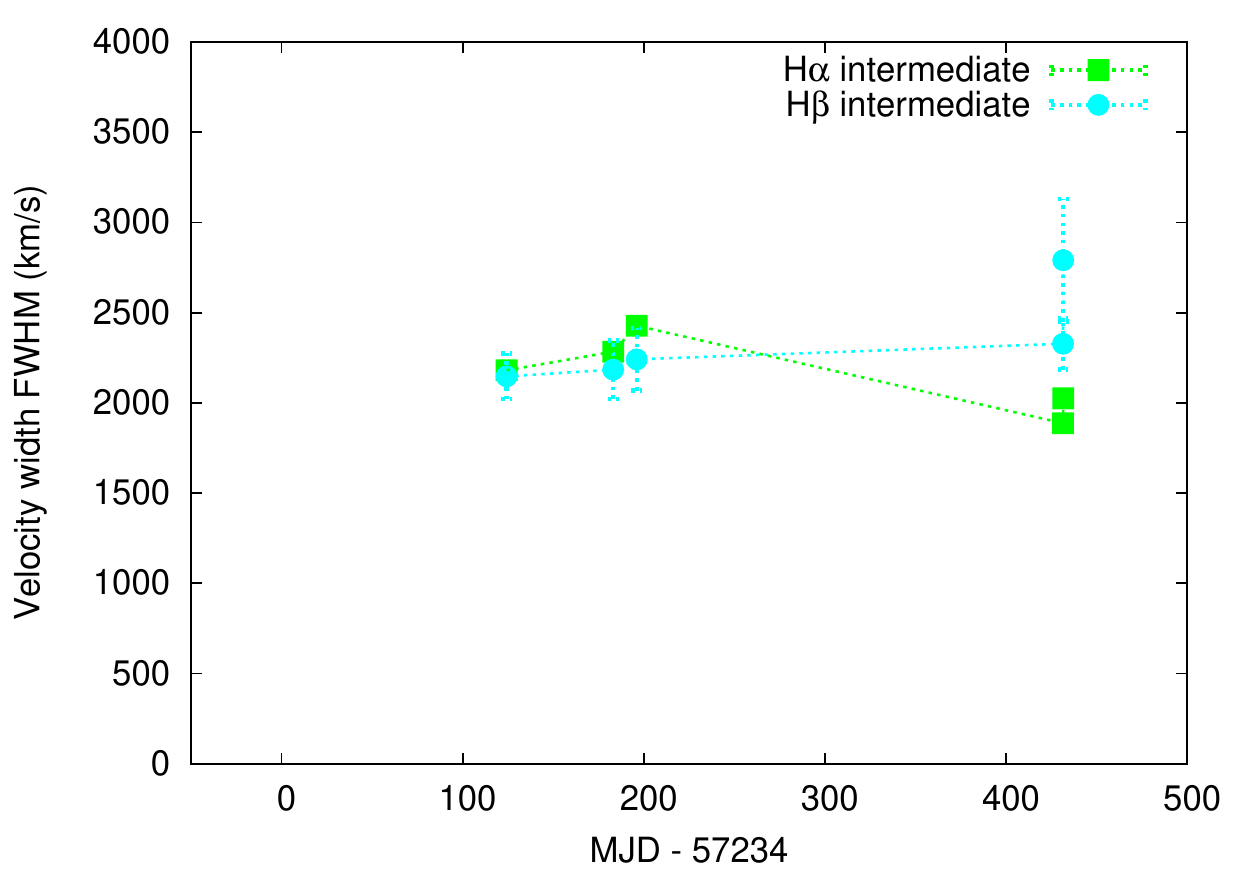}
\includegraphics[clip, width=3.4in]{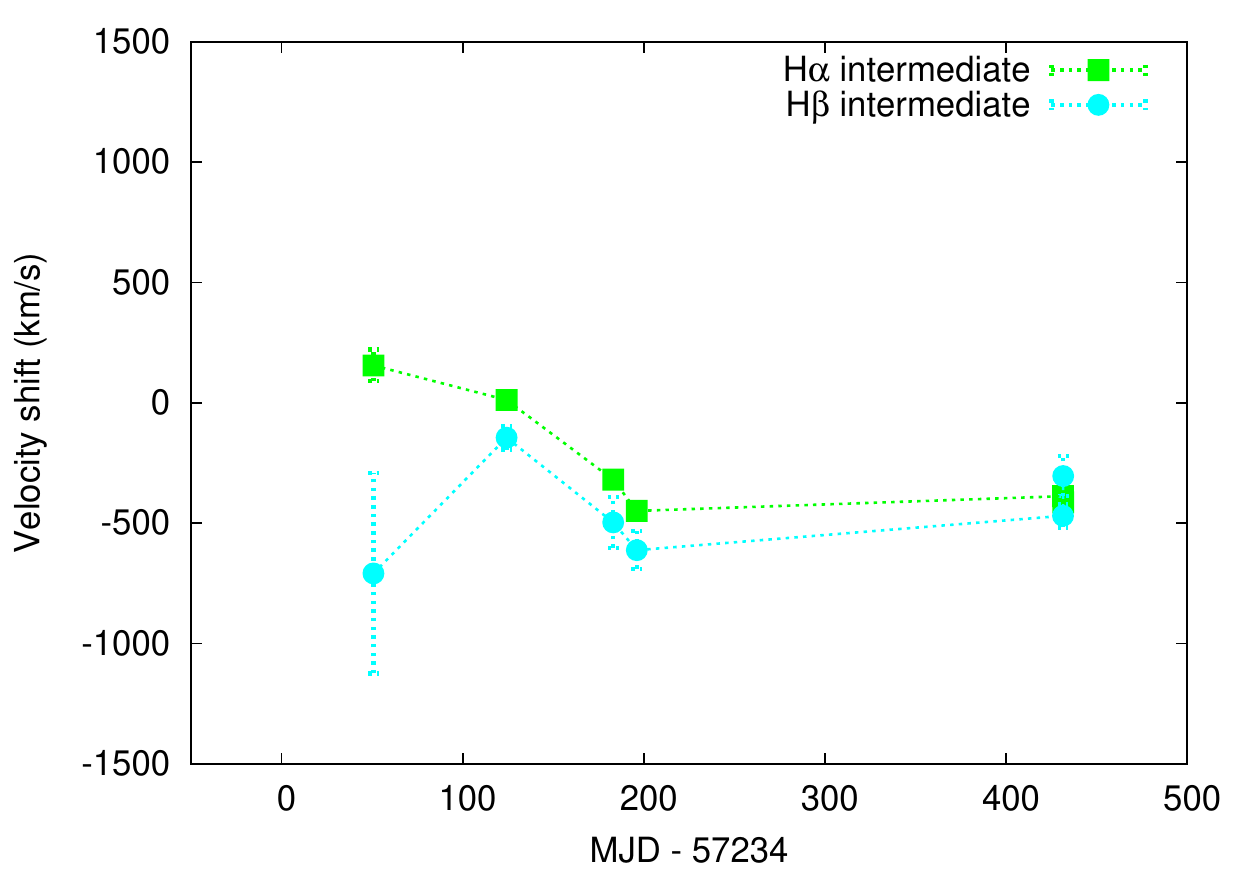}
}
 \caption{
Comparisons of the rest-frame intrinsic velocity width and the velocity shift of the decomposed intermediate Gaussian component of H$\beta$ and H$\alpha$ line profiles of KISS15s, as a function of observed time.
The velocity width of the intermediate component is unresolved in the Nayuta/LISS and is omitted from the plot.
 }
 \label{fig:spec_lightcurves_1_hbeta}
\end{figure}

\begin{figure}[tbp]
\center{
\includegraphics[clip, width=3.4in]{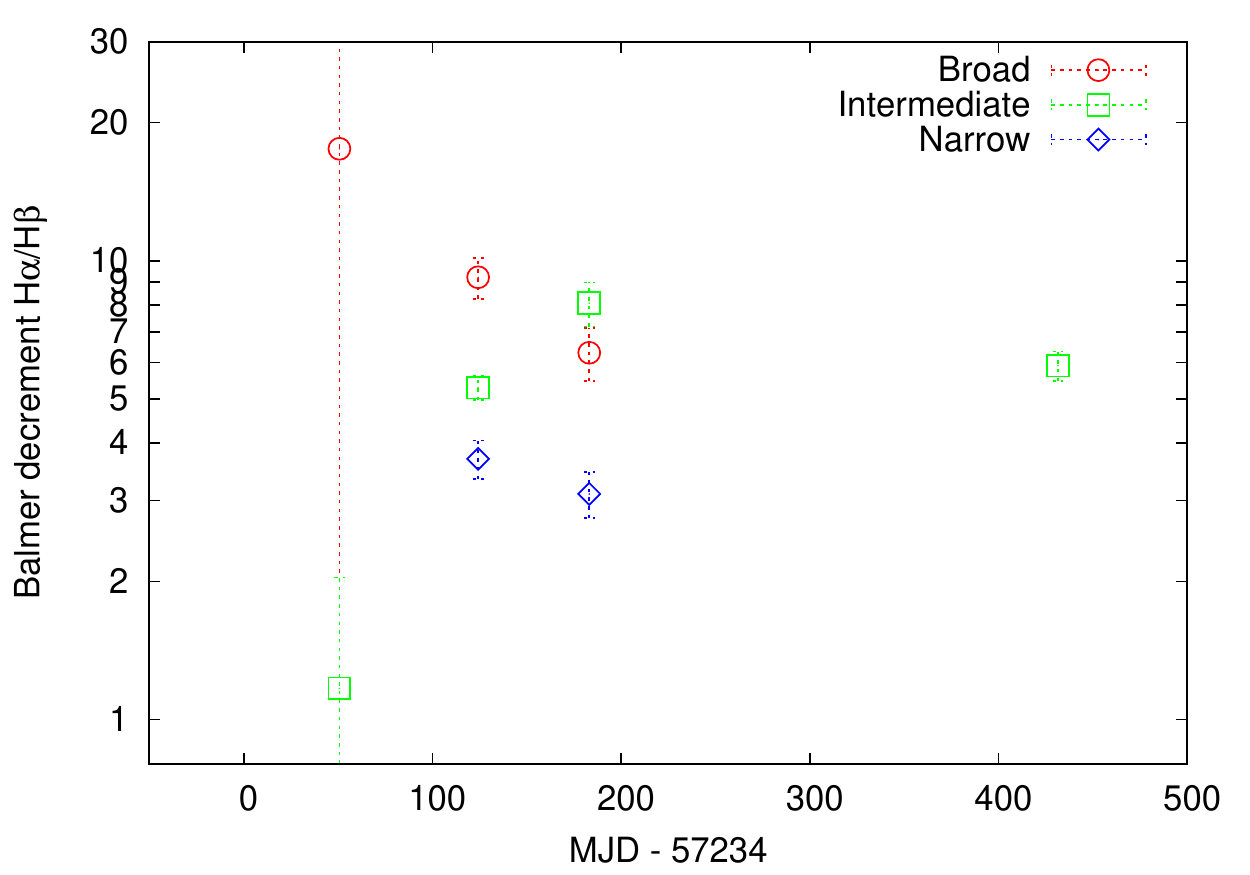}
}
 \caption{Balmer decrement calculated from the luminosity ratios H$\alpha$/H$\beta$ of the broad, intermediate, and narrow velocity width components.
 Only spectro-photometrically-calibrated LISS and low-resolution DIS data are used.
The host galaxy extinction of $E(B-V)_{\text{host}} = 0.6$~mag is corrected.
The broad H$\beta$ component is not detected in the final epoch DIS data, and the narrow H$\alpha$ component in the final epoch DIS data is significantly affected by complex instrumental broadening, as discussed in Section~\ref{append:inst_profile}; thus, these data are excluded from the plot.
 }
 \label{fig:balmer_decrement_1}
\end{figure}

\begin{deluxetable}{llcc}
\tablecolumns{5}
\tablewidth{0pc}
\tablecaption{Instrumental broadening $\sigma_{\text{inst}}$ in the H$\beta$ spectral region of Nayuta/LISS and ARC3.5-m/DIS blue-arm. \label{obslog_dis_hbeta}}
\tablehead{
\colhead{MJD} & \colhead{Date} & \colhead{Inst.-Grating} & \colhead{$\sigma_\text{inst}$}\\
\colhead{} & \colhead{} & \colhead{} & \colhead{(km~s${}^{-1}$)}
}
\startdata
57284.7 & 2015-09-19 & LISS-very low  & 1720 \\\hline
57358.2 & 2015-12-02 & DIS-B400/R300  & 173.5\\
57417.1 & 2015-07-31 & DIS-B400/R300  & 134.1\\
57665.4 & 2016-10-04 & DIS-B400/R300  & 178.5\\\hline
57430.1 & 2016-02-12 & DIS-B1200/R1200& 56.4 \\
57665.4 & 2016-10-04 & DIS-B1200/R1200& 35.5 \\
\enddata
\tablecomments{The velocity width of the instrumental line profile $\sigma_\text{inst}$ at the H$\beta$ wavelength region is evaluated by single Gaussian model fittings for arc lamp spectra in the wavelength ranges of $\lambda_{\text{obs}} = 5200-6000$~\AA\ (two lines) for LISS and  $\lambda_{\text{obs}} = 4600-5100$~\AA\ (three lines) for the DIS blue-arm.}
\end{deluxetable}

\begin{deluxetable*}{lllll|ll}
\tablewidth{700pt}
\tabletypesize{\scriptsize}
\tablecaption{Best-fit Gaussian model for the H$\beta$ line profile without the blue-shift excess component. \label{spectral_fit_hbeta_1}}
\tablehead{
\colhead{Parameter name} & \colhead{LISS} & \colhead{DIS low 1}  & \colhead{DIS low 2} & \colhead{DIS low 3} & \colhead{DIS high 1} & \colhead{DIS high 2}\\
\colhead{Date} & \colhead{2015-09-19} & \colhead{2015-12-02}  & \colhead{2016-01-30} & \colhead{2016-10-04} & \colhead{2016-02-12} & \colhead{2016-10-04}\\
\colhead{Days since discovery} & \colhead{50.7 d} & \colhead{124.2 d}  & \colhead{183.1 d} & \colhead{431.4 d} & \colhead{196.1 d} & \colhead{431.4 d}
}
\startdata
Norm. ($10^{-17}$ erg s${}^{-1}$ cm${}^{-2}$ \AA${}^{-1}$)                      &    8.42 $\pm$     0.17 &     4.34 $\pm$      0.02 &     3.07 $\pm$      0.03 &     2.81 $\pm$      0.02   & 	    4.27 $\pm$     0.10 &     2.40 $\pm$      0.10 \\
Index                       &   -0.14 $\pm$     0.38 &     0.27 $\pm$      0.08 &     0.71 $\pm$      0.17 &     0.64 $\pm$      0.14   & 	    0.33 $\pm$     0.38 &    -0.84 $\pm$      0.61 \\
Flux$_{b}$ ($10^{-17}$ erg s${}^{-1}$ cm${}^{-2}$)                  &  112.66 $\pm$   319.06 &   126.76 $\pm$     13.01 &   142.20 $\pm$     18.66 &   {\bf -30.90 $\pm$     13.68}   & 	  187.99 $\pm$    48.25 &   {\bf -46.10 $\pm$     43.01} \\
$\sigma_{\text{obs}, i}$ (km s$^{-1}$)    & 2686.27 $\pm$  1116.26 &   961.88 $\pm$     56.06 &   971.78 $\pm$     71.37 &  1041.09 $\pm$     63.18   & 	  989.29 $\pm$    76.22 &  1230.31 $\pm$    149.90 \\
${\lambda}_{\text{obs}, i}$ (\AA) & 5034.68 $\pm$     7.00 &  5044.17 $\pm$      0.83 &  5038.24 $\pm$      1.77 &  5038.71 $\pm$      0.82   & 	 5036.31 $\pm$     1.32 &  5041.49 $\pm$      1.41 \\
Flux$_{i}$ ($10^{-17}$ erg s${}^{-1}$ cm${}^{-2}$)                  &  310.68 $\pm$   229.86 &    73.74 $\pm$      4.45 &    56.67 $\pm$      6.18 &   111.15 $\pm$      7.89   & 	  114.68 $\pm$    10.25 &   112.13 $\pm$     18.27 \\
$\sigma_{\text{obs}, n}$ (km s$^{-1}$)    &     \nodata &   145.20 $\pm$      5.83 &   134.40 $\pm$      7.27 &   169.42 $\pm$      6.27   & 	   69.97 $\pm$     3.71 &    45.23 $\pm$      3.54 \\
${\lambda}_{\text{obs}, n}$ (\AA) &    \nodata &  5045.32 $\pm$      0.24 &  5046.05 $\pm$      0.27 &  5047.15 $\pm$      0.23   & 	 5045.29 $\pm$     0.08 &  5046.15 $\pm$      0.08 \\
Flux$_{n}$ ($10^{-17}$ erg s${}^{-1}$ cm${}^{-2}$)                  &     \nodata &    17.52 $\pm$      1.60 &    18.97 $\pm$      2.04 &    20.21 $\pm$      1.79   & 	   30.37 $\pm$     2.09 &    16.51 $\pm$      1.45 \\
${\lambda}_{\text{obs}, \text{[OIII]}4959}$ (\AA) & 5178.06 $\pm$    14.71 &  5146.52 $\pm$      0.26 &  5147.11 $\pm$      0.31 &  5148.44 $\pm$      0.33   & 	 5146.56 $\pm$     0.14 &  5148.08 $\pm$      0.10 \\
Flux$_{\text{[OIII]}4959}$ ($10^{-17}$ erg s${}^{-1}$ cm${}^{-2}$)                  &    56.84 $\pm$    27.57 &    15.08 $\pm$      1.22 &    16.16 $\pm$      1.74 &    16.06 $\pm$      1.53   & 	   22.37 $\pm$     1.94 &    12.70 $\pm$      1.35 \\
${\lambda}_{\text{obs}, \text{[OIII]}5007}$ (\AA) &    \nodata &  5196.09 $\pm$      0.11 &  5197.15 $\pm$      0.15 &  5198.26 $\pm$      0.16   & 	 5196.60 $\pm$     0.07 &  \nodata \\
Flux$_{\text{[OIII]}5007}$ ($10^{-17}$ erg s${}^{-1}$ cm${}^{-2}$)                  &    \nodata &    45.23 $\pm$      1.70 &    50.94 $\pm$      2.67 &    50.25 $\pm$      2.18   & 	   57.48 $\pm$     2.85 &     \nodata \\
\enddata
\tablecomments{
The fitting parameters are a normalization and a spectral index of the power-law continuum, observed-frame velocity widths, observed-frame central wavelengths, and integrated fluxes for the broad, intermediate, and narrow components.
 However, note that the line width and the central wavelength of the broad component are fixed to those constrained from the H$\alpha$ line profile fitting (Table~\ref{spectral_fit_2}).
The H$\beta$ and [\ion{O}{3}]4959/5007 narrow components are assumed to have the same line width. 
The broad component is undetected in late time spectra, as indicated by bold fonts.
The observation epochs of LISS, DIS low 1, DIS low 2, DIS low 3, DIS high 1, and DIS high 2 are MJD = 57284.7, 57358.2, 57417.1, 57665.4, 57430.1, and 57665.4, respectively.
The DIS high-resolution spectra are not spectro-photometrically calibrated; thus, the flux values suffer from absolute flux calibration uncertainties.}
\end{deluxetable*}

The observed spectra in the wavelength range of the H$\beta$ emission line have a lower S/N ratio than the H$\alpha$ emission line, as shown in Figure~\ref{fig:spec_halphahbeta}.
The broad component is barely seen in the H$\beta$ profile, and the blue-shifted intermediate component clearly detected in H$\alpha$ emission line profiles (Section~\ref{balmerfit}) is not seen in the H$\beta$ profile.

We modeled the H$\beta$ emission line profiles as for the H$\alpha$ emission line analyzed in Section~\ref{balmerfit} (Table~\ref{spectral_fit_2}).
The observed-frame central wavelength and the velocity width of the broad H$\beta$ line were constrained from the broad H$\alpha$ wavelength fitting result for each spectrum at each epoch.
The blue-shifted intermediate component was not included in the fitting.
For the LISS spectrum, the unresolved intermediate H$\beta$ and narrow [\ion{O}{3}] lines were assumed to have the same width, and the flux ratio and the central wavelength ratio of the [\ion{O}{3}] doublet lines were fixed to 1:3 and 4959:5007, respectively.

Figure~\ref{fig:spec_ana_hbeta} shows the profiles of the H$\beta$+[\ion{O}{3}] emission of KISS15s.
The best-fit model spectra and the fitting residuals are also shown in the same figure, and the best-fit model parameters are tabulated in Table~\ref{spectral_fit_hbeta_1}.
Note that the broad H$\beta$ component is not detected ($<1\sigma$) in the last epoch DIS low-/high-resolution spectra, as indicated by the bold font in Table~\ref{spectral_fit_hbeta_1}.
Figure~\ref{fig:spec_lightcurves_1_hbeta} compares the temporal evolutions of the rest-frame line FWHM and the velocity shift of the intermediate H$\beta$ and H$\alpha$ lines, where the H$\beta$ line widths are converted into intrinsic values using Equation~\ref{fwhm_rest}; the instrumental broadening factors are listed in Table~\ref{obslog_dis_hbeta}.
The velocity widths and the velocity shifts of H$\alpha$ and H$\beta$ intermediate components are consistent with each other within the error designated, suggesting that the temporal variation in the Balmer emission line profiles are not solely due to the chromatic extinction of newly formed dust in the CDS region.

Figure~\ref{fig:balmer_decrement_1} presents the Balmer decrements of the broad, intermediate, and narrow H$\alpha$ and H$\beta$ components calculated from the fitting results (Tables~\ref{spectral_fit_2} and \ref{spectral_fit_hbeta_1}).
We can see that the Balmer decrements of the broad and intermediate components are larger than the canonical case B value of H$\alpha$/H$\beta$ $\simeq 3$ \citep{ost89} at least $\sim 100$~days since discovery, even after correcting for the putative SMC-like dust extinction of $E(B-V)_{\text{host}} = 0.6$~mag.
The steep Balmer decrement is often observed in SNe IIn.
For example, \cite{sta91} showed that SN~1988Z had a steep Balmer decrement of H$\alpha$/H$\beta \sim 7$ \citep[see also][]{fil89,smi17}, and the Balmer decrement of SN~2005ip is larger than 10 \citep{smi09}.
The steep Balmer decrements in broad and intermediate components suggest that the emission regions responsible for the broad and intermediate lines are extremely dense, $n_{e}>10^{8}-10^{13}$~cm${}^{-3}$ and are optically thick up to the Balmer series emission lines \citep[e.g.,][]{dra80,che94,smi09,lev14}.
Such high densities may imply the presence of dense clumps (or asphericity) in the ejecta-CSM regions emitting broad and intermediate lines \citep{chu94,smi09,che17}.
It should also be noted that the high densities, leading to efficient cooling, may explain the absence of broad and intermediate coronal lines in the KISS15s spectra \citep[e.g.,][]{smi09}.
However, we have to keep in mind that the Balmer decrement estimates are directly affected by profile model uncertainties and the relative spectroscopic flux calibration accuracies between the wavelength ranges of H$\alpha$ and H$\beta$ lines.
Thus, detailed numerical comparison between the observed values and theoretical predictions is difficult with the current dataset.

\end{document}